\documentclass[11pt,rmp,onecolumn,aps]{revtex4-2}
\usepackage{graphicx} % Required for inserting images
\usepackage{amsfonts,amsmath,amssymb}
\usepackage{braket}
\usepackage{bm}
\usepackage{bbm}
\usepackage{setspace}
\usepackage[margin=0.75in]{geometry}
\usepackage{physics}
\usepackage{mathrsfs}
\usepackage[colorlinks, allcolors=blue]{hyperref}
\usepackage{mathtools}
\usepackage[T1]{fontenc}
\usepackage{multirow}
\usepackage[normalem]{ulem}
\usepackage[caption=false]{subfig}

\makeatletter
\DeclareFontFamily{OMX}{MnSymbolE}{}
\DeclareSymbolFont{MnLargeSymbols}{OMX}{MnSymbolE}{m}{n}
\SetSymbolFont{MnLargeSymbols}{bold}{OMX}{MnSymbolE}{b}{n}
\DeclareFontShape{OMX}{MnSymbolE}{m}{n}{
    <-6>  MnSymbolE5
   <6-7>  MnSymbolE6
   <7-8>  MnSymbolE7
   <8-9>  MnSymbolE8
   <9-10> MnSymbolE9
  <10-12> MnSymbolE10
  <12->   MnSymbolE12
}{}
\DeclareFontShape{OMX}{MnSymbolE}{b}{n}{
    <-6>  MnSymbolE-Bold5
   <6-7>  MnSymbolE-Bold6
   <7-8>  MnSymbolE-Bold7
   <8-9>  MnSymbolE-Bold8
   <9-10> MnSymbolE-Bold9
  <10-12> MnSymbolE-Bold10
  <12->   MnSymbolE-Bold12
}{}

\let\llangle\@undefined
\let\rrangle\@undefined
\DeclareMathDelimiter{\llangle}{\mathopen}%
                     {MnLargeSymbols}{'164}{MnLargeSymbols}{'164}
\DeclareMathDelimiter{\rrangle}{\mathclose}%
                     {MnLargeSymbols}{'171}{MnLargeSymbols}{'171}
\makeatother

\usepackage{xcolor}

\begin{document}

\singlespacing
\title{Theory of spin qubits and the path to scalability} 

\newcommand{\Basel}{Department of Physics, University of Basel, Klingelbergstrasse 82, 4056 Basel, Switzerland}
\newcommand{\Fahd}{Physics Department, King Fahd University of Petroleum and Minerals, 31261, Dhahran, Saudi Arabia}
\newcommand{\QCenter}{Quantum Center, KFUPM, Dhahran, Saudi Arabia}
\newcommand{\Chair}{RDIA Chair in Quantum Computing}

\author{Z.~M.~McIntyre}\affiliation\Basel
\author{Abhikbrata Sarkar}\affiliation\Basel
\author{Daniel Loss}\affiliation\Basel\affiliation\Fahd\affiliation\QCenter\affiliation\Chair

\date{\today}

\begin{abstract}
Spin qubits have emerged as a leading platform for quantum information processing due to their long coherence times, small footprint, and compatibility with the existing semiconductor industry. We first provide an introduction to the different qubit implementations currently being investigated, including single electron-spin qubits, hole-spin qubits, donor qubits, and multispin encodings. We discuss how the confinement and strain present in semiconductor heterostructures produce addressable levels whose spin degree of freedom can be used to encode a qubit. A large emphasis is placed on reviewing the theoretical foundations and recent experimental demonstrations of proposed mechanisms for long-range coupling, including hybrid approaches based on circuit QED and Andreev qubits, as well as spin shuttling. Finally, we review a recent proposal for linking spin qubits using topological spin textures.

\end{abstract}

\maketitle

\tableofcontents

\section{Introduction}
More than a century ago, the Stern-Gerlach experiment demonstrated the quantization of angular momentum and established `spin' as an intrinsic degree of freedom. In the following decades, spin became central to the formulation of quantum mechanics. Nuclear magnetic resonance and electron spin resonance demonstrated that spin could be manipulated using radio frequency and microwave pulses. 

In the latter half of the twentieth century, the meteoric rise of the semiconductor industry enabled the fabrication of microscopic structures such as quantum wells and quantum dots, where electrons could be confined to nanoscale regions. The idea of quantum computation added a new layer of urgency. Beginning with Feynman’s insight that quantum systems could efficiently simulate other quantum behaviors \cite{feynman2018simulating}, the discovery of Shor’s algorithm \cite{shor1994algorithms} and the development of the DiVincenzo criteria \cite{divincenzo2000physical} motivated a practical search for material platforms suitable for quantum computing. A variety of qubit platforms have been developed in pursuit of this goal, including quantum dot spins~\cite{loss1998quantum,burkard2023semiconductor}, superconducting circuits~\cite{kjaergaard2019superconducting,blais2021circuit}, trapped ions~\cite{blatt2012quantum,d.bruzewicz2019trappedion}, neutral atoms~\cite{henriet2020quantum}, and photonic networks~\cite{kok2007linear}. Each platform has made remarkable progress in coherence, control, and scaling~\cite{ladd2010quantum}. 
%In the current noisy intermediate-scale quantum (NISQ) era, quantum computers usually host a few tens to hundreds of noisy qubits. 
However, the central question is which technology can scale to thousands or millions of reliable qubits equipped with quantum error correction \cite{knill1998resilient,preskill1998reliable,kitaev2003fault}. 

Semiconductor spin qubits~\cite{burkard2000spintronics,awschalom2002semiconductor,cerletti2005recipes,hanson2007spins, kloeffel2013prospects, a.zwanenburg2013silicon,chatterjee2021semiconductor, scappucci2021germanium, fang2023recent,burkard2023semiconductor} offer the most natural path to scalability. These low-disorder device platforms feature high mobility charge carriers and benefit from advanced fabrication facilities used in the classical semiconductor industry.  Spin qubits have been realized in various semiconductor platforms, including self-defined and gate-defined quantum dots. A large variation is exhibited in geometry (e.g., planar quantum dots, nanowires, etc.), type of primary carriers (e.g., electrons or holes), and the number of spins used to encode a single qubit. The underlying control mechanisms for many of these qubits largely follow the initial proposal: the Loss-DiVincenzo (LD) qubit. 

Early experimental efforts demonstrated single-shot spin readout for quantum dots in planar GaAs/AlGaAs heterostructures~\cite{elzerman2004singleshot,hanson2005single}. Such III-V semiconductor platforms found initial success in the initialization, control, and measurement of single- and two-electron spin qubits, but the presence of a nonzero nuclear magnetic moment implied a short coherence time. Later group-IV semiconductor platforms demonstrated excellent spin coherence with donor-bound and gate-defined electron-spin qubits owing to novel engineering advances. Highly sophisticated complementary metal-oxide-semiconductor (CMOS) devices, particularly in silicon- and germanium-based platforms, have since emerged as the front runners of spin quantum computing. The need to eventually scale up such devices has motivated the investigation of hybrid systems compatible with long-range coupling, including superconducting spin qubits and spin qubits coupled to superconducting microwave resonators. Another promising strategy for realizing high-connectivity devices involves physically displacing the qubits themselves, opening up new avenues not typically available in architectures restricted to nearest-neighbor coupling.  

\begin{figure}
    \centering
    \vspace{-0.1in}  \includegraphics[width=0.5\linewidth]{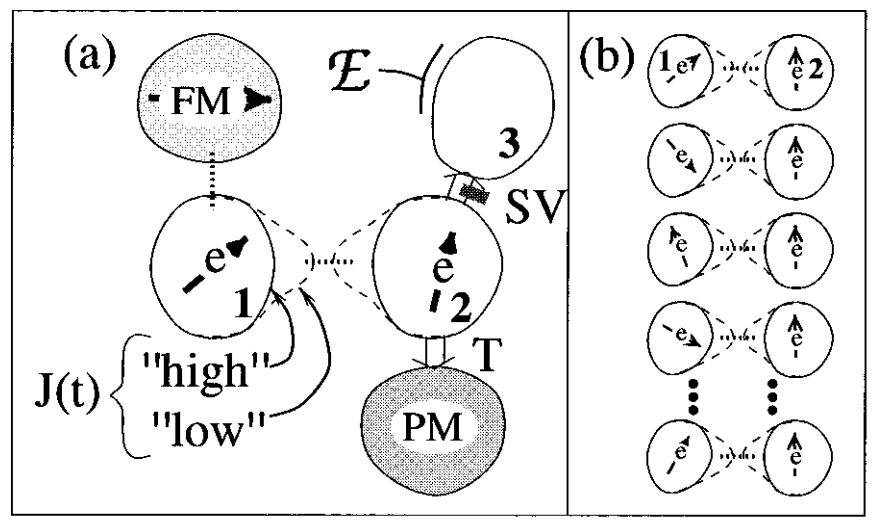}
    \caption{ The Loss-DiVincenzo proposal of a system of two coupled quantum dots \cite{loss1998quantum}: (a) The inter-dot tunneling is gate voltage controlled. In the `low' barrier case, virtual tunneling produces a time-dependent Heisenberg exchange $J(t)$. Single qubit rotations can be performed by hopping to an auxiliary ferromagnetic dot. Readout is achieved by tunneling to the paramagnetic dot; spin-dependent tunneling through ‘spin valve’ (SV) into dot 3 can lead to spin measurement via an electrometer E. (b) Proposed setup for SWAP-gate operations in an array of many non-interacting quantum dot pairs. Adapted from \textcite{loss1998quantum} with permission.}
    \label{fig:LDkane}
\end{figure}

A wealth of excellent review articles have documented the theoretical and experimental advances made over the last couple of decades in the field of spin qubits~\cite{hanson2007spins, kloeffel2013prospects, a.zwanenburg2013silicon,chatterjee2021semiconductor, scappucci2021germanium, burkard2023semiconductor}. Together, these works serve as a high-quality reservoir of information covering the underlying theoretical frameworks for electron- and hole-based quantum computing \cite{kloeffel2013prospects,fang2023recent,burkard2023semiconductor}; as well as mechanisms and techniques for electrical and magnetic spin manipulation \cite{morello2020donor,chatterjee2021semiconductor}, coherence properties \cite{paladino20141}, and quantitative benchmarking of diverse qubit platforms in terms of fidelities and characteristic timescales \cite{morello2020donor,stano2022review,chatterjee2021semiconductor}. The motivation for this review arises from the growing need to bridge these fundamental principles with recent advances that have the potential to push the field beyond the current era of noisy intermediate-scale quantum (NISQ) devices. In this article, we present a comprehensive overview of the various semiconductor spin qubit platforms and discuss recent experimental progress, novel concepts, and future prospects that may one day enable large-scale, fault-tolerant quantum information processing with semiconductor spin qubits.

\section{Spin qubit platforms in use}
Let us recall that the general description of a spin qubit in Dirac notation is $\left|\Psi\right\rangle=\alpha\left|\uparrow\right\rangle+\beta\left|\downarrow\right\rangle$, with the probabilities of the qubit being in the `up' ($\left|\uparrow\right\rangle$) and the `down' ($\left|\downarrow\right\rangle$) states summing to unity: $|\alpha|^2+|\beta|^2=1$. Thus, a qubit can be described mathematically as a normalized unit vector in a two-dimensional complex Hilbert space:
\begin{equation}\label{eq:Bloch}
\left|\Psi\right\rangle=\cos\frac{\theta}{2}\left|\uparrow\right\rangle+e^{i\phi}\sin\frac{\theta}{2}\left|\downarrow\right\rangle.
\end{equation}
Equation~\eqref{eq:Bloch} also implies that any (pure) qubit state can be represented by a point  $(\,\theta,\,\phi)$ on the surface of a three-dimensional unit sphere $S^2$, known as the {\it Bloch sphere}. Operations on a qubit state correspond to rotations on the Bloch sphere, which are described by the irreducible representation $\mathcal{D}_{1/2}$ of the full rotation group with the following matrices (Pauli matrices):
\begin{equation}
    \mathcal{I}{=}\begin{bmatrix}
        1 & 0\\
        0 & 1\\
    \end{bmatrix}\!;\sigma_x{=}\begin{bmatrix}
        0 & 1\\
        1 & 0\\
    \end{bmatrix}\!;\sigma_y{=}\begin{bmatrix}
        0 & -i\\
        i & 0\\
    \end{bmatrix}\!;\sigma_z{=}\begin{bmatrix}
        1 & 0\\
        0 & -1\\
    \end{bmatrix}.
\end{equation}

The physical realization of a spin qubit would involve energetically resolving the spin-$\uparrow$ and spin-$\downarrow$ states, which is usually achieved using the Zeeman splitting provided by an external static magnetic field. Spin-flip rotations, $\left|\uparrow\right\rangle\leftrightarrow\left|\downarrow\right\rangle$, can be achieved through electron-spin resonance (ESR) using a resonant transverse AC magnetic field delivered by an on-chip antenna. The simple description of this spin Hamiltonian would be as follows ($\hbar=1$):
\begin{eqnarray}\label{req:ultimateHamiltonian}
    \mathcal{H}_{s}=\frac{\omega_L}{2}\sigma_z+\Omega(t)\sigma_x ,
\end{eqnarray}
where $\omega_L$ and $\Omega(t)$ denote the precession frequency of the spin along the static magnetic field and the transverse AC magnetic field, respectively. 

In practice, a spin qubit can be encoded in the state of one or more electrostatically confined electrons or holes. Due to specific electronic properties, semiconductors allow individual electrons or holes to be spatially confined and coherently manipulated---something which is impossible in metals or insulators. In addition, semiconductor spin qubits are the most scalable due to their compatibility with complementary metal-oxide-semiconductor (CMOS) fabrication. Coherent spin qubits have been realized in many semiconductor architectures. In the following, we classify these systems into four major categories.

\begin{figure}
    \centering
    \includegraphics[width=0.5\linewidth]{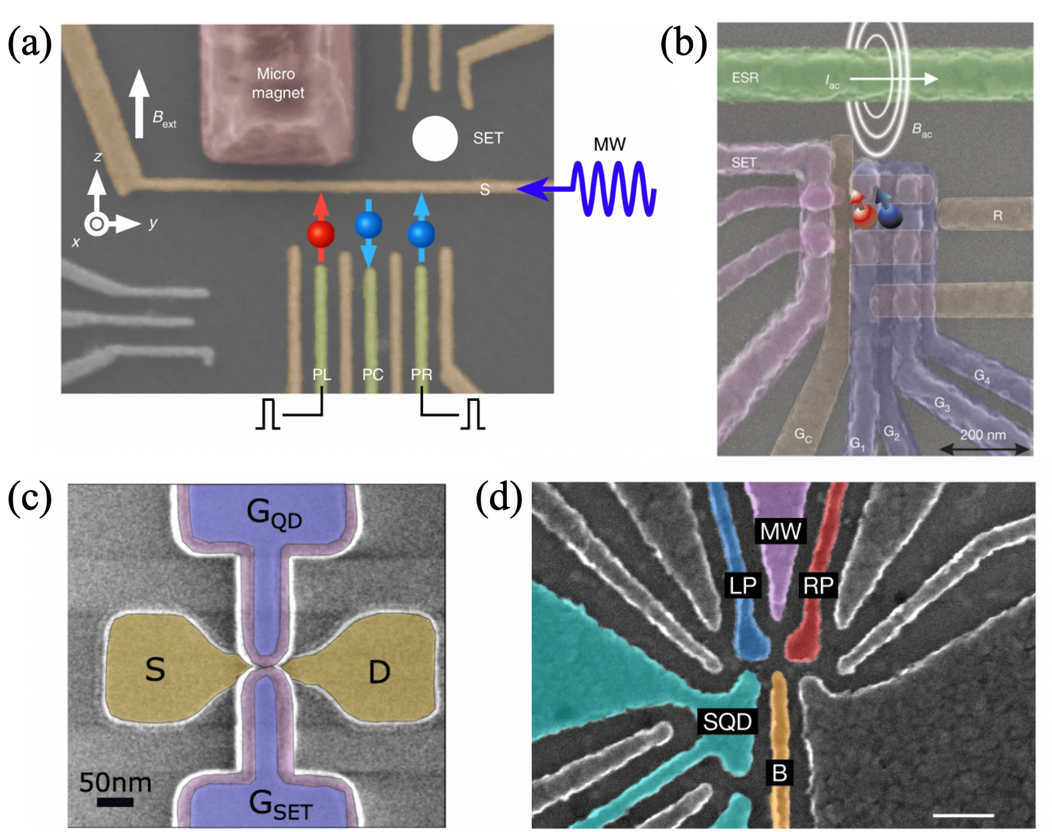}
    \caption{LD qubit architectures: (a) A hybrid LD and singlet-triplet (ST) qubit in a triple quantum dot (TQD) defined in a two-dimensional electron gas (2DEG) at the interface of a GaAs/AlGaAs heterostructure. Single qubit rotations are enabled by a nearby micromagnet. Reproduced from \textcite{noiri2018fast}, licensed under CC BY 4.0. (b) DQD in a Si/SiO$_2$ device, where single-qubit gates are driven by an AC magnetic field from an ESR antenna, and two-qubit gates are mediated via exchange. Reproduced from \textcite{veldhorst2015two} with permission. (c) Fully depleted silicon-on-insulator (FD-SOI) device hosting an electron spin qubit with micromagnet-generated synthetic spin-orbit coupling. Reproduced from \textcite{klemt2023electrical}, licensed under CC BY 4.0. (d) Si/SiGe heterostructure hosting qubits with high-fidelity (>99\%) single- and two-qubit gates used to demonstrate a variational quantum eigensolver algorithm. Reproduced from \textcite{xue2022quantum}, licensed under CC BY 4.0.}
    \label{fig:LD}
\end{figure}

\subsection{Loss-DiVincenzo qubits}
The original Loss-DiVincenzo (LD) proposal \cite{loss1998quantum} put forward the idea of manipulating spins in quantum dots via magnetic coupling to a nearby ferromagnetic island. Although a global magnetic field would resolve the $\left|\uparrow\right\rangle$ and $\left|\downarrow\right\rangle$ spin eigenstates by Zeeman splitting, the magnetic coupling could be varied over time to produce a transverse magnetic field. The net effect would be an oscillation between the $\left|\uparrow\right\rangle$ and $\left|\downarrow\right\rangle$ states, constituting a quantum $X$-gate. Early experiments focused on both optically-controlled and gate-defined quantum dots in III-V semiconductors \cite{foletti2009universal,gywat2010spins,latta2009confluence,urbaszek2013nuclear,greilich2006mode,atature2006quantum,berezovsky2008picosecond,vamivakas2010observation,de2011ultrafast,brunner2011twoqubit}. Initialization and readout of LD qubits were successfully demonstrated in GaAs \cite{elzerman2004single}, as well as the characterization and operation of single-qubit gates \cite{koppens2006driven,dutt2005stimulated,pioroladriere2008electrically,cnowack2007coherent}. Another important aspect of the LD proposal was the operation of two-qubit gates through the exchange interaction $J$ between neighboring quantum dots \cite{burkard1999coupled, schliemann2001double, petta2005coherent,nowack2011single,barthel2010fast}. The barrier potential between two quantum dots could be controlled to turn $J$ `on' or `off', enabling $\sqrt{\text{SWAP}}$ gates. Note that $\sqrt{\text{SWAP}}$ in combination with single-qubit gates can produce a controlled-NOT (CNOT) gate \cite{loss1998quantum,klinovaja2012exchange}: 
\begin{equation}
    U_{\text{CNOT}}= e^{i\pi(\sigma_1^z-\sigma_2^z)/4}\,U_{\sqrt{\text{SWAP}}}\,\,\,\,e^{i\pi\sigma_1^z/2}\,\,\,U_{\sqrt{\text{SWAP}}}.
\end{equation}

Initially, the magnetic manipulation of electron spins in quantum dots relied on ESR, but the high power dissipation associated with ESR hindered the qubit lifetime \cite{coish2007exchange}. This motivated the electrical manipulation of LD qubits, which can be achieved via electric dipole spin resonance (EDSR) \cite{golovach2006electric,tokura2006coherent,rashba2008theory} in the presence of a nearby micromagnet \cite{nowack2007coherent,pioro2008electrically,laird2007hyperfine,brunner2011two}. Despite early success, III-V semiconductor quantum dots are strongly affected by the hyperfine interaction between the confined electron and the nuclear spins of the host lattice, which produces a fluctuating Overhauser field and limits the spin-coherence time. This interaction has been extensively studied in self-assembled III-V quantum dots \cite{tartakovskii2007nuclear,xu2009optically,maletinsky2007dynamics,stepanenko2006enhancement,klauser2006nuclear,braunecker2009nuclear}, where nuclear-spin-induced dephasing was identified as a major obstacle to scalable spin-qubit implementations \cite{foletti2009universal,urbaszek2013nuclear}. Group IV semiconductors can efficiently circumvent this issue and have the potential for isotopic purification as well. High-fidelity initialization and readout, single-qubit gates, and two-qubit gates for LD qubits have been demonstrated with silicon quantum dots in Si/SiGe \cite{takeda2016fault,kawakami2014electrical,yoneda2018quantum,zajac2017quantum,watson2018programmable,xue2019benchmarking,croot2020flopping,petit2020high,xue2022quantum,mills2022two,thalakulam2010fast,shi2011tunable,simmons2011tunable} and Si CMOS devices \cite{veldhorst2014addressable,veldhorst2015two,vahapoglu2022coherent,klemt2023electrical}. Scaling up of LD qubit devices with electron spins have been progressing through efficient gate-stack design and uniformly high-yield industrial production~\cite{neyens2024probing,george202412,marcks2025valley,koch2025industrial,li2026tri}.

%noise and fidelity benchmarking
\subsection{Donor qubits}
Following the LD proposal, a quantum-computing architecture based on $^{31}$P donor nuclear spin arrays in Si with a gate-based sensing technique was subsequently proposed~\cite{kane1998silicon,vrijen2000electron}. In this setup, distant nuclear spins can couple via electron spins due to the hyperfine interaction, which can be enabled (or disabled) via `J'-gates by turning on (or off) electron wave function overlap. Another set of gates, known as `A'-gates, can electrically modify the electron wave function and, in turn, the resonance frequency of the nuclear spins. Scanning tunneling microscope (STM) lithography and masked ion-implantation methods have allowed the incorporation of shallow $^{31}$P donors on a silicon substrate \cite{morello2020donor,holmes2024improved}. The donor nucleus supplies the naturally occurring central 3D potential required to confine the electron spin, which is why these systems are also referred to as self-defined quantum dots. The nuclear gyromagnetic ratio is
$\sim2000$ times smaller than the electron gyromagnetic
ratio, allowing the nuclear spin qubit and the electron spin
qubit to be addressed individually at different frequencies.

Initial experiments implemented ESR-based spin rotations of donor-bound electrons \cite{pla2012single},  which, however, suffered from fast damping due to the hyperfine interaction. Isotopic enrichment \cite{muhonen2014storing}, along with dynamical decoupling schemes, improved the coherence time to the order of seconds \cite{laucht2015electrically,laucht2017dressed}. High-fidelity nuclear spin qubits \cite{pla2013high,wolfowicz2014conditional,laucht2015electrically} have exhibited  long coherence times of $\sim 30$ seconds \cite{muhonen2014storing}.

High-fidelity initialization and readout of single $^{31}$P donor-bound electrons \cite{morello2010single,tracy2016single,jamieson2017deterministic,koch2019spin,keith2019single} and nuclear spins \cite{pla2013high} have been demonstrated. Multi-donor dots with controlled exchange interactions have also been synthesized  \cite{fricke2021coherent}.  EDSR via engineered spin-orbit coupling, which can arise from the hyperfine interaction or an adjacent micromagnet, has been proposed in dot-donor or multi-donor systems \cite{krauth2022flopping,sarkar2022optimisation}. 
%In accordance with the original Kane proposal, 
High-fidelity two-qubit gates have also been successfully employed in coupled donor systems \cite{weber2014spin,broome2018two,he2019two,madzik2021conditional,madzik2022precision,stemp2024tomography}. 

\begin{figure}
    \centering
    \vspace{-0.1in}
    \includegraphics[width=0.7\linewidth]{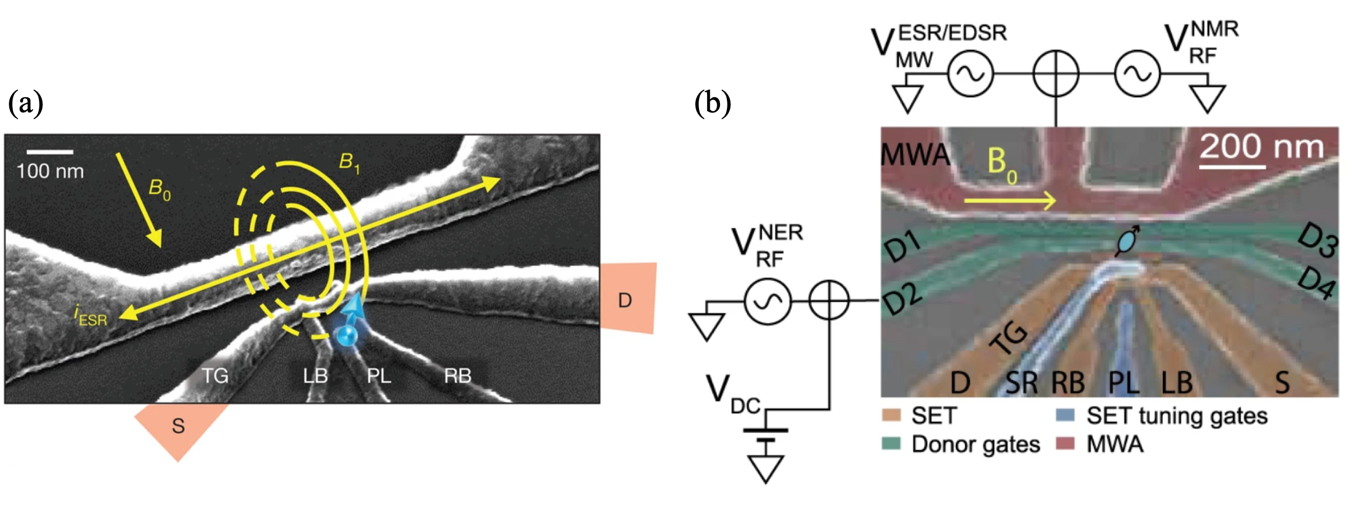}
    \caption{Donor qubit architectures: (a) Spin qubit encoded in a single shallow phosphorus ($^{31}$P) donor implanted in silicon CMOS device, with spin manipulation carried out via ESR. Reproduced from \textcite{pla2012single} with permission. (b) Antimony ($^{123}$Sb) donor device used to demonstrate ESR, EDSR, nuclear magnetic resonance (NMR) and nuclear electric resonance (NER). Reproduced from \textcite{fernandez2024navigating}, licensed under CC BY 4.0.}
    \label{fig:donors}
\end{figure}

Beyond phosphorus, several experiments have explored higher spin donor nuclei such as antimony (Sb), arsenic (As), and bismuth (Bi) \cite{morello2020donor,asaad2020coherent,fernandez2024navigating,yu2025schrodinger}. These spins' large spin quantum numbers (e.g.~$I{=}7/2$ for $^{121}$Sb, $I{=}9/2$ for $^{209}$Bi) provide multi-level Hilbert spaces (with dimension $>2$) that can be used to encode qudits or logical qubits equipped with quantum error correction~\cite{chiesa2020molecular,gross2021designing,gross2024hardware,lim2025demonstrating}. The Bi donor, with its exceptionally large hyperfine coupling ($A$ = 1.475 GHz), exhibits avoided crossings between the electron and nuclear spin states known as {\it clock transitions}, which produce excellent coherence times (at moderate magnetic-field strengths) without isotopic purification \cite{wolfowicz2013atomic}. Rare-earth dopants in silicon add an additional dimension by combining a highly coherent spin with an optical transition in the C-band (1.54 $\mu$m), enabling direct integration with fiber-based quantum networks \cite{yin2013optical,siyushev2014coherent}. The Er$^{3+}$ and Yb$^{3+}$ ions embedded in Si host extremely narrow optical linewidths and long lived electron-nuclear spin states and have been operated as spin-photon interfaces compatible with optical initialization, readout, and entanglement generation \cite{stevenson2022erbium,kindem2020control,ourari2023indistinguishable}.

%noise and fidelity benchmarking of donors
\subsection{Multispin encodings}
Encoding a qubit in the state of multiple spins can be beneficial in protecting against electric and magnetic noise~\cite{meier2003quantumPRL}. For example, two electrons in a DQD can be used to encode a singlet-triplet qubit, the qubit states being the singlet $\left|S\right\rangle=(1/\sqrt{2})(\left|\uparrow\downarrow\right\rangle-\left|\downarrow\uparrow\right\rangle)$ and the unpolarized triplet $\left|T_0\right\rangle=(1/\sqrt{2})(\left|\uparrow\downarrow\right\rangle+\left|\downarrow\uparrow\right\rangle)$. These states span a decoherence-free subspace that is immune to global magnetic-field noise (i.e.~noise impacting both spins in the same manner). 

Single-qubit rotations on the Bloch sphere can be described by $\mathcal{H}_{ST_0}=(J/2)\sigma_z+(\Delta B_z/2)\sigma_x$, where $\sigma_z$ and $\sigma_x$ are Pauli matrices defined in the subspace spanned by $\ket{S}$ and $\ket{T_0}$. Early works on two-qubit gates in gate-defined LD systems already identified the $ST_0$ qubit, achieved by adiabatically controlling the exchange interaction $J(\epsilon)$ as a function of the detuning between the dots \cite{levy2002universal}. In this all-electrical scheme, the transverse field gradient $\Delta B_z$ is ideally constant; however, local fluctuations to $\Delta B_z$ may arise from hyperfine fields, $g$ factor anisotropy, etc. Initialization and readout were first enabled by Pauli spin blockade (PSB), where pulsing to the (0,2) charge state loads a pure singlet while triplet states remain blocked and detectable through a charge sensor \cite{ono2002current,petta2005coherent,johnson2005triplet}. The PSB mechanism is analogous to the spin-to-charge conversion process initially demonstrated for single-spin qubits \cite{elzerman2004single}, but uses charge transitions instead of Zeeman splitting. High-fidelity coherent single-qubit gates for $ST_0$ qubits have been demonstrated in GaAs \cite{petta2005coherent,foletti2009universal,barthel2010interlaced,bluhm2010enhancing,barthel2009rapid,koppens2008spin,van2011charge}, a Si donor-dot system \cite{harvey2017coherent}, Si/SiGe heterostructures \cite{maune2012coherent,liu2021magnetic}, Si CMOS \cite{harvey2017alldelg,jock2018silicon,fogarty2018integrated,jock2022silicon}, and $^{31}$P donors \cite{broome2017high,he2019two,stemp2024tomography}. \textcite{petta2008dynamic} also demonstrated that hyperfine fields can drive rapid rotations of an $ST_+$ qubit by “flipping” an electron spin and “flopping” a nuclear spin. Two-qubit gates have been demonstrated by capacitive and exchange coupling between neighboring double dots \cite{taylor2005fault,shulman2012demonstration,nichol2017high,qiao2021floquet}. An important variation of the $ST_0$ qubit is the resonantly driven $ST_0$ qubit, which uses $\left|\uparrow\downarrow\right\rangle$ and $\left|\downarrow\uparrow\right\rangle$ as the qubit states; instead of adiabatic exchange pulses, a transverse field drives the qubit on resonance \cite{bluhm2011dephasing}. Under static magnetic fields, the Zeeman
energy competes with exchange, and a polarized triplet
($\left|T_+\right\rangle$ in GaAs, $\left|T_-\right\rangle$ in Si) may become degenerate with $\ket{S}$. Transverse magnetic-field gradients or spin-orbit coupling can lift this degeneracy, producing a $ST_+$ (or $ST_-$) qubit.

\begin{figure}
    \centering
    \includegraphics[width=0.55\linewidth]{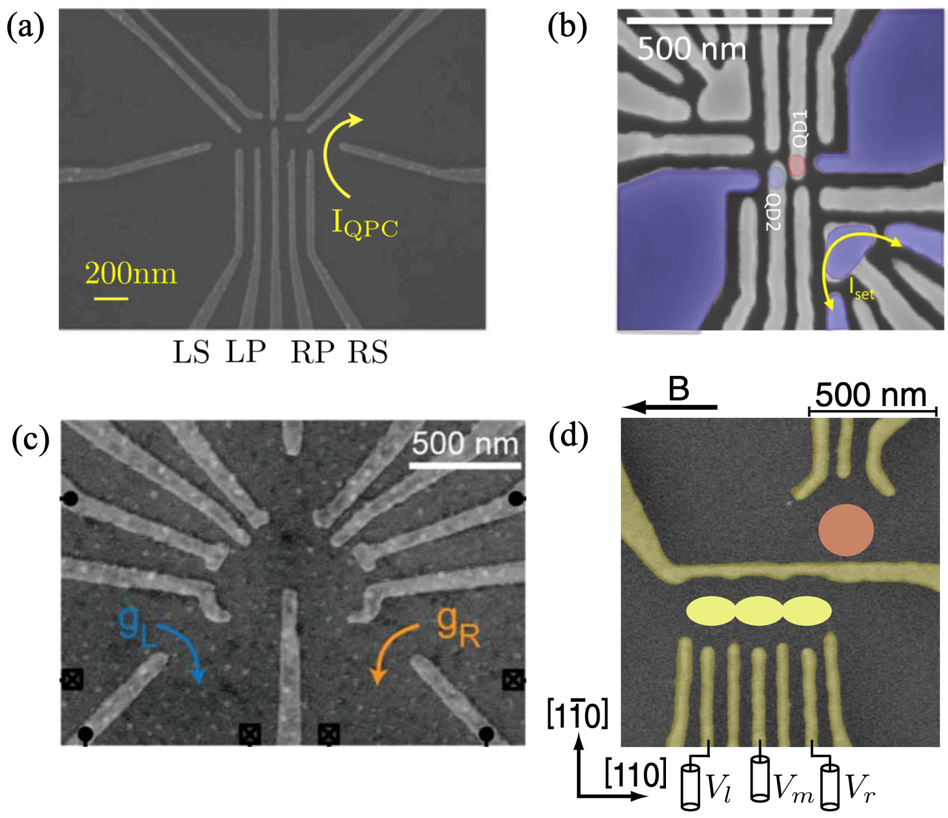}
    \caption{(a) Singlet-triplet qubit in a Si/SiGe heterostructure. A magnetic-field difference $\Delta B$ between the two dots is created by a nearby micromagnet, enabling two-axis control. Reproduced from \textcite{wu2014two}, licensed under CC BY 4.0. (b) Singlet-triplet qubit in a SiMOS device, where qubit rotations are mediated by spin-valley coupling. Reproduced from \textcite{jock2022silicon}, licensed under CC BY 4.0. (c) An exchange-only (EO) qubit in a GaAs/AlGaAs heterostructure incorporating a 2DEG 110 nm beneath the surface. This qubit encoding removes the need for an inhomogeneous field, and the exchange interactions between adjacent spins suffice for all single- and two-qubit operations. Reproduced from \textcite{laird2010coherent} with permission. (d) Triple-dot device hosting a resonant exchange (RX) spin qubit. Reproduced from \textcite{medford2013quantum} with permission. While EO qubits operate via time-dependent exchange pulses, RX qubits are driven by resonant modulation at a fixed, always-on exchange sweet spot.}
    \label{fig:multispin}
\end{figure}

The exchange only (EO) qubit is encoded in the state of  three electrons~\cite{divincenzo2000universal}. In a triple quantum dot with one electron per dot, the three-spin Hilbert space consists of one quadruplet with total spin $S=3/2$ and two doublets with total spin $S=1/2$. The EO qubit is encoded in one of the $S=1/2$ doublets and is immune to global magnetic field noise. For the doublet with $S_z=+1/2$, the qubit basis states are $\left|0\right\rangle=\left|S\right\rangle\left|\uparrow\right\rangle$ and $\left|1\right\rangle=\sqrt{1/3}\left|\uparrow\uparrow\right\rangle\left|\downarrow\right\rangle{-}\sqrt{2/3}\left|T_0\right\rangle\left|\uparrow\right\rangle$. This encoding allows all single- and two-qubit operations to be performed using only electrically tunable exchange interactions. Single-qubit gates can be achieved by electrically modulating the Heisenberg exchange $J_{12}(t)$ and $J_{23}(t)$ \cite{divincenzo2000universal,kempe2002exact,weinstein2005energetic,laird2010coherent,sala2017exchange,stastny2025singlet,medford2013self,bosco2026exchange}. Two-qubit gates can also be achieved via exchange, providing a scalable, ultrafast, fully electrical scheme, with charge noise and cross-talk limiting the gate fidelity \cite{russ2017three}. A similar three-electron scheme based on having two spins in one dot of a DQD uses the same basis states and is known as the hybrid qubit~\cite{shi2012fast}.

When $J_{12}{=}J_{23}{=}J$, a constant energy gap opens between $\left|0\right\rangle$ and $\left|1\right\rangle$, and an AC exchange pulse $\Delta J(t)$ enables single qubit rotations. This particular variation of the EO qubit is called the resonant exchange (RX) qubit \cite{meier2003quantum,medford2013quantum,taylor2013electrically,doherty2013two,zhang2018leakage,sala2020highly,Gaudreau2012coherent}. Recently, \textcite{foulk2025singlet} proposed a four-spin singlet-only always-on gapless exchange (SAGE) qubit, which is protected from magnetic field gradients with scalable baseband control and suppressed leakage to non-computational states. Another proposal for five-electron $p$-orbital ($pO$) spin qubits was given by \textcite{caporaletti2025proposed}. Due to a vanishing dipole moment between qubit states, these $pO$ qubits only interact with electric-field \textit{gradients} due to charge noise. They can also be operated at zero magnetic field for compatibility with superconducting interconnects.

\subsection{Hole qubits}

Theoretical proposals for hole-based quantum computing were made soon after the conception of electron spin qubits \cite{bulaev2005spin,bulaev2007electric,kloeffel2013prospects}. Since then, holes in the valence band of group-IV and group III-V semiconductors have emerged as viable spin qubit candidates. In contrast to electrons, holes exhibit {\it intrinsic} spin-orbit coupling due to the valence band having nonzero angular momentum ($l=1$). Hence, a spin-$1/2$ hole has a total angular momentum of $j=3/2$ in the topmost valence band. Hole spin qubits have achieved ultra-fast fully electrical gate control in planar germanium (e.g., Ge/Si$_{1-x}$Ge$_x$ heterostructures) and silicon (e.g., c-MOS). The motivation to develop high mobility p-channels led to the creation of Ge/Si$_{1-x}$Ge$_x$ heterostructures with a strained Ge layer supporting a two dimensional hole gas \cite{hendrickx2018gate,hendrickx2020fast,hendrickx2020single,lawrie2020spin,terrazos2021theory,liu2022gate}, and subsequently, to Ge/Si core-shell nanowires \cite{xiang2006ge,higginbotham2014hole,froning2021ultrafast} and Ge hut wires featuring a one-dimensional hole gas strongly confined in Ge \cite{watzinger2018germanium,gao2020site}. 

The intrinsic SOC for holes leads to large electrical tunability, and consequently, hole spin qubits exhibit large g-factor anisotropy~\cite{hutin2018si,liles2021electrical,geyer2024anisotropic}.  Operational {\it sweet spots} \cite{bosco2021hole,bosco2021fully,bosco2022fully,adelsberger2022hole,malkoc2022charge,hendrickx2024sweet,wang2021optimal,sarkar2023electrical} and {\it sweet lines} \cite{mauro2024geometry} have been explored, with the goal of maximizing gate speed while minimizing dephasing (predominantly) due to charge noise. Coherence can be further improved  by choosing certain external magnetic field orientations~\cite{hendrickx2024sweet}, or by preparing the nuclear environment in a ``narrowed'' spin distribution~\cite{fischer2008spin}, or by squeezing the dot potential~\cite{bosco2021fully}, allowing for the suppression of the hyperfine interaction of valence-band holes with nuclear spins. Note that in III-V semiconductor QDs, the hyperfine interaction for holes is weaker than that of electrons, but non-negligible \cite{maier2012effect,eble2009hole,fallahi2010measurement,chekhovich2011direct,brunner2009coherent}. The lack of micromagnets as well as high hole mobility means that Ge hole spin qubit architectures can be scaled efficiently~\cite{scappucci2021germanium}. Singlet-triplet qubits~\cite{jirovec2021singlet}, a four-qubit processor~\cite{hendrickx2021four}, a 16-qubit crossbar Ge array~\cite{borsoi2024shared}, vertically stacked DQDs~\cite{tidjani2023vertical,ivlev2024coupled}, and more recently, parallel operation in an 18-qubit $2\times N$ dot array have all been demonstrated~\cite{dijkema2026simultaneous}, the latter representing the largest spin-qubit array to date.

%Singlet-triplet qubits have been implemented in Ge \cite{jirovec2021singlet}, along with a four-qubit processor \cite{hendrickx2021four} and a 16-qubit crossbar Ge array \cite{borsoi2024shared}. Most recently, parallel operation in an 18-qubit $2\times N$ dot array has been demonstrated, representing the largest spin-qubit array to date~\cite{dijkema2026simultaneous}. Bi-layer structures with holes have also been explored in Ge~\cite{tidjani2023vertical,ivlev2024coupled}, with the prospect of efficient scaling.  The concept of operational {\it sweet spots} \cite{bosco2021hole,bosco2021fully,bosco2022fully,adelsberger2022hole,malkoc2022charge,hendrickx2024sweet,wang2021optimal,sarkar2023electrical} and {\it sweet lines} \cite{mauro2024geometry} has also emerged in Ge, with the goal of maximizing gate speed while minimizing dephasing (predominantly) due to charge noise. Hyperfine-induced hole spin decoherence can be suppressed for certain magnetic field orientations, or narrowed nuclear spin bath, or by squeezing \cite{bosco2021fully,fischer2008spin}. In III-V semiconductor QDs, the hyperfine interaction of valence-band holes with nuclear spins is weaker than that of electrons but non-negligible \cite{maier2012effect,eble2009hole,fallahi2010measurement,chekhovich2011direct,brunner2009coherent}. The large g-factor anisotropy for holes originates from the electrically tunable spin-orbit coupling and plays an important role in single- and two-qubit gates \cite{hutin2018si,liles2021electrical,geyer2024anisotropic}.

\begin{figure}
    \centering
    \vspace{-0.1in}
    \includegraphics[width=0.7\linewidth]{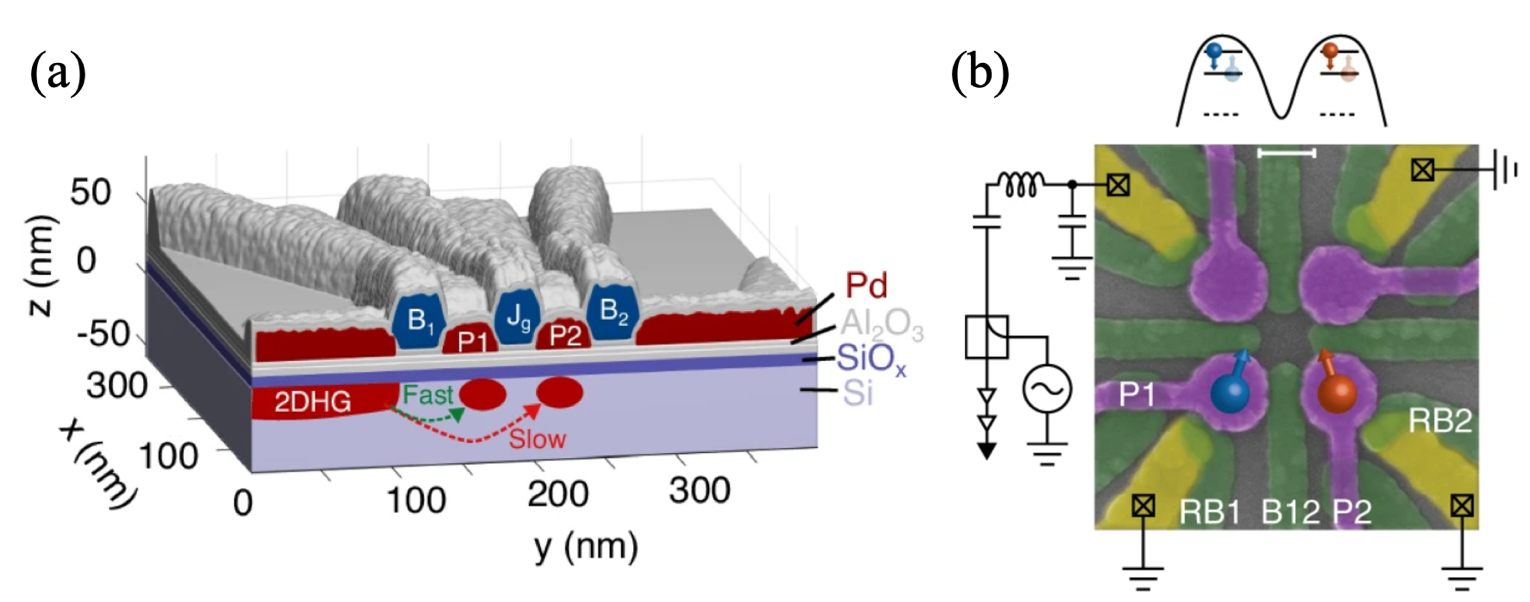}
    \caption{Hole spin qubit architectures: (a) Planar hole DQD in a Si/SiO$_2$ device, exhibiting exchange- and $\Delta g$-driven oscillation of a singlet-triplet qubit. Reproduced from \textcite{liles2024singlet}, licensed under CC BY 4.0. (b) Four dot system in a Ge/SiGe heterostructure, featuring high fidelity single- and two-qubit operation of Ge holes. Reproduced from \textcite{hendrickx2020single}, licensed under CC BY 4.0. The driving mechanism for holes is fully electrical, owing to strong intrinsic spin-orbit coupling in the topmost valence bands.}
    \label{fig:holes}
\end{figure}

Silicon cMOS planar devices have demonstrated a large tunability of spin qubit properties by electrical (gate) and mechanical (strain) means \cite{maurand2016cmos,liles2024singlet,piot2022single,bassi2025optimal}. Silicon has a smaller spin-orbit energy gap of $\Delta_{SO}=44$ meV, leading to nontrivial mixing with the split-off band \cite{wang2024electrical}. FinFET devices have demonstrated hole spin qubit operation at temperatures above 4 Kelvin \cite{bosco2021hole,camenzind2022hole, petit2020universal}.

As one can imagine, achieving true quantum advantage with spin qubits would mean coherently manipulating the spin and charge degrees of freedom of many confined particles with very high fidelity. The electronic structure of the underlying material plays a central role in this.

\section{Electronic structure of semiconductors}
Semiconductors are crystalline solids whose electronic behavior lies between metals and insulators. They have a {\it band gap} ($E_v$) of a few electron volts, and when an electron in the valence band of a semiconductor is supplied with energy $\geq E_v$, it contributes to electrical conduction \cite{kittel2018introduction}. Semiconductors form a highly versatile class of materials that includes elements (Si, Ge), binary compounds (GaAs, ZnS, PbS), oxides (Cu$_2$O, ZnO), and organic semiconductors ([CH$_2$]$_n$). Depending on composition and carrier density, some semiconductors can also exhibit magnetic behavior (Cd$_{1-x}$Mn$_x$Te) \cite{furdyna1988diluted,dietl2010ten} or even become superconducting (La$_2$CuO$_4$, when La is replaced by Ba/Sr) \cite{bednorz1986possible}. These properties are well understood from the electronic band structure of semiconductors, which dictates how electrons and holes move in the crystalline lattice \cite{harrison1989electronic}. 

\subsection{Band structure}
The most relevant crystal structures for spin qubits are the diamond lattice (Si, Ge, Si$_{1-x}$Ge$_x$) and the zincblende lattice (GaAs, ZnS). A unit cell of the diamond/zincblende lattice has the structure of two interlocking face centered cubic (FCC) lattices, with one basis at $(0,0,0)$ and the other at $\big(a/4,a/4,a/4\big)$, where $a$ is the side length of the FCC unit cell. While diamond has the same kind of atom constituting both bases, zincblende has two different atoms. As a result, diamond crystals have the symmetry of an Octahedron ($O_h$ symmetry), while zincblende crystals have tetrahedral symmetry ($T_d$ symmetry), which lacks a center of inversion \cite{kittel2018introduction}. 

The primitive lattice vectors of the FCC lattice can be taken to be $\mathbf{a}_1{=}(a/2)(0,1,1)$, $\mathbf{a}_2{=}(a/2)(1,0,1)$, and $\mathbf{a}_3{=}(a/2)(1,1,0)$, yielding the reciprocal lattice vectors $\mathbf{b}_1{=}(2\pi/a)(-1,1,1)$, $\mathbf{b}_2{=}(2\pi/a)(1,-1,1)$, and $\mathbf{b}_3{=}(2\pi/a)(1,1,-1)$. These reciprocal lattice vectors define the {\it Brillouin zone} (BZ), i.e., the smallest polyhedron confined by planes that perpendicularly bisect the reciprocal lattice vectors. Mathematically, the advantage of this description is two-fold: i) The wave vector $\mathbf{k}$ can be represented as a point in the BZ, and ii) the high symmetry of the BZ contains the rotational, translational, and reflection symmetries of the crystal structure. For the FCC lattice, the high symmetry directions in the BZ are $[100]$ (or $\Gamma-\Delta-X$), $[111]$ (or $\Gamma-\Lambda-L$), and $[110]$ (or $\Gamma-\Sigma-K$) \cite{bouckaert1936theory}.
%\begin{figure}
    %\centering
    %\vspace{-0.1in}
    %\includegraphics[width=0.98\linewidth,height=3.3 in]{Part_I/figures/reviewBandStrIllust.png}
    %\caption{(a) The diamond unit cell, (b) the zinc blende unit cell structure (in zinc blende, the four atoms at (1/4,1/4,1/4) and equivalent positions are of the second kind), c) primitive lattice vectors of diamond and zinc blende, (c) the first Brillouin zone of diamond and zinc blende, highlighting certain high-symmetry points and symmetry lines.}
   %%\end{figure}

Quantum mechanically, an explicit description of the crystal is given by the following Hamiltonian:
\begin{eqnarray}\label{req1:manybody}
    &&\mathcal{H}=\sum_i \frac{p_i^2}{2m_i}+\sum_j \frac{P_j^2}{2M_j}+\frac{1}{2}\sum_{j',j}'\frac{Z_jZ_{j'}e^2}{4\pi\epsilon_0|\mathbf{R}_j-\mathbf{R}_{j'}|}-\frac{1}{2}\sum_{j,i}\frac{Z_je^2}{4\pi\epsilon_0|\mathbf{r}_i-\mathbf{R}_{j}|}+\frac{1}{2}\sum_{i,i'}'\frac{e^2}{4\pi\epsilon_0|\mathbf{r}_i-\mathbf{r}_{i'}|},
\end{eqnarray}
where $\mathbf{r}_i$ ($\mathbf{p}_i$) are the $i$-th electron position (momentum) and $\mathbf{R}_j$ ($\mathbf{P}_j$) are the $j$-th nuclei position (momentum); $Z_j$ is the atomic number. Two key approximations can be made here: i) the filled orbitals of {\it core} electrons (e.g. $1s^22s^22p^6$ in Si) are localized at the given nucleus to form ion cores, and only the {\it valence} electrons (e.g. $3s^23p^2$ in Si) in the outermost shell contribute to the electronic motion, and ii) the {\it Born-Oppenheimer approximation} \cite{born1927quantentheorie}, which assumes that the ion cores move much more slowly than the valence electrons, making them effectively stationary in the electron frame of reference. These approximations allow for the following simplification,
\begin{equation}\label{req:fullcrysHam}
    \mathcal{H}=\mathcal{H}_{\text{ion}}(\mathbf{R}_j)+\mathcal{H}_{e}(\mathbf{r}_i,\mathbf{R}_{j0})+\mathcal{H}_{\text{ep}}(\mathbf{r}_i,\delta\mathbf{R}_j).
\end{equation}
The electronic motion is captured by $\mathcal{H}_{e}(\mathbf{r}_i,\mathbf{R}_{j0})$. 

The {\it mean-field approximation} states that each valence electron in the crystal experiences the same average potential $V(\mathbf{r})$ from the static nuclei at positions $\mathbf{R}_{j0}$, implying that the electronic structure can be described by a single-particle Hamiltonian 
\begin{equation}
    \mathcal{H}_{\text{1e}}\Phi_n(\mathbf{r})=\left(\frac{p^2}{2m_0}+V(\mathbf{r})\right)\Phi_n(\mathbf{r})=E_n\Phi_n(\mathbf{r}).
\end{equation}
The periodicity of the crystal structure then has an important consequence: The potential $V(\mathbf{r})$ is periodic. Hence, according to the Bloch theorem \cite{bloch1929quantenmechanik}, the eigenfunctions of $\mathcal{H}_{\text{1e}}$ can be written in the form $\Phi_{n\mathbf{k}}(\mathbf{r})=e^{i\mathbf{k}\cdot\mathbf{r}}u_{n\mathbf{k}}(\mathbf{r})$, where $\mathbf{k}$ is a wave vector and $u_{n\mathbf{k}}(\mathbf{r})$ are {\it Bloch functions} in the $n$-th {\it band}. Notably, these Bloch functions inherit the periodicity of the lattice [see p.~167 of \textcite{kittel2018introduction}].
The {\it band structure} is a plot of the electronic energies $E_n$ with respect to $\mathbf{k}$.

The valence electrons in semiconductors are in the $s$ and $p$ orbitals, which, from nearby lattice sites, overlap to create discrete $\sigma$ and $\pi$ states. Using the scenario of a crystal with many orbital overlaps, the energy gap between these discrete states would be $\sim10^{-27}$ eV, effectively forming a continuous energy band. Note that the band structure is determined by the geometry, as well as the orbital and chemical bonding, and spin-orbit coupling. The conduction band is composed of antibonding states, whereas the valence band is composed of bonding states \cite{kittel2018introduction,peter2010fundamentals}. The {\it band gap}, i.e., the energy difference between the conduction band minima and valence band maxima, is a consequence of the finite energy difference between the bonding and antibonding states. The valence band maxima in semiconductors always sit at the $\Gamma$ point of the Brillouin zone, since the bonding $p$-orbitals across the entire lattice minimize the total energy at $k=0$ and the valence band is filled. Zincblende semiconductors (GaAs, InSb) feature conduction band minima at the $\Gamma$ point, i.e., a direct band gap, due to the absence of inversion symmetry and the heteropolar bond allowing strong mixing of $s{-}p$ already at $k=0$, lowering the energy. In diamond (Si, Ge), the conduction band is purely $s$ antibonding at the $\Gamma$ point ($s{-}p$ mixing is prohibited in homopolar bonds), but at $k\neq 0$, the finite $s{-}p$ mixing lowers the energy, inducing an indirect band gap \cite{marder2010condensed}.

\begin{figure}
    \centering
    \vspace{-0.1in}
    \includegraphics[width=0.7\linewidth]{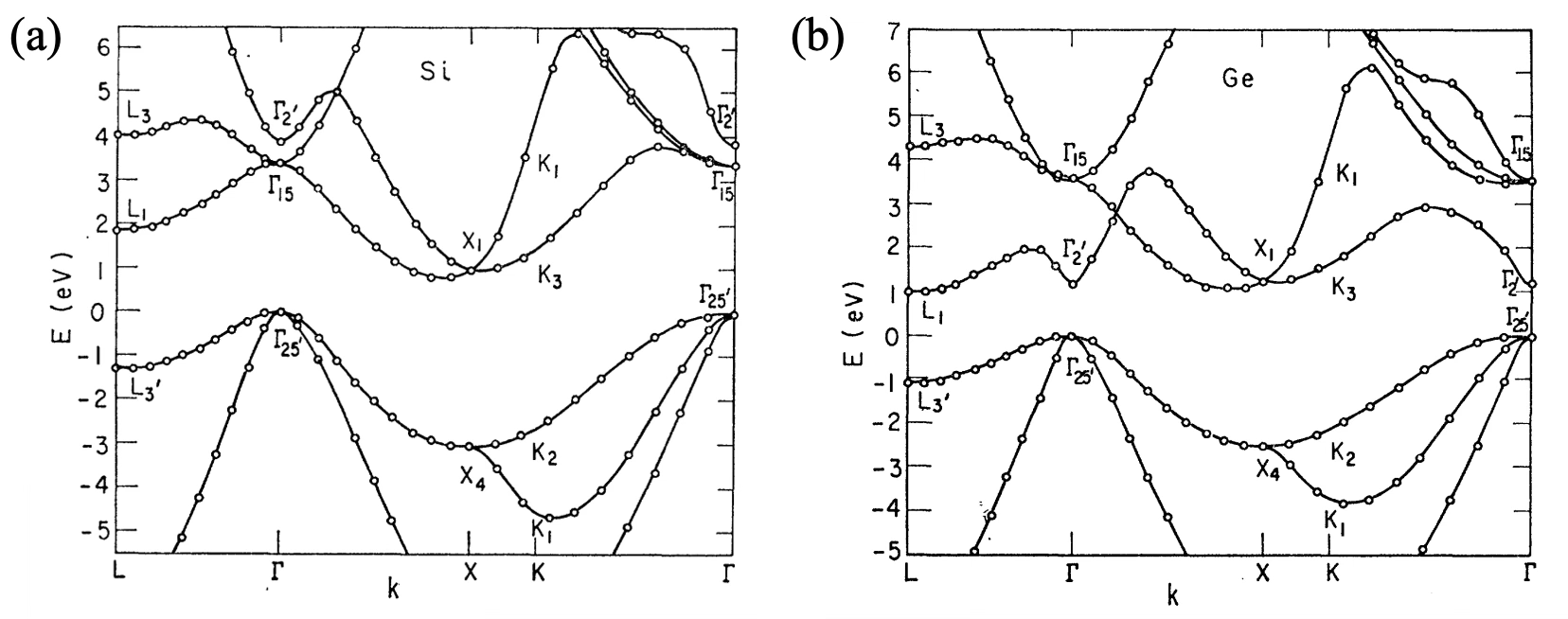}
    \caption{(a) Band structure of Si, with the energy dispersion calculated along the high-symmetry directions in the BZ. (b) Band structure of Ge. Notably, the energy at the top of the filled valence band is set to zero. The double group notation gives the symmetry of the electronic wave function in the crystal at high symmetry points of the BZ. Reproduced from \textcite{cohen1966band} with permission.}
    \label{fig:bands}
\end{figure}

\begin{table}
    \centering
    \begin{tabular}{l|c|c|c|c}
       \hline 
       \hline
       Parameters & Si & Ge & GaAs & InSb  \\
        \hline
        Band gap, $E_g$ (eV) & 1.12 & 0.66 & 1.42 & 0.17\\
        \hline
        Spin-orbit gap, $\Delta_{\text{SO}}$ (eV) & 0.04 & 0.29 &0.34 & 0.80\\
        \hline
        effective mass: & & & & \\
        \begin{tabular}{p{3.1cm} p{0.7cm}}
           $m_e^*$ ($m_0$)  & $m_l$ \\
        \end{tabular}  & 0.98 & 1.59  & \multirow{2}{*}{0.06} & \multirow{2}{*}{0.01}\\
        \begin{tabular}{p{3.1cm} p{0.7cm}}
            & $m_t$ \\
        \end{tabular}  & 0.19 & 0.08  &  & \\
        \begin{tabular}{p{3.1cm} p{0.7cm}}
           $m_h^*$ ($m_0$)  & $m_{hh}$ \\
        \end{tabular}  & 0.49 & 0.33  & 0.51 & 0.43\\
        \begin{tabular}{p{3.1cm} p{0.7cm}}
            & $m_{lh}$ \\
        \end{tabular}  & 0.16 & 0.04  & 0.08 & 0.02\\
        \hline
        CB minima at, $k=$ & 0.85$k_0$ & $\sqrt{3}k_0$ & 0 & 0\\
        \hline 
        \hline
    \end{tabular}
    \caption{Band structure parameters of group IV (Si, Ge) and binary III-V (GaAs, InSb) semiconductors. Here, $m_0$ is the free electron mass of $9.11\times 10^{-31}$ kg, and $k_0=2\pi/a$, where $a$ is the lattice constant.}
    \label{tabI1:bandparameters}
\end{table}

Since the system Hamiltonian commutes with the point group of the crystal symmetry,  electronic properties can also be efficiently described in the language of invariants \cite{Kostergrouptheoryultimate}. The invariance of the electronic Hamiltonian $\mathcal{H}_e$ under a symmetry operation $g\in \mathcal{G}$ can be expressed as
\begin{equation}
    \mathcal{D}(g)\mathcal{H}_e(g^{-1}\mathcal{K})\mathcal{D}^{-1}(g)=\mathcal{H}_e(\mathcal{K}),
\end{equation}
where $\mathcal{G}$ denotes the point group of crystal symmetry, $\mathcal{K}$ represents a general tensor operator, and $\mathcal{D}(g)$ is the matrix representation of the operation/group element $g$ \cite{winkler2003spin,dresselhaus2007group}. The symmetry operations of a point group $\mathcal{G}$ consist of reflection, rotation, or translation and can be described by transformation matrices (e.g., a 3D rotation matrix). All such transformation matrices also form a group, known as the {\it representation} of $\mathcal{G}$. While infinitely many such representations can be constructed, {\it irreducible representations} cannot be decomposed into smaller transformation matrices, nor can they be mapped to another representation by applying similarity transformations. By definition, eigenstates of $\mathcal{H}_e$ must form the irreducible representations of its symmetry group $\mathcal{G}$. Since the bands are precisely the eigenstates of the crystal Hamiltonian, the band structure can be described as a collection of the irreducible representations of $\mathcal{G}$. Group III-V semiconductors exhibit tetrahedral (point group $T_d$ or $F\overline{4}3m$) symmetry, while group IV semiconductors exhibit octahedral (point group $O_h$ or $Fd3m$) symmetry \cite{bradley2009mathematical}. 

As an example, let us consider the 48 symmetry operations of the silicon unit cell ($O_h$):
\vspace{-0.1in}
\begin{itemize}
       \setlength{\itemsep}{-3pt}
    \item $E$: identity
    \item 8$C_3$: clockwise and counterclockwise rotations of 120$^\circ$ about the [111], [$\overline{1}$11], [1$\overline{1}$1], and [11$\overline{1}$] axes, 
    \item 3$C_2$: 180$^\circ$ rotations about the $[100], [010], [001]$ axes,
    \item 6$S_4$: clockwise and anticlockwise 90$^\circ$ improper rotations about the [100], [010], and [001] axes,
    \item 6$\sigma_d$: reflections with respect to the (110), (1$\overline{1}$0), (101), (10$\overline{1}$), (011), and (01$\overline{1}$) planes,
    \item $i$: inversion,
    \item 3$\sigma_h$:  reflections with respect to the $(100), (010), (001)$ planes,
    \item 6$C_4$: clockwise and counterclockwise rotations of 90$^\circ$ about the [100], [010], and [001] axes,
    \item 6$C_2'$: rotations of 180$^\circ$ about the [110], [1$\overline{1}$0], [101], [10$\overline{1}$], [011], and [01$\overline{1}$] axes, and
    \item 8$S_6$: improper clockwise and anticlockwise rotations of 120$^\circ$ about the [111], [$\overline{1}$11], [1$\overline{1}$1] and [11$\overline{1}$] axes.
\end{itemize}
\vspace{-0.1in}
The irreducible representations of $O_h$, i.e., the silicon bands, can be interpreted in terms of this group of symmetry operations (or {\it classes}) through the {\it character table}. %see band structure and character table.
At the $\Gamma$ point $\Phi_{n\mathbf{0}}(\mathbf{r})=u_0(\mathbf{r})$, the electronic wavefunction feels the full point group symmetry of the crystal, and the bands are denoted by irreducible representations $\Gamma_\nu$. 
At nonzero $\mathbf{k}$, the bands follow the symmetry of the little groups of $O_h$. For example, at the $L$ point, the Si band structure is denoted by the irreducible representation of the point group $D_{3d}$ \cite{Kostergrouptheoryultimate}. 

Putting this in the context of spin qubit modeling, the tensor operator $\mathcal{K}$ in $\mathcal{H}(\mathcal{K})$ can depend on components of the wave vector $\mathbf{k}$, the electric field $\mathbf{E}$, the magnetic field $\mathbf{B}$, or the strain $\mathbf{\varepsilon}$. The symmetrically allowed terms in the system Hamiltonian can be identified without intricate microscopic calculations, and a great deal of optimization regarding qubit operation and performance can be predicted in this way. We now turn to practical modeling approaches for computing and analyzing band structures quantitatively.

\subsection{$\mathbf{k.p}$ theory and the effective mass approximation}
The $\mathbf{k}\cdot\mathbf{p}$ formalism is a powerful tool that extrapolates the band structure over the entire Brillouin zone, using zone center energy gaps and optical matrix elements as parameters. These are often obtained from experiments. As described in the previous section, the electronic structure is effectively given by the single particle Hamiltonian $\mathcal{H}_{1e}$. Using Bloch's theorem, the action of this Hamiltonian on a Bloch function $u_{n\mathbf{k}}$ can be written as
\begin{eqnarray}
    &&\bigg[\frac{p^2}{2m_0}+V(\mathbf{r})\bigg]e^{i\mathbf{k}\cdot\mathbf{r}}u_{n\mathbf{k}}(\mathbf{r})=E_{n\mathbf{k}}e^{i\mathbf{k}\cdot\mathbf{r}}u_{n\mathbf{k}}(\mathbf{r}),\nonumber\\&&\bigg[\frac{p^2}{2m_0}+V(\mathbf{r})+\frac{\hbar}{m_0}\mathbf{k}\cdot\mathbf{p}\bigg]u_{n\mathbf{k}}(\mathbf{r})=\bigg[E_{n\mathbf{k}}-\frac{\hbar^2k^2}{2m_0}\bigg]u_{n\mathbf{k}}(\mathbf{r}).
\end{eqnarray}
The formalism involves first solving for the orthonormal solutions of $\big[p^2/(2m_0){+}V(\mathbf{r})\big]u_{n\mathbf{k_0}}(\mathbf{r}){=}E_{n\mathbf{k_0}}'u_{n\mathbf{k_0}}(\mathbf{r})$ for any $\mathbf{k}{=}\mathbf{k_0}$ in the first BZ, and then treating the $(\hbar/m_0)\mathbf{k}\cdot\mathbf{p}$ term as a perturbation. For a non-degenerate band, the eigenfunction $u_{n\mathbf{k}}$ and eigenenergy $E_{n\mathbf{k}}$ for $\mathbf{k}$ near $\mathbf{k_0}$ become, to leading order,
\begin{eqnarray}
    &&u_{n\mathbf{k}}(\mathbf{r})=u_{n\mathbf{k_0}}(\mathbf{r})+\frac{\hbar}{m_0}\!\sum_{n'\neq n}\!\frac{\left\langle u_{n\mathbf{k_0}}(\mathbf{r})|\mathbf{k}\cdot\mathbf{p}|u_{n'\mathbf{k_0}}(\mathbf{r})\right\rangle}{E_{n\mathbf{k_0}}-E_{n'\mathbf{k_0}}}u_{n'\mathbf{k_0}},\nonumber\\
    && E_{n\mathbf{k}}=E_{n\mathbf{k_0}}+\frac{\hbar^2k^2}{2m_0}{+}\frac{\hbar^2}{m_0^2}\!\sum_{n'\neq n}\!\frac{|\left\langle u_{n\mathbf{k_0}}(\mathbf{r})|\mathbf{k}\cdot\mathbf{p}|u_{n'\mathbf{k_0}}(\mathbf{r})\right\rangle|^2}{E_{n\mathbf{k_0}}-E_{n'\mathbf{k_0}}}.
\end{eqnarray}
\begin{figure}
    \centering
    \includegraphics[width=0.55\linewidth]{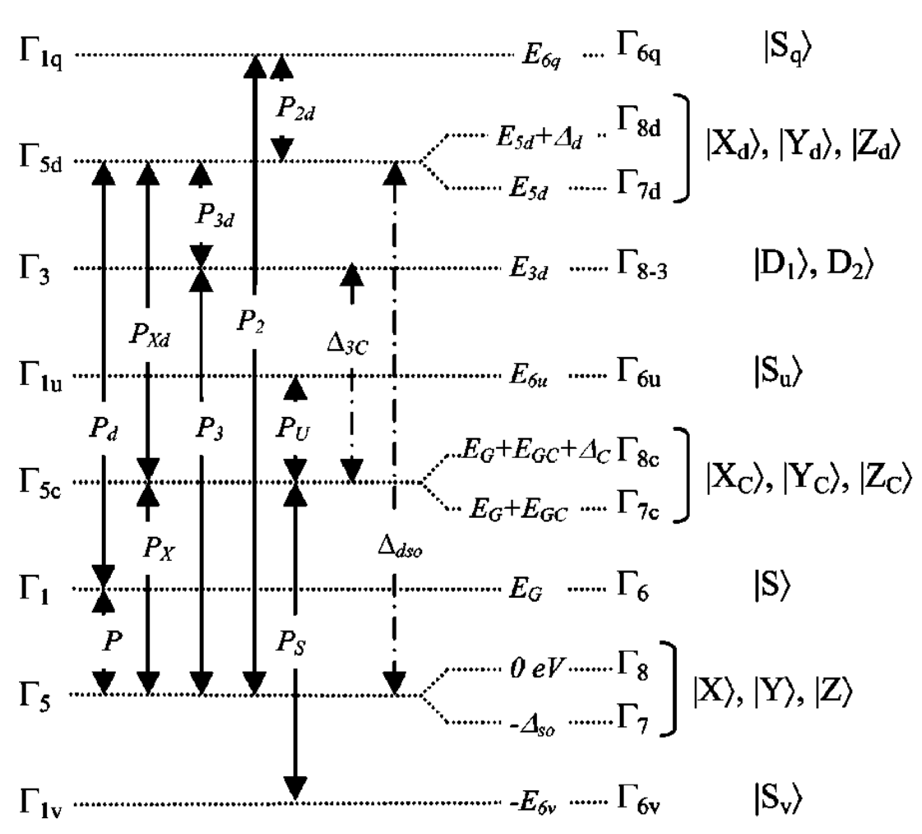}
    \caption{Thirty-band $\mathbf{k}\cdot\mathbf{p}$ model: For the $O_h$ group (applicable to Si and Ge), the various $\mathbf{k}\cdot\mathbf{p}$ couplings at $k=0$, e.g., $P{=}\left\langle S|p_x|iX\right\rangle$, $P_d{=}\left\langle S|p_x|iX_d\right\rangle$, and $P_X{=}\left\langle X_c|p_y|iZ\right\rangle$, are outlined. Reproduced from \textcite{richard2004energy} with permission. The resultant $\mathbf{k}\cdot\mathbf{p}$ Hamiltonian is diagonalized in the basis of the wavefunctions of the bands (denoted on the right) to evaluate the electronic band structure in the vicinity of $k=0$ \cite{chow2013semiconductor}.}
    \label{fig:kdotp}
\end{figure}

In principle, the band structure near any $\mathbf{k}{=}\mathbf{k_0}$ can be calculated in this way, given a large basis set of $u_{n\mathbf{k_0}}(\mathbf{r})$ \cite{richard2004energy}. In the vicinity of high symmetry points of the BZ, the $\mathbf{k}\cdot\mathbf{p}$ method is very effective in deducing the band structure $E_{n\mathbf{k}}$ \cite{pfeffer1996five,peter2010fundamentals}. An important parameter emerges near the high symmetry points, namely the {\it effective mass} $m^*$ of the band. Close to the $\Gamma$ point, for example, we have $E_{n\mathbf{k}}=E_{n\mathbf{0}}{+}\hbar^2k^2/(2m^*)$, where
\begin{equation}
    \frac{1}{m^*}\approx\frac{1}{m_0}+\frac{2}{m_0^2k^2}\!\sum_{n'\neq n}\!\frac{|\left\langle u_{n\mathbf{0}}(\mathbf{r})|\mathbf{k}\cdot\mathbf{p}|u_{n'\mathbf{0}}(\mathbf{r})\right\rangle|^2}{E_{n\mathbf{0}}-E_{n'\mathbf{0}}}.
\end{equation}
The significance of the effective mass is that an electron in a crystal has a mass different from that of a free electron due to the $\mathbf{k}\cdot\mathbf{p}$ coupling between electronic states in different bands. 

\subsection{Density functional theory}
Density functional theory (DFT) is widely implemented for first-principle calculations of semiconductor band structures \cite{sholl2022density}. DFT attempts to solve Eq.~\eqref{req1:manybody} under the key assumption that the properties of the ground state of a system of interacting electrons are determined by its ground state density $n_{\sigma}(\mathbf{r})$ 
%$n_{\sigma}(\mathbf{r}){=}\Sigma_{\sigma_1...\sigma_N}\int d^3r_1...d^3r_N|\Psi(\mathbf{r}\sigma,\mathbf{r}_1\sigma_1,...,\mathbf{r}_N\sigma_N)|^2$ 
only \cite{hohenberg1964inhomogeneous,martin2020electronic}. Further simplification is achieved in two steps: i) the interacting system is mapped onto an auxiliary (fictitious) noninteracting Kohn-Sham system of electrons \cite{kohn1965self} having the same ground state density $n_
{\sigma}(\mathbf{r})$, and ii) it is assumed that the electron-electron interactions are well approximated by an effective exchange-correlation energy term \cite{becke1993new}. The result is the self-consistent Kohn-Sham equations, which minimize the total energy with respect to the density. These equations are given by
\begin{equation}\label{req:KS}
    \bigg[\frac{p^2}{2m_0}+V(\mathbf{r})+U\big([n];\mathbf{r}\big)+V_{xc}^\sigma\big([n];\mathbf{r}\big)\bigg]\psi_{\alpha\sigma}(\mathbf{r})=\epsilon_{\alpha\sigma}\psi_{\alpha\sigma}(\mathbf{r}),
\end{equation}
where $\psi_{\alpha\sigma}$ are the Kohn-Sham orbitals satisfying $n_\sigma(\mathbf{r})=\Sigma_{\alpha} \theta(\mu-\epsilon_{\alpha\sigma})|\psi_{\alpha\sigma}(\mathbf{r})|^2$ with $\mu$ denoting the chemical potential, $\theta$ being the Heaviside function, and $\sigma=(\uparrow,\downarrow)$ denoting the spin. In Eq.~\eqref{req:KS}, $V(\mathbf{r})$ is the ionic potential coming from the core electrons and the nucleus, $U([n];\mathbf{r})=\int d^3r'[n_\uparrow(\mathbf{r}')+n_\downarrow(\mathbf{r}')]/|\mathbf{r}-\mathbf{r}'|$, and $V_{xc}^\sigma\big([n];\mathbf{r}\big)=\delta E_{xc}/\delta n_\sigma(\mathbf{r})$ is the exchange-correlation potential, given by the functional derivative of the exchange-correlation energy $E_{xc}$, which, in practice, is estimated using the local spin density (LSD) approximation, generalized gradient approximation (GGA), or exact exchange (EXX), etc. 

Iteratively minimizing the total energy $E[n]$ of the Hamiltonian in Eq.~\eqref{req:KS} using a plane wave expansion of Kohn-Sham orbitals according to the Bloch theorem produces the band energies along high-symmetry directions in the BZ \cite{grosso2013solid}. Notably, DFT calculations in semiconductors face significant complexity since the valence wave-functions oscillate strongly in the vicinity of the atomic core, and a large number of plane waves are required to capture this. However, the approximation of the inner electrons being part of the ionic core and the valence electrons determining the electronic properties means that the ionic potential $V(\mathbf{r})$ `feels' both the nuclear attraction and the repulsion of the inner electrons. This is known as the pseudopotential method. In this method, the Kohn Sham wavefunctions for the $m^{\mathrm{th}}$ band are expanded in a plane-wave basis, $\psi_{\mathbf{k},m}(\mathbf{r})=e^{i\mathbf{k}\cdot\mathbf{r}}\sum_\mathbf{G}c_{\mathbf{k},m}(\mathbf{G})\,e^{i\mathbf{G}\cdot\mathbf{r}}$, and the ionic potential is simply the sum of the pseudopotentials of all the atoms in the system, denoted by $w(\mathbf{r},\mathbf{r}')=w(r)\delta(r-r')$. A few well-known forms of the pseudopotentials are the Kleinman and Bylander potential \cite{kleinman1982efficacious}, the Heine-Abarenkov potential \cite{heine1964new}, and the Hamann potential \cite{hamann1979norm}. In the plane wave basis, Eq.~\eqref{req:KS} then becomes $\sum_{\mathbf{G}'}\mathcal{H}_{\mathbf{G},\mathbf{G}'}(\mathbf{k})c_{\mathbf{k},m}(\mathbf{G}')=\epsilon_{\mathbf{k},m}c_{\mathbf{k},m}(\mathbf{G})$, where 
\begin{eqnarray}
    \mathcal{H}_{\mathbf{G},\mathbf{G}'}(\mathbf{k})&=&\frac{1}{2}|\mathbf{k}+\mathbf{G}|^2\delta_{\mathbf{G},\mathbf{G}'}+w(\mathbf{k}+\mathbf{G},\mathbf{k}+\mathbf{G}')+v_{\text{Hartree}}(\mathbf{G}-\mathbf{G}')+v_{\text{xc}}(\mathbf{G}-\mathbf{G}').
\end{eqnarray}
The last three terms depend on the Fourier transform of the potential energies, e.g., $v_{\text{Hartree}}(\mathbf{G}-\mathbf{G}')=4\pi n(\mathbf{G}-\mathbf{G}')/|\mathbf{G}-\mathbf{G}'|^2$. The advantage of the pseudopotential method is that the effect of the ionic potential is contained in the Fourier transform of $w(r)$, known as {\it form factors} \cite{grosso2013solid,cohen1966band}. The symmetry of semiconductor lattices implies that only a few form factors (three in Si and Ge) are needed to calculate this term.

\subsection{Tight-binding approach}
%\begin{figure}[htbp!]
    %\centering
    %\includegraphics[width=\linewidth]{Part_I/figures/reviewtightbindingdrawing.png}
   % \caption{Linear combination of atomic orbitals leading to bond integrals for $s$, $p$ or $d$ type orbitals joined along a bond. Radial symmetry along the bond is assumed, leading to the designation of the bond as $\sigma$, $\pi$ or $\delta$. These fundamental bond integrals are assembled through the Slater–Koster table \cite{slater1954simplified} to construct a tight-binding Hamiltonian.}
    %\label{fig:TBLCAO}
%\end{figure}
While the methods discussed so far have assumed that core electrons are inert and that valence electrons mostly determine the band structure, the tight-binding (TB) approach assumes the opposite, i.e., that electrons are tightly bound to their nuclei, as in atoms. In a lattice, individual wave functions will overlap, and electronic wave functions will be approximated by the linear combinations of atomic orbitals (LCAO) \cite{kittel2018introduction}. 

Suppose $\mathbf{r}_{jl}= \mathbf{R}_{j}+ \mathbf{r}_{l}$ is the position of an atom in the lattice, where $\mathbf{R}_{j}$ is the position of the $j$-th primitive cell of the Bravais lattice and $\mathbf{r}_{l}$ is the position of the $l$-th atom within the primitive cell. Diamond has one atom per unit cell, while zincblende has two. The atomic orbital wave function is given by $h_l\psi_{ml}(\mathbf{r}-\mathbf{r}_{jl})=e_{ml}\psi_{ml}(\mathbf{r}-\mathbf{r}_{jl})$, where $h_l(\mathbf{r}-\mathbf{r}_j)$ is the isolated Hamiltonian for the $l$-th atom in the $j$-th unit cell. The total Hamiltonian $\mathcal{H}$ of the crystal is then approximated by the sum of the atomic Hamiltonians and of interactions allowing for orbital overlap of nearby atoms: $\mathcal{H}\simeq\sum_{jl}h_l(\mathbf{r}-\mathbf{r}_j)+\mathcal{H}_{\text{int}}$. Due to the translational symmetry of the crystal, the unperturbed lattice wave function is given by $\psi_{ml\mathbf{k}}=\frac{1}{\sqrt{N}}\sum_je^{i\mathbf{k}\cdot\mathbf{r}_{jl}}\psi_{ml}(\mathbf{r}-\mathbf{r}_{jl})$, where $N$ is the number of unit cells in the crystal \cite{jonestheory}. The solution to the total Hamiltonian can then be written as a linear combination of $\psi_{ml\mathbf{k}}$: $\psi_\mathbf{k}=\sum_{ml}c_{ml}\psi_{ml\mathbf{k}}$. This procedure essentially produces an eigenvalue problem:
\begin{equation}\label{req:TB}
    \sum_{ml}\big(\mathcal{H}_{ml,m'l'}-E_k\delta_{mm'}\delta_{ll'}\big)c_{m'l'}(\mathbf{k})=0.
\end{equation}

The simplification of Eq.~\eqref{req:TB} can be achieved by implementing the symmetry arguments relevant to semiconductors and restricting the interaction to nearest neighbors. The unperturbed matrix element is $\mathcal{H}_{ml,m'l'}=\sum_{j=1}^5e^{i\mathbf{k}\cdot(\mathbf{R}_{j}+\mathbf{r}_{l'}-\mathbf{r}_{l})}\langle\psi_{ml}(\mathbf{r}-\mathbf{r}_{jl})|\mathcal{H}|\psi_{m'l'}(\mathbf{r}-\mathbf{r}_{jl'})\rangle$, where the sum of $j$ runs from $j=1$ to $j=5$ as the atom in the unit cell of diamond and zincblende has four neighboring atoms. The interaction matrix elements can also be simplified, taking into account that the overlap of neighboring orbitals depends only on certain linearly independent parameters: $\langle s|\mathcal{H}|s\rangle$, $\langle s|\mathcal{H}|p_z\rangle$, $\langle p_z|\mathcal{H}|p_z\rangle$, and $\langle p_x|\mathcal{H}|p_z\rangle$ \cite{slater1954simplified}. In addition, the Bloch phase factor $e^{i\mathbf{k}\cdot(\mathbf{R}_{j}+\mathbf{r}_{l'}-\mathbf{r}_{l})}$ only assumes certain values due to the nearest neighbors \cite{paxton2009introduction}. Taking the example of a diamond crystal (e.g., Si) with one atom in a unit cell consisting of $sp^3$ orbitals, the tight-binding approach produces a $8\times 8$ matrix that can be diagonalized to extract the band structure. Remarkably, this matrix depends only on a few {\it tight-binding parameters}, which are often evaluated by fitting the TB result to the band structure obtained from ab-initio calculations. 

Simple TB models often capture the valence band with excellent agreement, but not the conduction band. This is because, in the valence band, electrons are mostly concentrated in the bonds, and the atomic description of being bound to the nucleus fits well. However, in the conduction band, electrons are more delocalized. A TB approach would therefore require taking more orbitals into account; see, e.g., \textcite{vogl1983semi} ($sp^3s^*$) and \textcite{jancu1998empirical} ($sp^3d^5s^*$). 

\subsection{Spin-orbit coupling}
The final ingredient in studying the electronic structure of semiconductors is to include the spin angular momentum, which introduces spin-orbit coupling (SOC). Intrinsic SOC originates due to relativity, since a particle moving in an ionic potential $V_0(\mathbf{r})$ with velocity $\mathbf{v}$ `feels' a magnetic field in its frame of reference according to the Lorentz transformation: $\mathbf{B'}\propto(\mathbf{v}\times\mathbf{\nabla}V_0)/c^2$ \cite{kittel2018introduction}. The resultant interaction is
\begin{equation}\label{eq:intro15}
\mathcal{H}_{\text{SO}}\propto\mathbf{S}\cdot\mathbf{v}\times\left(\frac{dV_0}{dr}\frac{\mathbf{r}}{r}\right)=\lambda_{SO}\mathbf{S}\cdot\mathbf{r}\times\mathbf{p}=\lambda_{SO}\mathbf{S}\cdot\mathbf{L}\,,
\end{equation} 
where $\lambda_{SO}$ scales inversely to $c^2$ and is proportional to the atomic number \cite{slater1930atomic}. Note that the relativistic {\it Dirac equation} produces a generic spin-orbit coupling term, and the Lorentz transformation argument is rather simplistic, as it neglects the contribution of the atomic core to SOC \cite{winkler2001spin,luttinger1956quantum,bychkov1984oscillatory,tahan2005rashba}.

In $\mathbf{k}\cdot\mathbf{p}$ theory, SOC can be included directly, as the interaction is local within the unit cell. However, the $\mathbf{k}\cdot\mathbf{p}$ formalism is most useful when the effect of spin-orbit coupling is relevant near the high-symmetry points (e.g., holes in Ge). For example, at $\mathbf{k}{=}0$, the atomic orbitals are angular momentum eigenstates: for instance, the $s$ orbitals correspond to $l{=}0$, and the $p$ orbitals have nonzero angular momentum $l{=}1$. Without SOC, the orbitals are degenerate $|l,m_l\rangle$ states. Including SOC reduces this degeneracy, as orbitals with spin angular momentum are eigenstates of the total angular momentum $\mathbf{J}=\mathbf{L}+\mathbf{S}$. For the upper valence bands at the $\Gamma$-point, the six-fold degeneracy of $l{=}1,\,s{=}\frac{1}{2}$ is lifted to produce a spin-orbit gap $\Delta_{SO}$ between four-fold degenerate $j{=}l{+}s{=}\frac{3}{2}$ states and two-fold degenerate $j{=}l{-}s{=}\frac{1}{2}$ states \cite{dresselhaus1955spin,luttinger1955motion}. The new total angular-momentum eigenstates can be written as combinations of $s$ and $p$ orbitals, according to the Clebsch–Gordan coefficients. The $\mathbf{k}\cdot\mathbf{p}$ formalism captures spin-orbit coupling very well near high-symmetry points, since the perturbative $\mathbf{k}\cdot\mathbf{p}$ matrix elements with SOC can easily be evaluated in the original basis without SOC \cite{kane1956energy,birpikus}. In DFT, spin-orbit coupling is either incorporated implicitly in the nonlocal pseudopotential projectors, or via a second variation method, or it is included variationally in the atomic-sphere approximation. In the tight binding approach, SOC is explicitly included as an on-site operator $\mathcal{H}_{SO}{=}\lambda_{SO}(L_xS_x{+}L_yS_y{+}L_zS_z)$. Improved TB models (e.g., $sp^3d^5s^*$ with spin, $40\times 40$ matrix) can therefore include the effect of spin-orbit coupling in the band structure at all $\mathbf{k}$ values.

\subsection{From theory to qubit architectures}
A key step in using semiconductor electronic structure to understand spin qubits is the construction of an effective low energy Hamiltonian in the presence of the quantum dot confinement. The resultant qubit properties can be understood by considering a reduced Hilbert space, where the complex multiband structure is projected onto one or a few relevant bands \cite{friesen2003practical}. The $\mathbf{k}\cdot\mathbf{p}$ formalism is often efficient in providing an analytical, closed-form description of the qubit properties. Firstly, it allows for an effective mass description of the qubit dynamics, and the governing physics is often well captured in the envelope function approximation (EFA). The EFA expresses a crystal electron (or hole) wavefunction as a slowly varying envelope function determined by the QD confinement, and multiplied by a rapidly oscillating Bloch state at the band edge. Secondly, SOC is easily incorporated into descriptions based on the effective mass approximation near high symmetry points in the BZ. The sub-band Kane model \cite{kane1956energy} can be written using quasi-degenerate perturbation theory, which provides effective masses, g-factors, and spin–orbit coefficients that incorporate interband mixing beyond a single-band approximation. For hole qubits, the dynamics of holes in the valence band and the intrinsic nature of the SOC can be captured by the Luttinger-Kohn Hamiltonian $\mathcal{H}_{LK}$ \cite{luttinger1955motion}. Hole qubits can then be described by $\mathcal{H}_{LK}{+}V(\mathbf{r})+\mathcal{H}_Z$, where the Zeeman energy of the total angular-momentum eigenstates is $\mathcal{H}_Z{=}-2\kappa\mu_B\mathbf{B}\cdot\mathbf{J}$. 

Due to the indirect band gap, electrons in the Si conduction band have energy valleys, which can be  described by a multivalley effective mass approach. For example, a single donor-dot qubit Hamiltonian is $\mathcal{H}_D+g\mu_B\mathbf{B}\cdot\mathbf{\sigma}$, where
\begin{equation}\label{req:donorHamil}
\mathcal{H}_{\text{D}} =\sum_{i\in\{x,y,z\}} \!\left[\frac{k_{i,\parallel}^2}{2m^*_\parallel}+\frac{k_{i,\perp1}^2{+}k_{i,\perp2}^2}{2m^*_\perp}\right] \!+\!V(\mathbf{r}{-}\mathbf{R}_D) + \mathcal{H}_{\sigma\tau}.
\end{equation}
In Si/SiGe heterostructures grown in $\hat{z}$ direction, the uniform strain partially lifts the valley degeneracy, and often the valley-orbit picture is well described by the $\pm z$ valley-spin Hamiltonian $\mathcal{H}_{\text{eff}}+g\mu_B\mathbf{B}\cdot\mathbf{\sigma}$, where
\begin{eqnarray}\label{req:SiSiGeElec}
    &&\mathcal{H}_{\text{eff}} =\biggl[\frac{\hbar^2}{2}\biggl(\frac{(k_z+eA_z(x,y)/\hbar)^2}{m_l}+\frac{k_x^2+k_y^2}{m_t}\biggr)+ V(\mathbf{r})+ \Xi_u\varepsilon_{zz}\biggr]\,\mathbb{I}_{4\times 4}-2\Xi_u'\varepsilon_{xy}\,\mathbb{I}_{2\times 2}\,\tau_z{+}\mathcal{H}_{\sigma\tau}.
\end{eqnarray}
%\textcolor{red}{$\mathcal{H}_{\sigma\tau}$ undefined?} 
Here, the spin-valley Hamiltonian $\mathcal{H}_{\sigma\tau}$ contains the interplay of spin, valley, and orbital degrees of freedom. Using models based on the effective mass approximation also simplifies multi-spin studies at a modest computational cost. For example, the calculated single-particle wavefunctions can be used to calculate exchange interactions (cf.~the Heitler-London and Hund-Mulliken models); this can be extended to study singlet-triplet, exchange only, and resonant exchange qubits.

\begin{figure}
    \centering
    \includegraphics[width=0.75\linewidth]{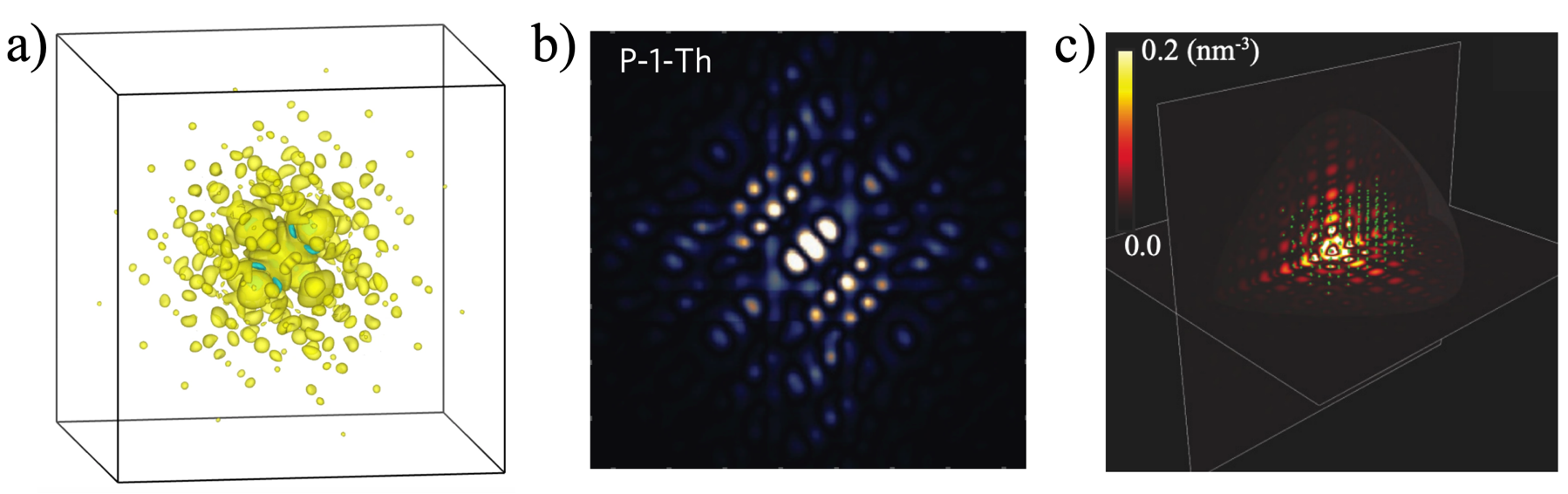}
    \caption{Application of theoretical tools for qubit description, with single shallow donors in silicon as examples. (a) Density functional theory (DFT) simulation of the charge density of a single electron on a bismuth ($^{209}$Bi) donor projected onto the $A_1$ representation, correctly predicting the ground-state symmetry characteristics of the electron wavefunction. Reproduced from \textcite{swift2020first}, licensed under CC BY 4.0. (b) Tight-binding-based multi-million atom simulation of the charge density of a single electron on a phosphorus ($^{31}$P) donor. Reproduced from \textcite{usman2016spatial} with permission. This result accurately pinpoints the location of the donor quantum dot. (c) Ground state of a $^{31}$P donor electron, calculated based on a multivalley effective mass approximation. Reproduced from \textcite{gamble2015multivalley} with permission.}
    \label{fig:theorytoqubitpipeline}
\end{figure}

DFT supplies first-principle microscopic calculations of defect levels, hyperfine constants, spin–orbit splittings, valley splittings, interface effects, etc., which cannot be measured directly but are essential inputs to spin qubit Hamiltonians \cite{yin1982theory,kresse1996efficient,swift2020first}. %DFT is often implemented as parameter generator for other models and is also a key tool in understanding qubit decoherence. 
Atomistic TB approaches enable large scale and highly resolved simulations, providing realistic device modeling of spin qubits beyond the continuum limit \cite{boykin2004valence,klimeck2007atomistic1,klimeck2007atomistic2}. These models can be used to move seamlessly from the electronic and symmetry properties of the semiconductor lattice to the study of realistic spin qubits under external fields.

%EFFECTIVE SPIN HAMILTONIAN!!!
\section{Physics of semiconductor spin qubit operation}
\subsection{Initialization and readout}
The initialization and readout of semiconductor spins typically rely on spin-to-charge conversion, in which the spin states are mapped onto charge configurations that can be differentiated electrically \cite{Field1993measurements}. Two foundational schemes are (i) energy-selective tunneling to a reservoir (Elzerman readout), in which the dot level is pulsed such that only the higher-energy spin state can tunnel out, producing a single shot signal in a nearby quantum point contact (QPC) or single electron transistor (SET) \cite{elzerman2004single}, and (ii) Pauli spin blockade (PSB) in DQDs \cite{ono2002current,borselli2011pauli,prance2012single}, where only the singlet can occupy the (0,2) charge state while triplets remain blockaded. This spin-selective blocking enables both rapid initialization into (0,2)S and high-fidelity single-shot readout via charge sensing \cite{johnson2005triplet,petta2005coherent}. Readout fidelities >99\% have been achieved on microsecond timescales using latched PSB, which maps the spin state onto a long-lived metastable charge configuration, producing a larger electrical signal \cite{harvey2018high}.

The readout toolbox expanded significantly as DC QPC/SET detectors were embedded into radio frequency (RF) resonant circuits, creating RF-QPC/RF-SET sensors with MHz–to-GHz bandwidth and dramatically improved fidelity \cite{reilly2007fast,morello2010single,cassidy2007single}. Gate-based RF reflectometry further simplified device architectures by enabling readout of the qubit state through the detection of dispersive shifts on an RF-coupled gate, eliminating the need for a dedicated charge sensor and allowing for fast readout in dense Si/CMOS arrays \cite{colless2013dispersive,connors2020rapid}. Readout can also be achieved by coupling the qubit to a superconducting resonator \cite{petersson2010charge}, as discussed in detail in Sec.~\ref{sec:spin-circuit}.

\subsection{Control }
The first proposed spin-qubit control mechanism is based on electron-spin resonance (ESR). This resonant driving mechanism remains an important strategy for realizing single- and two-qubit gates in experiments. The Hamiltonian describing the ESR was given previously in Eq.~\eqref{req:ultimateHamiltonian} and reads 
\begin{equation}
    \mathcal{H}_{s}=\frac{\omega_L}{2}\sigma_z+\Omega(t)\sigma_x,
\end{equation}
where $\omega_L=g\mu_B B_0$ is the qubit splitting and $\Omega(t)=g\mu_B B_{ac}\cos{(\omega t+\phi)}$ is an AC magnetic field oscillating at frequency $\omega$. In an interaction picture defined with respect to $U=e^{-i (\omega/2) \sigma_z t}$, $\mathcal{H}_s$ is given by $\mathcal{H}_s^R=U^\dagger \mathcal{H}_sU-i U^\dagger \dot{U}$, which can be evaluated as 
\begin{equation}
    \mathcal{H}_{s}^R
    %=-\frac{\hbar\omega}{2}\sigma_z+R_z(-\omega t)\mathcal{H}_sR_z(-\omega t)^\dagger\nonumber\\
    = \frac{\Delta}{2}\sigma_z+\frac{\Omega_0}{2}(\sigma_x\cos{\phi}+\sigma_y\sin{\phi})+\text{counter-rot},
\end{equation}
where $\Delta=\omega_L-\omega$ is the detuning of the drive, $\Omega_0=g\mu_B B_{ac}$, and where ``counter-rot'' designates terms oscillating at frequencies $\pm 2\omega$. These terms can be neglected within a rotating-wave approximation valid for $2\omega\ll \Omega_0$. For resonant driving ($\Delta=0$), rotations around any axis in the $xy$ plane can then be achieved by varying the phase $\phi$ of the drive, allowing virtual $Z$ rotations as well~\cite{mckay2017efficient}.

Later experiments achieved resonant driving based on electric dipole spin resonance (EDSR) \cite{nadj2010spin,schroer2011field,nadj2012spectroscopy,golovach2006electric}, where an AC electric field replaces $\mathbf{B}_{ac}$. This strategy requires SOC, which can be intrinsic or synthesized via a micromagnet-induced magnetic field gradient, $B_{ac}=\frac{1}{2}E_{ac}\partial B_m(z)/\partial z$. In the presence of SOC, the AC electric field produces an AC magnetic field in the rest frame of the electron, which can then be used to rotate the spin as in conventional ESR. Another prominent  control mechanism is adiabatic exchange pulsing, pioneered in singlet-triplet qubits and widely used in two-qubit $\sqrt{\text{SWAP}}$ gates and exchange only qubits. Under adiabatic evolution, a quantum state evolves slowly and continuously while remaining in an instantaneous eigenstate. For $ST_0$ qubits, this is commonly implemented by adiabatically pulsing the interdot detuning $\epsilon(t)$, which changes the tunnel coupling and hence controls the exchange interaction  $J(\epsilon)\simeq 4t_c^2/(U-\epsilon)$, where $U$ is the charging energy and $t_c$ is the tunnel coupling. When the usual adiabatic condition $\langle m(t)|\dot{H}(t)|n(t)\rangle/(E_m(t)-E_n(t))^2\ll 1$ is satisfied, the system follows a single exchange-eigenstate trajectory, enabling phase accumulation gates in the computational basis. %pic- petta
This type of qubit control is comprised of electric pulses and is therefore robust against magnetic noise, as well as cryogenically compatible. Importantly, resonant driving is also used in DQD~\cite{khomitsky2012spin} and TQD qubits, where $J(t)$ or the magnetic field gradient $\Delta B$ are modulated by alternating voltage. 

In the context of all-electrical universal quantum computing, anisotropic exchange interactions have motivated significant work. In the presence of spin-orbit coupling, the exchange interaction no longer remains Heisenberg-like due to the inversion asymmetry \cite{bonesteel2001anisotropic}. Accessed by $g$-tensor modulation between two quantum dots, anisotropic exchange can drive coherent Rabi oscillations without any magnetic field gradient \cite{khomitsky2012spin,stepanenko2003spin,geyer2024aniso,bence2020exchange,saezmollejo2024exchange}.

\subsection{Qubit decoherence}
Spin qubit decoherence arises due to the effects of an uncontrolled noisy environment and encompasses both energy relaxation and dephasing. It is important to achieve a microscopic understanding of spin (de)coherence processes so that fidelity can be controlled and maximized either through materials-based strategies or through the design of efficient dynamical-decoupling pulse sequences that achieve substantial coherence enhancements. We can classify noise into two caterogies: (i) {\it Markovian noise}: random processes without memory (e.g., phonon relaxation, Johnson–Nyquist noise), in which the probability of a future state depends only on the present state, not on the sequence of states that led to that state; and (ii) {\it non-Markovian noise}: processes in which noise correlations persist over time, introducing a history dependence of the present state on the past dynamics. Examples of non-Markovian noise sources include charge noise~\cite{malkoc2022charge,bosco2021fully} and hyperfine interactions with the nuclear-spin bath~\cite{khaetskii2002electron,khaetskii2003electron,fischer2009dealing,cywinski2009pure, coish2009nuclear}.  Important metrics in the context of decoherence are the qubit relaxation time $T_1$, the pure dephasing time $T_\phi$, and the total (echo) dephasing time $T_2$. For Markovian noise, we have $T_2^{-1}=(2T_1)^{-1}{+}(T_\phi)^{-1}$ \cite{bulaev2005spin,golovach2004phonon}.

The relaxation time $T_1$ corresponds to the time at which the probability of the qubit being in its excited state reaches its $1/e$ decay point \cite{feher1959electron,pines1957nuclear,roth1960g,hasegawa1960spin}. Relaxation processes are typically driven by phonons: Spin-orbit coupling produces a small admixture of spin-up and spin-down states, and the electron-phonon interaction couples these admixtures, leading to relaxation \cite{khaetskii2001spin,glazov2004magnetic,bulaev2005spin,golovach2008spin}. The electron-phonon interaction is modeled by the last term of the full crystal Hamiltonian in Eq.~\eqref{req:fullcrysHam}.  For spin qubits, this is ultimately the interaction between the spin levels and the low energy, long wavelength ($k\sim \Gamma$) acoustic phonon modes \cite{tahan2002decoherence}. Phonons are quantized collective excitations of the lattice and are bosonic modes that describe the displacement of ions from their equilibrium positions. For an ion at position $\mathbf{R}$ being displaced by a phonon with wave vector $\mathbf{q}$ and polarization $\alpha$, the ion displacement vector $\mathbf{u}^{\alpha\mathbf{q}}(\mathbf{R})$ can be quantized as 
\begin{equation}
\mathbf{u}^{\alpha\mathbf{q}}=\sqrt{\frac{1}{2V_c\rho \omega_{\mathbf{q},\alpha}}}\left(a^\dagger_{-\mathbf{q},\alpha}+a_{\mathbf{q},\alpha}\right)e^{i\mathbf{q}\cdot\mathbf{R}}\hat{\mathbf{c}}^\alpha,
\end{equation}
where $a_{\mathbf{q},\alpha}$ annihilates a phonon with wave vector $\bm{q}$ and polarization $\alpha$. Here, $\hat{\mathbf{c}}$ is the polarization unit vector, $V_c$ is the volume of the crystal, $\rho$ denotes the crystal mass density, and $\omega_{\mathbf{q},\alpha}$ is the frequency of the phonon mode. The plane-wave factor $e^{i\mathbf{q}\cdot\mathbf{R}}$ is a consequence of the translational symmetry of the crystal.

The electron-phonon Hamiltonian $\mathcal{H}_{ep,\alpha\mathbf{q}}$ can be written in terms of the symmetrized tensor of the displacement gradient, i.e.,~the strain tensor $\varepsilon_{i,j}^{\alpha\mathbf{q}}{=}\frac12\big(\partial u^{\alpha\mathbf{q}}_i/\partial r_j+\partial u^{\alpha\mathbf{q}}_j/\partial r_i\big)$ \cite{tahan2014relaxation}. The result is $\mathcal{H}_{ep,\alpha\mathbf{q}}{=}\left.\partial E_{nk}\right|_{i,j}\,\varepsilon_{i,j}^{\alpha\mathbf{q}}$, where the change in the electronic band structure $\partial E_{nk}$ is parametrized by the deformation potentials $\mathcal{D}_{ij}$. Putting everything together, 
\begin{equation}
    \mathcal{H}_{ep,\alpha\mathbf{q}}=\sqrt{\frac{1}{2V_c\rho \omega_{\mathbf{q},\alpha}}}\mathcal{D}_{ij}\hat{c}^\alpha_iq_j \big(e^{-i\mathbf{q}\cdot\mathbf{R}}a^\dagger_{\mathbf{q},\alpha}+h.c.\big).
\end{equation}
Relaxation causes spin transitions  $\left|\uparrow\right\rangle\leftrightarrow\left|\downarrow\right\rangle$, which creates or annihilates a phonon in the process. These processes occur on a time scale $T_1$ given by 
\begin{equation}
    \frac{1}{T_1}=\sum_\alpha\bigg[2\pi\sum_{\mathbf{q}}\left|\left\langle\uparrow\right|H_{ep,\alpha\mathbf{q}}\left|\downarrow\right\rangle\right|^2\delta\left(\omega_L{-}\omega_{\alpha,\mathbf{q}}\right)\bigg],
\end{equation}
where $\omega_L$ is the energy splitting of the qubit ($\hbar{=}1$). Note that the relaxation rate scales as $1/T_1\propto B^5$~\cite{khaetskii2001spin,tahan2014relaxation,wang2021optimal}. At moderate magnetic fields, this scaling is observed in experiments \cite{dresier2008optical,kroutvar2004optically,amasha2008electrical}. However, in stronger fields, the relaxation rate decreases rather than increases. This is because at large $B$, the wavelengths of phonons with energy $g\mu_BB$ are much smaller than the size of the quantum dot, and the phonon-induced effects are averaged out \cite{kloeffel2013prospects}. As a result, the relaxation rate is maximal at field strengths $B$ where the phonon wavelength matches the size of the QD. %The $B$-exponent varies when the thermal energy $k_BT$ becomes comparable to $g\mu_BB$. 
At very low magnetic fields, the trend $T_1\propto B^{-5}$ due to single-phonon processes would suggest a divergence in the relaxation time. However, in reality, two-phonon processes \cite{trif2009relaxation} and hyperfine interactions with surrounding nuclear spins \cite{erlingsson2002hyperfine,koppens2007universal,coish2004hyperfine,fischer2009spin,coish2009nuclear,merkulov2002electron} limit $T_1$ in this regime. For holes, the $B$-exponent can differ depending on the primary spin-orbit coupling mechanism \cite{wang2021optimal,sarkar2025effect,gerardot2008optical}. A detailed discussion of the effects of nuclear spins on the coherence of electron and hole spin qubits was provided by~\cite{kloeffel2013prospects,stano2022review} and references therein.

Pure dephasing originates from phase uncertainty and, unlike relaxation, does not require the transmission of energy to the lattice. A timescale commonly reported in the literature is $T_2^*$, which is the time scale associated with the decay of the total in-plane magnetization of an ensemble of spins separated in space or time (e.g., a group of spins separated in space, or the time-averaged magnetization of a single spin). In experiments, this is measured through a Ramsey experiment \cite{ramsey1950molecular}, in which an initial $\pi/2$ pulse applied to, e.g., $\ket{\downarrow}$ has the effect of preparing spin coherence. The spin, now in state $\ket{\uparrow}+\ket{\downarrow}$, is then allowed to precess freely for a time $t_w$. In the presence of noise, the spin's precession frequency will deviate from its ideal value of $\omega_L$, and we can write the Hamiltonian generating spin evolution during this free-precession window as
\begin{eqnarray}
\mathcal{H}(t)=\frac{1}{2}[\omega_L+\delta\omega_L(t)]\sigma_z,
\end{eqnarray}
where $\delta\omega_L(t)$ denotes the qubit-splitting variation due to a random noise realization. Following the free-evolution time $t_w$, another $\pi/2$ pulse is applied to the qubit, which is subsequently measured in the computational basis \cite{petta2005coherent,veldhorst2014addressable}.

The noise $\delta \omega_L(t)$ leads to a random relative phase $\delta \phi(t)=\int_0^t dt'\delta \omega_L(t')$ between the qubit basis states. Since this relative phase differs slightly for each shot of the Ramsey experiment, the average qubit coherence decays as a function of time. This is known as free induction decay (FID). Contributions to FID from static or slow noise sources can be mitigated by applying dynamical-decoupling pulse sequences, which partially refocus the spin coherence \cite{koppens2008spin,press2010ultrafast,barthel2010interlaced,bluhm2011dephasing,tyryshkin2005spin}. Examples of such dynamical decoupling sequences are Hahn echo sequences [$\pi/2{-}t_w{-}\pi{-}t_w-$measure] and  Carr-Purcell Meiboom-Gill (CPMG) sequences [$\pi/2({-}t_w{-}\pi{-}t_w)^N-$measure, with $N$ denoting the number of $\pi$ pulses]. \textcite{stano2022review} have provided an extensive tabulation of the experimentally measured $T_2^*$ values, as well as the dephasing times $T_2^\mathrm{H}$ and $T_2^{\mathrm{CPMG}}$ obtained under the Hahn-echo and CPMG sequences.

For classical noise, the noise term can generally be modeled as $\delta \omega_L(t)=\sum_i  b_i(t)$, where $b_i(t)$ is the contribution from the $i^{\mathrm{th}}$ noise process, here assumed to arise from, e.g., different environmental spins or charge fluctuators. For stationary noise processes, one can then define noise correlation functions $g_{ij}(\tau)=\llangle b_i(\tau)b_j\rrangle$ depending on a single delay time $\tau$, where $\llangle \rrangle$ denotes an average over noise realizations. When $g_{ij}(\tau)=\delta_{ij}\delta(\tau)$, the noise is uncorrelated; otherwise, it is correlated spatially ($g_{ij}\neq 0$ for $i\neq j$), temporally ($g(\tau)\neq 0$ for $\tau \neq 0$), or both. 

For Gaussian noise, the decay of the qubit coherence $C(t)=\langle \sigma_+(t)\rangle/\langle \sigma_+(0)\rangle$ can be written as 
\begin{align}
    C(t)=e^{-\chi(t)},\,\,
    \chi(t)=\int \frac{d\omega}{2\pi} \frac{S(\omega)}{\omega^2} \mathcal{F}(\omega,t).
\end{align}
Here, $S(\omega)=\int dt \: e^{i\omega t}\llangle \delta \omega_L (t)\delta \omega_L\rrangle$ is the noise spectral density, and $\mathcal{F}(\omega,t)$ is a filter function depending on the pulse sequence applied to the spin. For FID, $\mathcal{F}(\omega,t)=2\mathrm{sin}^2(\omega t/2)$; different dynamical decoupling sequences produce different filter functions~\cite{cywinski2008enhance}. For white noise, for which $S(\omega)=\gamma_\phi/2=\mathrm{const}.$, we recover an exponential decay of $C(t)$ with rate $\gamma_\phi/2$, which can equivalently be obtained through a dissipative term in a Lindbladian master equation. In the opposite limit of quasistatic noise, modeled via a spectral density peaked at low frequencies, $S(\omega)\propto \delta(\omega)$, the decay is instead Gaussian rather than exponential (and therefore non-Markovian). 

The dephasing from the nuclear Overhauser field can be drastically reduced via isotopic purification and optimization of the global magnetic field orientation \cite{coish2004hyperfine,reilly2008suppressing,tyryshkin2012electron,piot2022single,mauro2024geometry}; hence, the primary dephasing mechanism for spin qubits is often charge noise, especially in dense qubit arrays. A nearby two-level charge fluctuator (TLF) tunneling between two metastable states may produce a random telegraph signal on the qubit splitting, described by $b_i(t)=\bar{b}_i\xi_i(t)$. Here, $\xi_i(t)=\pm 1/2$ is a switching function with auto-correlation function $\llangle \xi_i(t)\xi_i\rrangle=(1/4)e^{-2 \kappa_i \vert t\vert}$, where $\kappa_i$ is the rate of random jumps between the discrete values $\pm 1/2$~\cite{cywinski2008enhance}.  The effect of a distribution of TLFs with a log-uniform distribution of switching rates $\kappa_i$ is described by $1/f$ noise, for which $S(\omega)\propto 1/\omega$~\cite{paladino20141, cywinski2008enhance}. 

Various assumptions made above in writing $C(t)$ may break down in an experimental setting. For non-Gaussian noise, higher cumulants of the noise are generically non-zero and may modify the decay curve~\cite{galperin2006non,bergli2009decoherence,norris2016qubit}. Noise may also be treated quantum-mechanically, requiring that the classical fields $b_i(t)$ be promoted to quantum operators $\hat{b}_i$ evolving under their own bath Hamiltonian~\cite{paz2017multiqubit,kwiatkowski2020influence,wang2021intrinsic,mcintyre2022non}. In the quantum case, the classical average $\llangle \rrangle$  appearing in the spectral density $S(\omega)$ is replaced by a quantum average with respect to the initial state of the bath.  This can lead to observable consequences on the qubit coherence depending on whether or not the initial state of the bath is stationary under the Hamiltonian generating evolution of $\hat{b}_i$~\cite{kwiatkowski2020influence,wang2021intrinsic,mcintyre2022non}. For instance, $C(t)$ may acquire a nonzero phase [$\mathrm{Im}\:C(t)\neq 0$] arising due to the non-commutation of $\hat{b}_i(t)$ at different times. This phase has been experimentally measured for an NV-center spin subjected to a CPMG decoupling sequence~\cite{zhao2012sensing}.

In the multiqubit setting, spatial noise correlations on different qubits have also attracted significant attention in recent years~\cite{yoneda2023noise,rojas2023spatial,rojas2025origins}. Such correlations can be used to infer the locations of TLFs coupling to more than one qubit~\cite{rojas2026inferring}, which constitute a form of long-range correlated noise that is typically not included in the circuit-level simulations used to estimate the thresholds of quantum error correcting codes~\cite{fowler2012surface}. Simulations of the scaling of the logical error rate with code distance under a model of perfectly correlated noise were performed by \textcite{rojas2026scaling}, who also identified global magnetic-field drifts as another notable source of correlated noise. A theoretical analysis of correlated noise in the context of driven qubits was also carried out by \textcite{zou2024spatially}.

\section{Spin-circuit QED}\label{sec:spin-circuit}

As established in the preceding sections, many strategies for manipulating the joint state of two spins require that they have a finite exchange interaction. The short-range nature of the exchange interaction, together with the wiring bottlenecks associated with dense qubit layouts, therefore motivates the adoption of hybrid architectures enabling long-range coupling between distant qubits. Prominent among such architectures are those that harness the interaction of qubits with spectrally isolated standing-wave modes of the electromagnetic field housed in a microwave cavity. This approach is derived from the field of cavity quantum electrodynamics (QED)~\cite{mabuchi2002cavity,miller2005trapped,walther2006cavity}, which examines the interaction of atoms with optical cavities. Circuit QED with superconducting qubits is similarly descended from cavity QED; for a recent review, see~\textcite{blais2021circuit}. 

\begin{figure}
    \centering
    \includegraphics[width=0.65\linewidth]{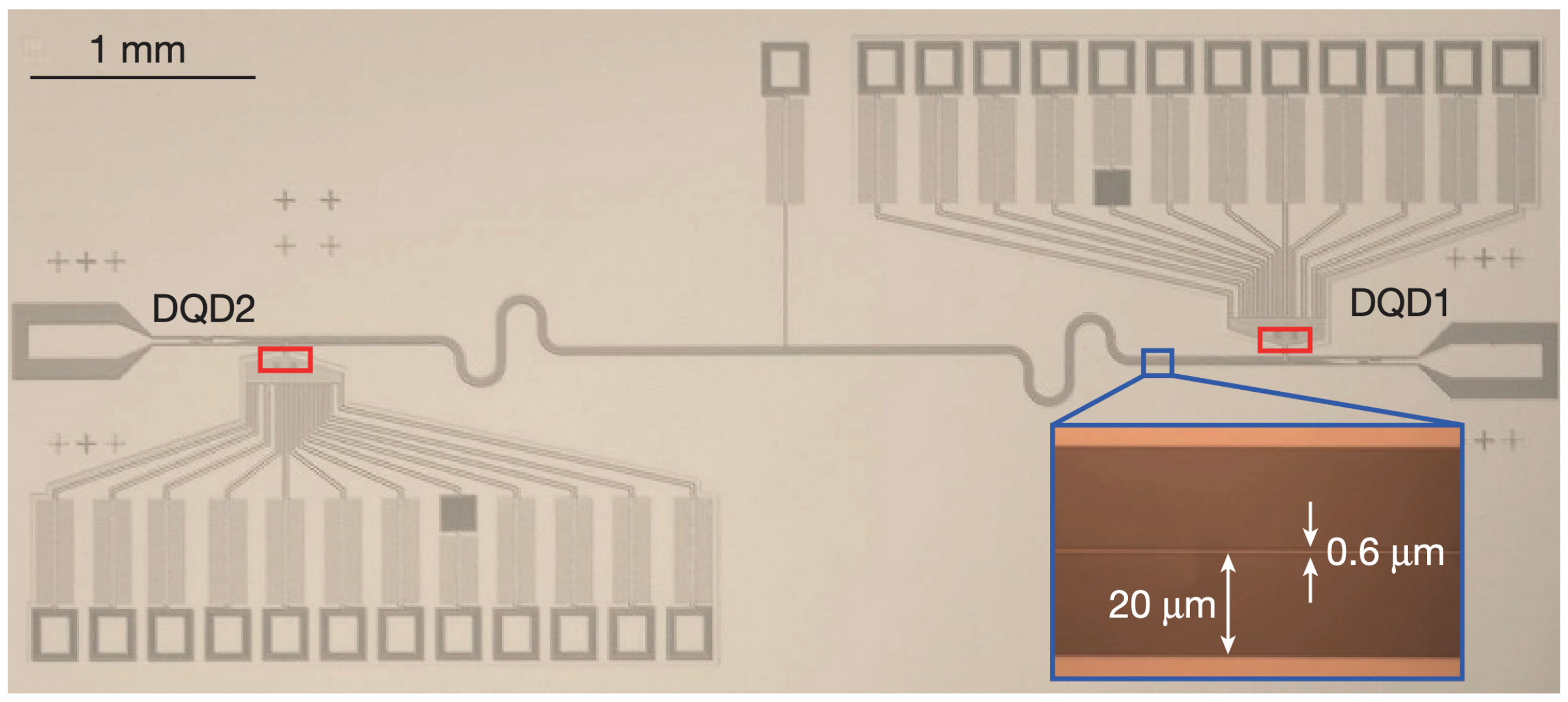}
    \caption{Optical image of a superconducting stripline resonator coupled to two double quantum dots DQD1 and DQD2. The inset shows the central superconductor (0.6 $\mu$m) and the edges of the ground planes (separated from the central strip by 20 $\mu$m). Reproduced from \textcite{mi2018coherent} with permission.}
    \label{fig:microwave-cavity}
\end{figure}

Early proposals for coupling spin qubits to light considered the use of all-optical Raman transitions involving hole states~\cite{imamog1999quantum}. Typically, however, spin qubits have splittings in the GHz regime and therefore interact most strongly with microwave-frequency radiation, motivating their integration into hybrid systems involving microwave cavities. In circuit QED architectures, the electromagnetic field is generally confined by a superconducting stripline resonator~\cite{childress2004mesoscopic,burkard2006ultra,hu2012strong,jin2012strong,tosi2014circuit} consisting of a center conductor and ground planes made of a superconducting material like Nb or Al, deposited on an insulating substrate like sapphire (Fig.~\ref{fig:microwave-cavity}). The length of the center conductor is chosen so that the fundamental mode is in the GHz range. While these resonators are distinct from the Fabry-Perot cavities used in many optical setups, their fields can be quantized in the same way, and we refer to them as cavities throughout this section for simplicity. 

\subsection{Electric-dipole coupling}

\begin{figure}
    \centering
    \includegraphics[width=0.5\linewidth]{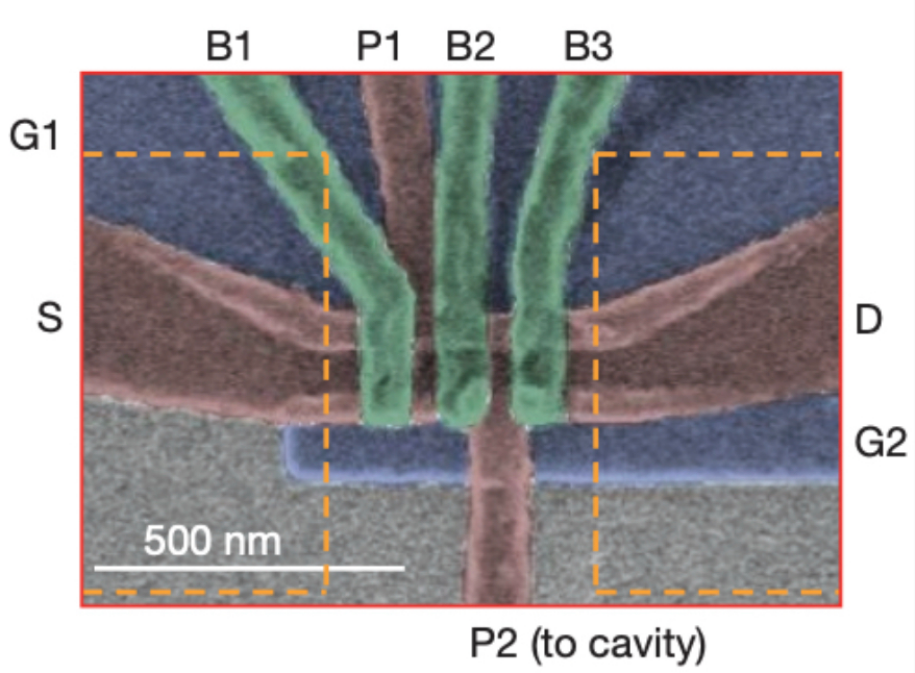}
    \includegraphics[width=0.45\linewidth]{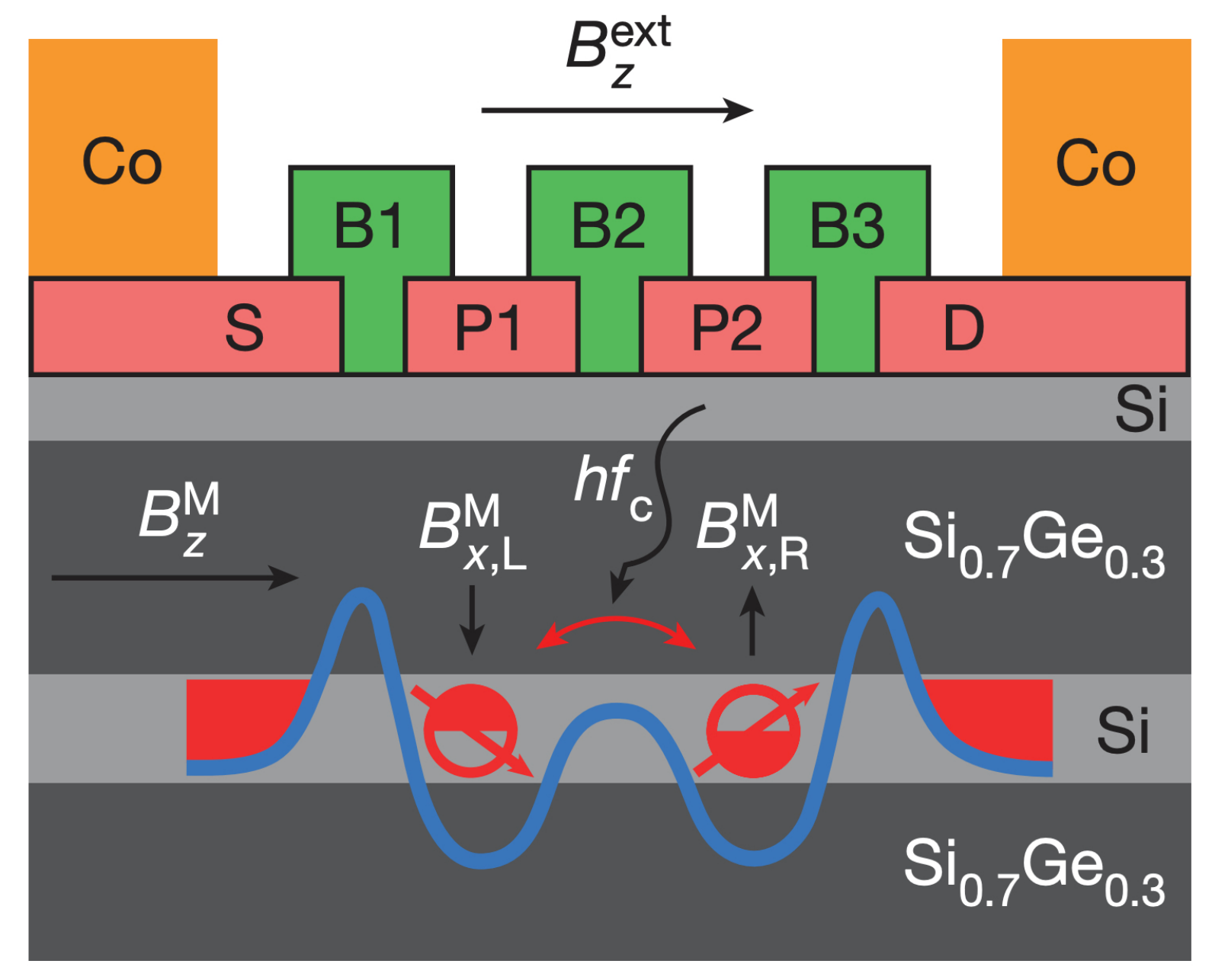}
    \caption{Left: Capacitive coupling to the cavity field can be achieved by connecting one of the plunger gates (P2) controlling the dot potentials to the cavity's central conductor. Right: The spin-charge hybridization required to define a flopping-mode qubit can be realized by applying a magnetic field gradient across the double quantum dot. Reproduced from \textcite{mi2018coherent} with permission.}
    \label{fig:device}
\end{figure}

In principle, it is possible to couple a spin directly to the magnetic field of a cavity. However, the coupling strengths associated with magnetic-dipole coupling ${-} \bm{\mu}\cdot\bm{B}$ tend to be much smaller than those associated with electric-dipole coupling. For instance, it was estimated by~\textcite{schoelkopf2008wiring} that the coupling of an electron's magnetic moment to the magnetic field of a cavity is generally on the order of around 100 Hz. For this reason, DQD-based qubits are almost always coupled to cavities via electric-dipole interactions, which typically lead to coupling strengths $\gtrsim 10$ MHz (Table~\ref{tab:strong-coupling}). In practice, such coupling is typically realized by connecting one of the plunger gates used to define the quantum dots to the central conductor of the cavity (Fig.~\ref{fig:device}).

Cavity-qubit couplings are typically derived starting from an electric dipole interaction $-e\bm{E}(\bm{r}_0)\cdot\bm{d}$, where $\bm{E}(\bm{r}_0)$ is the electric field of the cavity at the position $\bm{r}_0$ of the qubit, and $\bm{d}$ is the dipole operator of the qubit. Here, we take a slightly modified approach [following the supplement of \textcite{mcintyre2024photonic}] that accounts for possible spatial variation of the cavity field on the length scale of the qubit. We start from a cavity-qubit interaction of the form
\begin{equation}\label{cavity-qubit-interaction-1}
    H_{\mathrm{int}}=\int d^3{\bm{r}}\rho(\bm{r})V(\bm{r}),
\end{equation}
where $V(\bm{r})$ is the electric potential of the cavity mode and $\rho(\bm{r})$ is the charge density operator of the qubit. As a side note, when $V(\bm{r})$ varies slowly on the scale of $\rho(\bm{r})$, it can be expanded in a Taylor series about the position $\bm{r}_0$ of the qubit, giving
\begin{equation}\label{dipole-approximation-hamiltonian}
    H_{\mathrm{dipole}}\simeq Q V(\bm{r}_0)-\bm{d}\cdot\bm{E}(\bm{r}_0),
\end{equation}
where 
\begin{align}\label{dipole-approximation}
\begin{aligned}
    &Q=\int d^3\bm{r} \rho(\bm{r}),\,\,\,\bm{d}=\int d^3\bm{r}\rho(\bm{r})(\bm{r}-\bm{r}_0),\,\,\,\bm{E}(\bm{r}_0)=-\nabla V(\bm{r}_0).
\end{aligned}
\end{align}  
This recovers the expressions most commonly encountered in the literature. An effective cavity-qubit coupling Hamiltonian can then be obtained by quantizing the electric field and taking matrix elements of the dipole operator with respect to the qubit eigenstates.

For concreteness, we now assume that the qubit is encoded in the state of an electron (or hole) in a double quantum dot with orbital wavefunctions $\psi_{\mathrm{L}}$ and $\psi_{\mathrm{R}}$ associated with the ground states $\ket{\mathrm{L}}$ and $\ket{\mathrm{R}}$ of the left and right dots, respectively. To include potential spatial variation of the cavity field on the length scale of the DQD, we project the charge density operator onto the lowest dot orbitals and quantize the cavity field under the assumption that the qubit couples predominantly to a single cavity mode with annihilation operator $a$. This gives
\begin{align}
    &V(\bm{r})=\phi_0(\bm{r})V_{\textsc{rms}}(a + a^\dagger),\label{cavity-voltage-quantize}\\
    &\rho(\bm{r})=-e \lvert \psi_{\mathrm{L}}(\bm{r})\rvert^2\ketbra{\mathrm{L}}-e\vert \psi_{\mathrm{R}}(\bm{r})\rvert^2\ketbra{\mathrm{R}},\label{charge-density-quantize}
\end{align}
where $V_{\textsc{rms}}$ is the amplitude of the zero-point voltage fluctuations of the resonator and $\phi_0(\bm{r})$ is a dimensionless mode function that solves Poisson's equation with the boundary conditions imposed by the specific device geometry. Substituting Eqs.~\eqref{cavity-voltage-quantize} and \eqref{charge-density-quantize} back into Eq.~\eqref{cavity-qubit-interaction-1} gives~\cite{childress2004mesoscopic,burkard2020superconductor} 
\begin{equation}
    H_{\mathrm{int}}=\bar{g}\tau_0(a+a^\dagger)+g_{\mathrm{c}}\tau_z(a+a^\dagger),
\end{equation}
where $\tau_0=\ketbra{\mathrm{L}}+\ketbra{\mathrm{R}}$ and $\tau_z=\ketbra{\mathrm{L}}-\ketbra{\mathrm{R}}$ are pseudospin Pauli operators, and where 
\begin{align}
    \bar{g}=\frac{g_{\mathrm{L}}+g_{\mathrm{R}}}{2},\,\,\,
    g_{\mathrm{c}}=\frac{g_{\mathrm{L}}-g_{\mathrm{R}}}{2}.
\end{align}
The charge-qubit coupling $g_{\mathrm{c}}$ is proportional to the difference in the lever arms $\alpha_{\mathrm{L,R}}$ of the cavity on the left and right dots. This follows from writing the couplings $g_{\mathrm{L,R}}$ of the left and right dots to the cavity as  $g_{\mathrm{L,R}}=-e \alpha_{\mathrm{L,R}}V_{\textsc{rms}}$ with
\begin{equation}
    \alpha_{\mathrm{L,R}}=\int d^3{\bm{r}}\:\phi_0(\bm{r})\lvert \psi_{\mathrm{L,R}}(\bm{r})\rvert^2.
\end{equation}
If $\phi_0(\bm{r})$ is vanishingly small wherever, e.g., $\lvert \psi_\mathrm{R}(\bm{r})\rvert^2$ has significant weight, then an asymmetric coupling $\propto \ketbra{\mathrm{L}}$ like the one considered by~\textcite{childress2004mesoscopic} is recovered. In the literature, the term proportional to the identity on the qubit is often transformed away implicitly through a displacement transformation on the cavity, leaving a $\tau_z$-like coupling~\cite{burkard2020superconductor}. 
However, the consequences of the $\tau_0$-like term should be considered carefully in the case where the lever arms are modulated through, e.g., an applied gate voltage designed to introduce time-dependence into $\psi_{\mathrm{L,R}}$.  

We model the charge qubit using the Hamiltonian $H_\mathrm{c}=\varepsilon\tau_z/2+t_c\tau_x$, where $\varepsilon$ is the DQD detuning, $t_c$ is the tunnel coupling between the left and right dots, and $\tau_x=\ketbra{\mathrm{L}}{\mathrm{R}}+\mathrm{h.c}$. At the charge degeneracy point $\varepsilon=0$, the basis states of the charge qubit are given by the bonding and anti-bonding orbital states $\ket{\pm}=(\ket{\mathrm{L}}\pm \ket{\mathrm{R}})/\sqrt{2}$. With $\sigma_x=\ketbra{+}{-}+\mathrm{h.c.}$ denoting the Pauli-X operator of the qubit, the $\tau_z$-like coupling in $H_{\mathrm{int}}$ then yields a transverse cavity-qubit coupling. Neglecting the $\tau_0$-like term under the assumption that both lever arms are static, the effective Hamiltonian of the cavity and charge qubit is then described by the well-known Rabi model~\cite{rabi1936process,rabi1937space,braak2011integrability},
\begin{equation}\label{charge-qubit-rabi}
    H_{\mathrm{R}}=\frac{\omega_{\mathrm{q}}}{2}\sigma_z+\omega_{\mathrm{c}}a^\dagger a +g_{\mathrm{c}}\sigma_x(a+a^\dagger),
\end{equation}
where for $\varepsilon=0$, the qubit splitting is given by $\omega_{\mathrm{q}}=2t_c$.

Ultimately, the susceptibility of charge qubits to charge noise provides a strong motivation for coupling the cavity to the electron or hole’s \textit{spin} degree of freedom instead. One way to couple an electron spin qubit to a cavity is to apply a transverse magnetic field gradient $\Delta B_x$ across the DQD to realize a flopping-mode qubit (Fig.~\ref{fig:device})~\cite{hu2012strong,beaudoin2016coupling,benito2017input,benito2019electric}. (For holes, the same effect can be achieved using the intrinsic spin-orbit coupling associated with valence band states~\cite{kloeffel2013circuit,mutter2021natural}.) The role of the magnetic field gradient is to generate a synthetic spin-orbit coupling, which hybridizes the electron's spin and charge degrees of freedom according to~\cite{hu2012strong,beaudoin2016coupling,benito2017input}
\begin{equation}\label{flopping-mode}
    H_{\mathrm{FM}}=\frac{\varepsilon}{2}\tau_z+t_c\tau_x+\frac{B_z}{2}\sigma_z'+\frac{\Delta B_x}{2}\sigma_x'\tau_z,
\end{equation}
where notably, the magnetic field $B_z$ setting the spin quantization axis is perpendicular to the field gradient $\Delta B_x$. We denote by $\sigma_{x,z}'$ the Pauli operators associated with the spin degree of freedom, reserving the unprimed operators for the qubit itself. At the charge degeneracy point, the ground and first-excited states encoding the flopping-mode qubit are given by $\ket{g}\simeq \ket{-,\downarrow}$ and $\ket{e}=\cos{(\Phi_{\textsc{so}}/2)}\ket{-,\uparrow}+\sin{(\Phi_{\textsc{so}}/2)}\ket{+,\downarrow}$, where $\ket{\pm}\propto \ket{\mathrm{L}}\pm \ket{\mathrm{R}}$ are the bonding and anti-bonding orbital states of the DQD, and where $\Phi_{\textsc{so}}\in(0,\pi)$ is an angle quantifying the degree of spin-orbit mixing according to $\Phi_{\textsc{so}}=\arctan{\Delta B_x/(2t_c-B_z)}$~\cite{benito2017input}.  In this case, the strength of the coupling between the cavity and the flopping-mode spin qubit is controlled by the transition-dipole matrix element $\langle e\vert \tau_z\vert g\rangle\approx \sin{(\Phi_{\textsc{so}}/2)}$. Neglecting the effects of higher excited states, the Hamiltonian of the cavity and flopping-mode qubit is again of Rabi form:
\begin{equation}\label{flopping-mode-rabi}
    H_{\mathrm{R}}=\frac{\omega_{\mathrm{q}}}{2}\sigma_z+\omega_{\mathrm{c}}a^\dagger a + g_{\mathrm{s}}\sigma_x(a+a^\dagger),
\end{equation}
where $g_{\mathrm{s}}\approx g_{\mathrm{c}}\sin{(\Phi_{\textsc{so}}/2)}$ is the spin–cavity coupling strength and $\omega_{\mathrm{q}}=[\sqrt{(2t_c+B_z)^2+\Delta B_x^2}-\sqrt{(2t_c-B_z)^2+\Delta B_x^2}]/2$ is the qubit splitting~\cite{benito2017input}.

\begin{figure}
    \centering
    \includegraphics[width=0.6\linewidth]{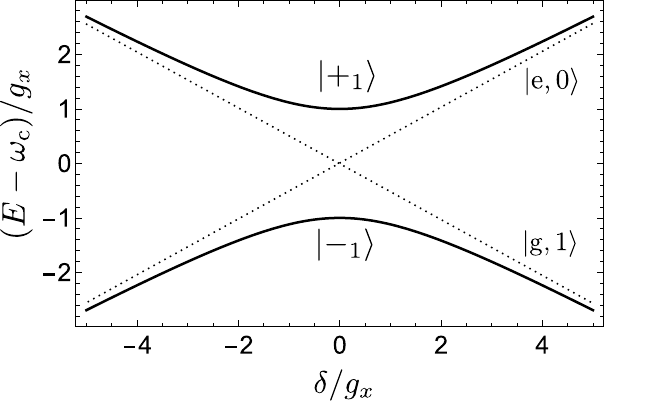}
    \caption{Hybridization of $\ket{g,1}$ and $\ket{e,0}$ results in an avoided crossing at zero detuning with an energy splitting equal to $2g_x$. The spectrum of the Jaynes-Cummings eigenstates $\ket{\pm_1}$ is plotted in black, while the decoupled eigenstates ($g_x=0$) are plotted as dotted lines.}
    \label{fig:jaynes-cummings}
\end{figure}

\subsection{The Jaynes-Cummings Hamiltonian}

We now consider a Rabi Hamiltonian of the form given in Eqs.~\eqref{charge-qubit-rabi} and \eqref{flopping-mode-rabi}, with the coupling strength $g_{\mathrm{c}}$ or $g_{\mathrm{s}}$ now denoted as $g_x$, with the understanding that a transversal coupling $\sigma_x(a+a^\dagger)$ can be generated for both charge-like and spin-like degrees of freedom. For coupling strengths $g_x\ll\lvert \omega_{\mathrm{q}}+\omega_{\mathrm{c}}\rvert$, the U(1)-symmetry-breaking (number-non-conserving) terms $a^\dagger \sigma_+$ and $a\sigma_-$ can be neglected within a rotating-wave approximation. The resulting Jaynes-Cummings Hamiltonian $H_{\mathrm{JC}}$ is then given by
\begin{equation}
    H_{\mathrm{JC}}=\frac{\omega_{\mathrm{q}}}{2}\sigma_z+\omega_{\mathrm{c}}a^\dagger a + g_x(\sigma_+ a + \sigma_- a^\dagger).
\end{equation}
The Jaynes-Cummings Hamiltonian exhibits a continuous U(1) symmetry related to the conservation of the total number $\hat{n}_{\mathrm{tot}}=a^\dagger a +\ketbra{e}$ of excitations in the combined qubit-cavity system: $[H_{\mathrm{JC}},\hat{n}_{\mathrm{tot}}]=0$. Formally, this conservation law corresponds to invariance under the unitary transformation generated by $U(\phi)=e^{-i\phi\hat{n}_{\mathrm{tot}}}$ for any $\phi\in[0,2\pi)$. The existence of two degrees of freedom and two conserved quantities—the Hamiltonian itself together with $\hat{n}_{\mathrm{tot}}$—implies integrability in the sense of Liouville~\cite{arnol2013mathematical}. The Hilbert space $\mathscr{H}=\ket{g,0}\oplus\sum_{n=1}^\infty \mathscr{H}_n$ of the qubit-cavity system can consequently be decomposed as a direct sum of dynamically invariant subspaces $\mathscr{H}_n$, where each $\mathscr{H}_n$ is a subspace of $\mathscr{H}$ spanned by the two states $\ket{e,n-1}$ and $\ket{g,n}$ having the same eigenvalue $n$ of $\hat{n}_{\mathrm{tot}}$. The Jaynes-Cummings Hamiltonian can therefore be solved by diagonalizing a family of $2 \times 2$ matrices $H_n$, each describing the restriction of $H_{\mathrm{JC}}$ to $\mathscr{H}_n$. Written in the basis $\{\ket{e,n-1},\ket{g,n}\}$,
\begin{equation}\label{h-n}
    H_n=\begin{bmatrix}
       (n-1)\omega_{\mathrm{c}}+\frac{\omega_{\mathrm{q}}}{2} & g_x\sqrt{n} \\
       g_x\sqrt{n} & n\omega_{\mathrm{c}}-\frac{\omega_{\mathrm{q}}}{2}
    \end{bmatrix}.
\end{equation}
This matrix can be diagonalized to solve for the spectrum of $H_{\mathrm{JC}}$, yielding the well-known Jaynes-Cummings ladder characterized by the eigenvalues
\begin{equation}\label{jaynes-cummings-eigenvalues}
    \lambda_{\pm,n}=\omega_{\mathrm{c}}\left(n-\frac{1}{2}\right)\pm \frac{1}{2}\sqrt{\delta^2+4 g_x^2n}
\end{equation}
and corresponding eigenstates
\begin{align}\label{jaynes-cummings-eigenstates}
\begin{aligned}
    &\ket{+_n}=\cos{\Theta_n}\ket{e,n-1}+\sin{\Theta_n}\ket{g,n},\,\,\,\ket{-_n}=-\sin{\Theta_n}\ket{e,n-1}+\cos{\Theta_n}\ket{g,n}.
\end{aligned}
\end{align}
Here, we have introduced the qubit-cavity detuning $\delta=\omega_{\mathrm{q}}-\omega_{\mathrm{c}}$, in terms of which the mixing angle $\Theta_n$ is defined by $\tan{2\Theta_n}=2g_x\sqrt{n}/\delta$. The spectrum of the Jaynes-Cummings Hamiltonian is therefore given by a non-degenerate ground state $\ket{g,0}$ together with an infinite ladder of disconnected doublets $\ket{\pm_n}$, which, at zero detuning, are separated in energy by $2g_x\sqrt{n}$ (Fig.~\ref{fig:jaynes-cummings}).

Under evolution generated by $H_{\mathrm{JC}}$, a state initialized in $\mathscr{H}_n$ will remain confined to $\mathscr{H}_n$ for all times. For example, for an initial product state $\ket{e,0}\in\mathcal{H}_1$, the probability $P(e)$ of measuring the qubit in $\ket{e}$ at time $t$ can be solved using Eqs.~\eqref{jaynes-cummings-eigenvalues} and \eqref{jaynes-cummings-eigenstates}, giving
\begin{equation}
    P(e)=\lvert \langle e,0\vert e^{-iH_{\mathrm{JC}}t}\vert e,0\rangle\rvert^2=\cos^2{(g_x t)}.
\end{equation}
These oscillatory dynamics---known as vacuum Rabi oscillations---describe a single excitation being coherently traded between the qubit and cavity with a frequency $2g_x$. In a realistic implementation, however, decoherence will invariably dampen the visibility of these oscillations. The system is said to be in the strong coupling regime when the time scale $\pi/g_x$
for vacuum Rabi oscillations is short relative to the time scale on which decoherence occurs. Typical sources of decoherence are photon loss and decoherence of the qubit itself. Certain sources of qubit decoherence may be specific to spin cQED setups, arising from the interaction between the qubit and a structured electromagnetic environment. An example is Purcell decay resulting from the modified local density of states near the qubit frequency: Strong coupling of the qubit to a cavity mode with a frequency resonant with or close to the qubit frequency will lead to an enhancement in the qubit relaxation rate. The effects of Purcell decay have been observed and characterized in a spin cQED setup operated with a hole-spin flopping mode qubit~\cite{noirot2025coherence}.

\begin{table*}
    \centering
    \begin{tabular}{|l||c|l|c|c|}
    \hline
         \textbf{Reference} & \textbf{Material} & \textbf{Qubit modality} &  \textbf{Key parameters} & \textbf{Cooperativity} \\[4pt]
         &&&$\left(\dfrac{2g_x}{2\pi},\dfrac{\kappa}{2\pi},\dfrac{\gamma}{2\pi}\right)$ & \\\hline
         \textcite{mi2017strong} & Si & Electron charge & (13.4, 1.0, 2.6) MHz & 68\\\hline
         \textcite{samkharadze2018strong} & Si & Electron flopping mode & (26, 5.4, 2.5) MHz & 52\\\hline
         \textcite{stockklauser2017strong} & GaAs & Electron charge & (238, 12, 40) MHz & 120 \\\hline
         \textcite{landig2018coherent} & GaAs & Resonant exchange & (62.8, 47.1, 19.6) MHz  & 4\\\hline
         \textcite{mi2018coherent} & Si & Electron flopping mode & (11, 1.8, 2.4) MHz & 5\\\hline
         \textcite{cubaynes2019highly} & C & Electron flopping mode & (4.0, 1.44, 0.25) MHz & 44 \\\hline
         \textcite{yu2023strong} & Si & Hole flopping mode & (570, 14, 14) MHz  & 1660\\\hline
         \textcite{de2024strong} & Ge & Hole charge & (330, 19, 57) MHz & 100 \\\hline
        \textcite{jiang2025coupling} & Si & Resonant exchange & (132, 5.1, 21.4) MHz & 160     \\\hline 
    \end{tabular}
    \caption{Demonstrations of strong coupling achieved across various semiconductor platforms. The cooperativity reported here is defined as $C=4g_x^2/\kappa\gamma$. }
    \label{tab:strong-coupling}
\end{table*}

\begin{figure}
    \centering
    \includegraphics[width=0.7\linewidth]{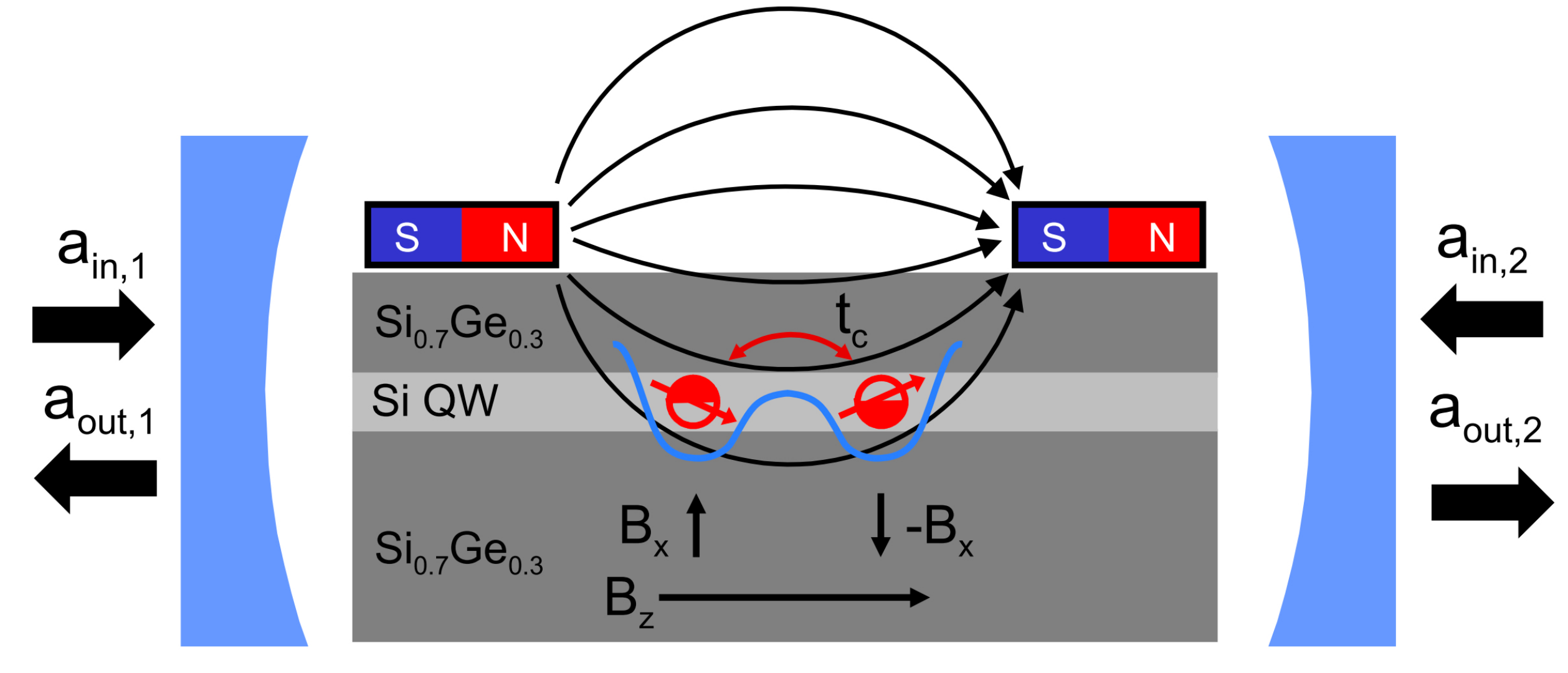}
    \caption{The cavity transmission spectrum is obtained by driving the cavity input port with an input drive $a_{\mathrm{in},1}$ and measuring the field $a_{\mathrm{out},2}$ at the output port. Reproduced from \textcite{benito2017input} with permission.}
    \label{fig:transmission}
\end{figure}

Vacuum Rabi dynamics can be probed through the cavity transmission as a way of determining whether a qubit-cavity system is in the strong coupling regime. This is accomplished by driving the cavity input port with a weak input tone $a_{\mathrm{in},1}(t)$ and spectroscopically analyzing the field $a_{\mathrm{out},2}(t)$ transmitted through its output port (Fig.~\ref{fig:transmission}). The experimentally measured transmission spectrum is typically analyzed by fitting to the spectrum derived analytically on the basis of a master-equation description of the evolution of the joint state $\rho$ of the qubit and cavity. Such a master equation is given by
\begin{equation}\label{transmission-master-equation}
    \dot{\rho}=-i[H_{\mathrm{JC}}+H_{\mathrm{in}}(t),\rho]+\frac{\gamma}{2}\mathcal{D}[\sigma_z]\rho+\kappa\mathcal{D}[a]\rho, 
\end{equation}
where $H_{\mathrm{in}}(t)=i\sqrt{\kappa_1}(a_{\mathrm{in},1}^*(t) a-\mathrm{h.c.})$ describes cavity driving due to the input field, and where the damping superoperator $\mathcal{D}[\mathcal{O}]$ acts according to $\mathcal{D}[\mathcal{O}]\rho=\mathcal{O}\rho\mathcal{O}^\dagger-\{\mathcal{O}^\dagger\mathcal{O},\rho\}/2$. In Eq.~\eqref{transmission-master-equation}, the term $\gamma\mathcal{D}[\sigma_z]/2$ models Markovian qubit dephasing with rate $\gamma$, while $\kappa\mathcal{D}[a]$ models cavity damping at the total rate $\kappa=\kappa_1+\kappa_2+\kappa_{\mathrm{int}}$. Here, $\kappa_{1,2}$ are the damping rates through ports 1 and 2, and $\kappa_{\mathrm{int}}$ accounts for intrinsic losses. 

In terms of $\dot{\rho}$, the equations of motion for the cavity annihilation operator $a$ and the spin-lowering operator $\sigma_-$ can be found as $\langle \dot{\mathcal{O}}\rangle_t=\mathrm{Tr}\{\mathcal{O}\dot{\rho}\}$, giving the coupled equations
\begin{align}
    &\langle\dot{a}\rangle_t=-\left(i\omega_{\mathrm{c}}+\frac{\kappa}{2}\right)\langle a \rangle_t-i g_x\langle \sigma_-\rangle_t-\sqrt{\kappa_1}a_{\mathrm{in},1}(t),\\
    &\langle\dot{\sigma}_-\rangle_t=-\left(i\omega_{\mathrm{q}}+\gamma\right)\langle \sigma_-\rangle_t+i g_x\langle a \sigma_z\rangle_t.
\end{align}
For sufficiently weak input driving, the cavity will contain at most one photon at any given time, and the dynamics of the cavity-qubit system can be restricted to the subspace $\ket{g,0}\oplus \mathscr{H}_1$. For any state in this subspace, the expectation value $\langle a \sigma_z\rangle$ is equal to $-\langle a \rangle$ since $a\sigma_z$ has nonzero matrix elements only when the qubit is in $\ket{g}$. For weak input driving, the equations of motion for $\langle a \rangle_t$ and $\langle \sigma_-\rangle_t$ can consequently be decoupled from the equation of motion for $\langle a \sigma_z\rangle_t$ to a good approximation.

\begin{figure}
    \centering
    \includegraphics[width=0.65\linewidth]{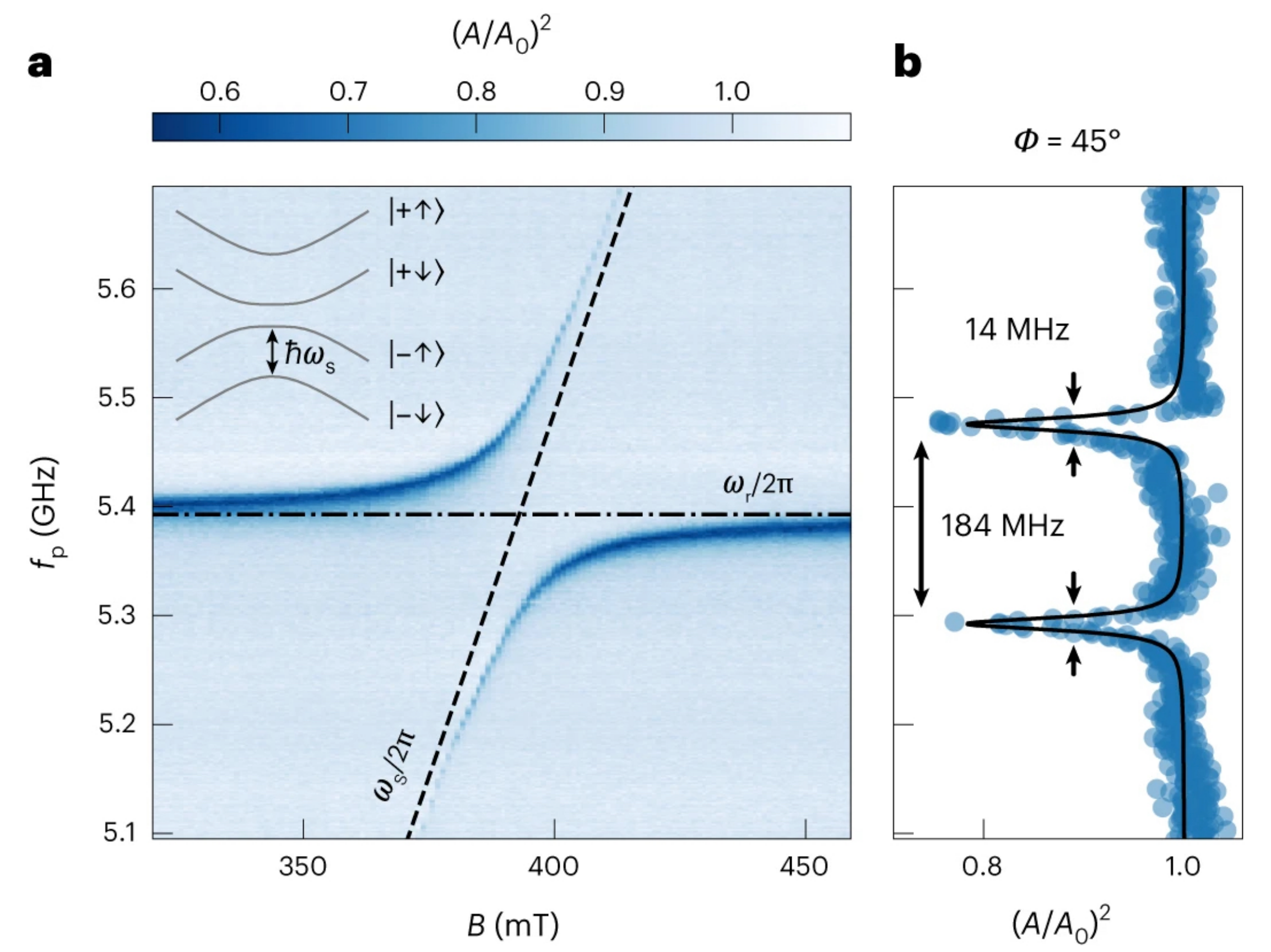}
    \caption{(a) Cavity transmission probability as a function of the input-field frequency $f_p$ and amplitude $B$ of a magnetic field oriented at an angle of $\phi=45^\circ$ relative to the nanowire housing a hole spin. The data shows a clear avoided crossing similar to that depicted in Fig.~\ref{fig:jaynes-cummings}(b).  (b) A frequency linecut of (a) at zero detuning exhibits a pair of Lorentzian peaks separated by $2g_x/2\pi=184$ MHz. Reproduced from \textcite{yu2023strong} with permission. }
    \label{fig:strong-coupling}
\end{figure}

For a cavity coupled to a transmission line with a bandwidth that is large relative to the time scale for cavity damping, the output field $a_{\mathrm{out},2}(t)$ can be related to the average cavity field $\langle a \rangle_t$ by input-output theory~\cite{gardiner1985input}, giving $a_{\mathrm{out},2}(t)=\sqrt{\kappa_2}\langle a\rangle_t$. By invoking the restricted subspace approximation, $\langle a\sigma_z\rangle=-\langle a \rangle$, the average qubit coherence $\langle \sigma_-\rangle_t$ can be eliminated from the equation of motion for $\langle a \rangle_t$. We can then straightforwardly solve for $\langle a \rangle_\omega=\int dt\: e^{i\omega t}\langle a \rangle_t$, giving
\begin{equation}\label{cavity-field-with-susceptibility}
    a_{\mathrm{out},2}(\omega)=\sqrt{\kappa_2}\langle a\rangle_\omega=T(\omega) a_{\mathrm{in},1}(\omega).
\end{equation}
In Eq.~\eqref{cavity-field-with-susceptibility}, we have introduced the transmission amplitude $T(\omega)$, which can be written in terms of the charge susceptibility $\chi_{\mathrm{q}}(\omega)=g_x(\omega-\omega_{\mathrm{q}}+i\gamma)^{-1}$ as 
\begin{equation}\label{cavity-susceptibility-strong-coupling}
   T(\omega)=\frac{i\sqrt{\kappa_1\kappa_2}}{\omega_{\mathrm{c}}-\omega-i\kappa/2+g_x\chi_{\mathrm{q}}(\omega)}.
\end{equation}
At zero detuning  and in the strong coupling regime $2g_x>\kappa,\gamma$, the transmission probability $\lvert T(\omega)\rvert^2$ features a pair of Lorentzian peaks centered at approximately $\omega_{\mathrm{c}}\pm g_x$, which are widely separated relative to their full-widths-at-half maximum (Fig.~\ref{fig:strong-coupling}). This spectroscopic signature of strong coupling has been observed in a variety of systems, with relevant parameters summarized in Table \ref{tab:strong-coupling}. One metric often reported in the context of strong-coupling experiments is the cooperativity $C=4g_x^2/(\kappa \gamma)$. A cooperativity $C\gg 1$ indicates that many vacuum Rabi oscillations will take place before the visibility of the oscillations is lost to decoherence. It should be remarked that for a spin qubit, the relevant dephasing rate $\gamma$ will typically fall somewhere in between the spin and charge dephasing rates (due to the spin-charge hybridization required to achieve coupling to the cavity). For this reason, it is possible for a spin-cavity system to reach the strong-coupling regime despite having a large charge dephasing rate. 

The cavity transmission can be used for readout by ensuring that $g_x\chi_{\mathrm{q}}(\omega_{\mathrm{c}})$ is large for one spin state and small for another. This is typically accomplished using Pauli spin blockade in the vicinity of a charge transition and has been used for single-shot readout of singlet and triplet states~\cite{petersson2012circuit,zheng2019rapid,borjans2021spin}. Nuclear spin readout has also been demonstrated using the dependence of the cavity transmission on the state of a phosphorus nuclear spin hyperfine-coupled to a flopping-mode electron spin qubit~\cite{mielke2021nuclear}. Charge stability diagrams can be mapped out by measuring the amplitude or phase of the cavity transmission or reflection as a function of applied gate voltages, as has been demonstrated for DQDs in GaAs~\cite{frey2012dipole}, InAs~\cite{petersson2012circuit}, carbon nanotubes~\cite{viennot2014out}, and Si~\cite{mi2017strong}.  The acquisition of a DQD charge stability diagram in just 20 ms has been achieved via measurements of the cavity transmission, with the speed of the measurement enabled by the integration of a Josephson parametric amplifier into the readout chain~\cite{stehlik2015fast}. Such rapid data acquisition is compatible with real-time tuning of the double-dot parameters~\cite{stehlik2015fast}.

There have also been theoretical proposals for leveraging transient (non-stationary) dynamics in the cavity field for qubit noise spectroscopy. Transient features in the cavity transmission can be used to extract spectral information about the noise impacting the qubit's free-induction-decay dynamics~\cite{mutter2022fingerprints}.  Transient qubit coherence revivals obtained under dynamical decoupling can also be used to identify qubit dynamics due to low-frequency classical or quantum fluctuations (the latter arising from the non-commutation of bath operators) via measurements of the cavity field~\cite{mcintyre2022non}.

\subsection{The dispersive regime}\label{sec:dispersive-regime}

On or near resonance, a transversally coupled cavity and qubit can coherently exchange excitations. This dynamics can be used to swap information between the qubit and cavity, but there are also advantages to operating in the dispersive regime, where the magnitude $\lvert \delta\rvert$ of the qubit-cavity detuning far exceeds $g_x$, and where the direct exchange of excitations is strongly suppressed. In this regime, a Schrieffer-Wolff transformation of the Jaynes-Cummings Hamiltonian can be used to derive an effective Hamiltonian 
\begin{equation}\label{dispersive-shrieffer-wolff}
    H_{\mathrm{disp}}=e^{S}H_{\mathrm{JC}}e^{-S}
\end{equation}
describing the coupling of the qubit and cavity, where the generator $S=g_x(\sigma_+ a -\sigma_- a^\dagger)/\delta$ of the transformation is chosen to eliminate the interaction $\propto g_x$ to first order. Up to a constant shift in energy and neglecting $O(g_x^3)$~\cite{blais2021circuit},
\begin{equation}\label{dispersive-hamiltonian}
    H_{\mathrm{disp}}=\frac{1}{2}(\omega_{\mathrm{q}}+\chi)\sigma_z+\omega_{\mathrm{c}}a^\dagger a + \chi\sigma_z a^\dagger a,
\end{equation}
where $\chi=g_x^2/\delta$ gives the strength of the dispersive coupling. We have kept the same notation for the cavity and qubit operators, but it should be noted that they are now in a dressed basis: Although the eigenstates of $H_{\mathrm{disp}}$ are typically labeled using the same quantum numbers $\ket{e(g),n}$ as the decoupled (product) states, these eigenstates still exhibit weak hybridization as a result of the cavity-qubit coupling. To see this, note that from Eq.~\eqref{dispersive-shrieffer-wolff}, it follows that an eigenstate $\ket{\psi}$ of $H_{\mathrm{disp}}$ with eigenvalue $\epsilon$ ($H_{\mathrm{disp}}\ket{\psi}=\epsilon\ket{\psi}$) is related via $\ket*{\Tilde{\psi}}=e^{-S}\ket{\psi}$ to the eigenstate $\ket*{\Tilde{\psi}}$ of $H_{\mathrm{JC}}$ having the same eigenvalue: $H_{\mathrm{JC}}\ket*{\Tilde{\psi}}=\epsilon \ket*{\Tilde{\psi}}$. However, since $S$ is small in the perturbative parameter $g_x/\lvert\delta\rvert\ll 1$, the cavity-qubit interaction only weakly dresses the decoupled eigenstates. 

One advantage of working in the dispersive regime is that it enables a form of approximately quantum non-demolition (QND) readout known as dispersive readout~\cite{blais2004cavity,blais2021circuit}, which was first demonstrated for a spin qubit by \textcite{mi2018coherent}. To implement dispersive readout, the cavity is driven by an input field $a_{\mathrm{in},1}(t)$ resonant with the bare cavity frequency $\omega_{\mathrm{c}}$. For a qubit with a $\sigma_z$ eigenvalue of $s=\pm 1$, the equation of motion for the cavity field is then given by
\begin{equation}\label{eom-dispersive-readout}
    \langle \dot{a}\rangle_t=-\left(i\omega_{\mathrm{c}}+is\chi +\frac{\kappa}{2}\right)\langle a \rangle_t-\sqrt{\kappa_1}a_{\mathrm{in},1}(t).
\end{equation}
This can be derived from a master equation similar to Eq.~\eqref{transmission-master-equation}, but with $H_{\mathrm{JC}}$ approximated by the dispersive Hamiltonian given in Eq.~\eqref{dispersive-hamiltonian}. Since a cavity-qubit coupling of the form $\sigma_za^\dagger a$ commutes with the dissipator $\mathcal{D}[\sigma_z]$, we can replace $\sigma_z$ with its eigenvalue $s$ in Eq.~\eqref{eom-dispersive-readout}. Solving for the cavity field in the frequency domain then gives
\begin{equation}
    \langle a \rangle_{\omega,s}=\frac{-\sqrt{\kappa_1}a_{\mathrm{in},1}(\omega)}{i(\omega_{\mathrm{c}}-\omega+s\chi)+\kappa/2},
\end{equation}
where here, the subscript $s$ indicates a dependence on the qubit state. Hence, under the assumption that $a_{\mathrm{in},1}(\omega)$ is monochromatic with weight only at $\omega_{\mathrm{c}}$, the amplitude $\alpha_s\equiv \langle a \rangle_{\omega_{\mathrm{c}},s}$ gives the steady-state displacement of the cavity under continuous input driving, conditioned on the qubit being in the $\sigma_z$-eigenstate with $s=\pm 1$. The distinguishability of the two states can be maximized by maximizing their phase-space (Euclidean) distance, given by
\begin{equation}\label{dispersive-distinguish}
    \lvert \alpha_+-\alpha_-\vert=\frac{2\chi}{\chi^2+\kappa^2/4}.
\end{equation}
The state of the qubit can then be measured by measuring the phase shift of the field reflected from or transmitted through the cavity. For instance, assuming the optimal value of $\chi=\kappa/2$ obtained from an optimization of Eq.~\eqref{dispersive-distinguish} over $\chi$, the phase shift of the field transmitted through the cavity is $\pm \pi/4$, depending on the qubit state. The signal-to-noise ratio for dispersive readout of a spin qubit in a DQD was optimized by \textcite{d2019optimal}.

\subsection{Cavity-mediated qubit-qubit coupling}

\begin{figure}
    \centering
    \includegraphics[width=0.55\linewidth]{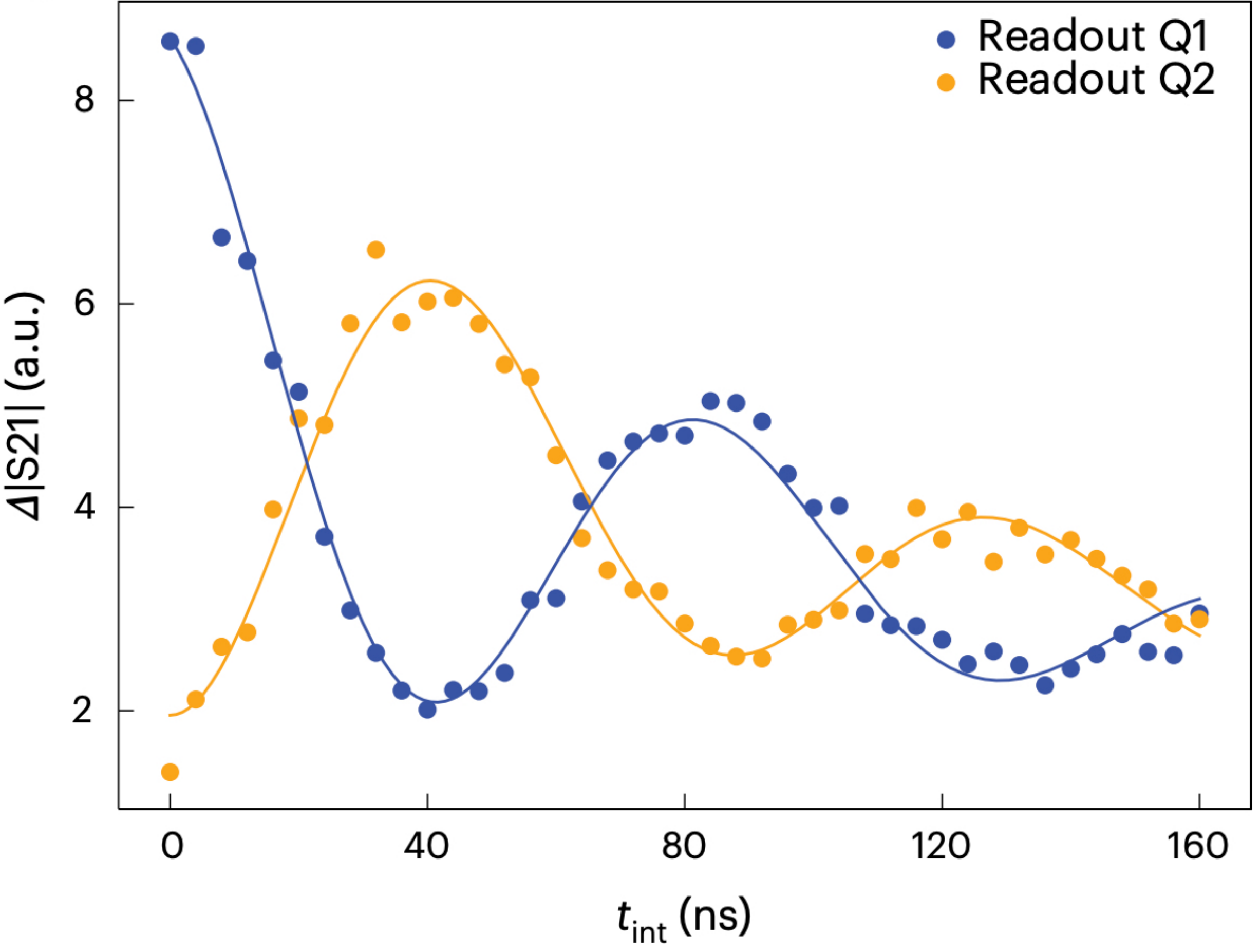}
    \caption{Virtual-photon-mediated iSWAP oscillations between distant flopping-mode spin qubits. Starting from the state $\ket{10}$ of qubits Q1 and Q2, the populations of Q1 and Q2 were measured via measurements of the cavity transmission, revealing oscillations with a frequency consistent with a coupling strength of $2J/2\pi=11.6$ MHz. Figure reproduced from \textcite{dijkema2025cavity}, licensed under CC BY 4.0.}
    \label{fig:iswap}
\end{figure}

Having discussed readout, we now turn our attention to the question of generating entanglement between qubits.
One way of realizing a cavity-mediated entangling gate is by transversally coupling two qubits to the same cavity mode. Neglecting counter-rotating terms, the Hamiltonian of the two qubits and cavity is then given by 
\begin{equation}
    H=\omega_{\mathrm{c}}a^\dagger a +\sum_{j=1,2}\left[\frac{\omega_j}{2}\sigma_z^{j}+g_j(\sigma_-^{j}a^\dagger + \mathrm{h.c.})\right],
\end{equation}
where the superscript $j$ on $\sigma_z^j$ indexes the qubit $j=1,2$.
When the two qubits are both resonant with the cavity, the vacuum Rabi splitting $2g$ will be collectively enhanced according to $2g\mapsto 2\sqrt{g_1^2+g_2^2}$. For $g_1=g_2$, this recovers the familiar $\sqrt{2}$ enhancement expected from the Tavis-Cummings model. Such a collective enhancement has been observed experimentally for both charge qubits~\cite{van2018microwave} and spin qubits~\cite{borjans2020resonant,harvey2022coherent} via measurements of the cavity transmission. 

In the dispersive regime $g_j\ll \lvert \delta_j\rvert$, where $\delta_j=\omega_j-\omega_{\mathrm{c}}$ denotes the detuning of qubit $j$ from the cavity, a Schrieffer-Wolff transformation with generator $S=\sum_j g_j(\sigma_+^{j}a-\mathrm{h.c.})/\delta_j$ can be used to derive an effective Hamiltonian $H'$ revealing a virtual-photon-mediated transverse interaction between the two qubits, appearing alongside the dispersive coupling already familiar to us from the single-qubit case~\cite{blais2007quantum,burkard2006ultra,trif2007spin,lalumiere2010tunable,benito2019optimized,warren2019long}:
\begin{equation}\label{hxy}
    H'=\sum_j \left(\frac{\Tilde{\omega}_j}{2}+\chi_ja^\dagger a\right)\sigma_z^{j}+J(\sigma_+^{1}\sigma_-^{2}+\mathrm{h.c.}).
\end{equation}
Here, $\Tilde{\omega}_j=\omega_j+\chi_j$ is the Lamb-shifted splitting of qubit $j$, $\chi_j=g_j^2/\delta_j$ is the dispersive coupling of qubit $j$ to the cavity, and $J$ sets the strength of the transversal qubit-qubit interaction according to
\begin{equation}
    J=\frac{g_1g_2}{2}\left(\frac{1}{\delta_1}+\frac{1}{\delta_2}\right).
\end{equation}
When the qubits are on resonance ($\Tilde{\omega}_1=\Tilde{\omega}_2$), their spectrum exhibits an avoided crossing of size $2J$ due to qubit-qubit hybridization. This was first observed with superconducting qubits in 2007~\cite{majer2007coupling} and has since been observed with charge qubits~\cite{van2018microwave} and flopping-mode electron-spin qubits~\cite{harvey2022coherent}.   

On resonance, the transversal qubit-qubit coupling can be used to mediate a two-qubit entangling gate. In an interaction picture defined with respect to the free qubit evolution (and neglecting the dispersive couplings under the assumption that the cavity is never populated), the time-evolution operator $U_{\textsc{swap}}$ generated by the interaction $\propto J$ in Eq.~\eqref{hxy} is given by 
\begin{equation}
    U_{\textsc{swap}}(t)=\begin{pmatrix}
        1 & 0 & 0 & 0\\
        0 & \cos{(Jt)} & i\sin{(Jt)} & 0\\
        0 & -i\sin{(Jt)} & \cos{(Jt)} & 0\\
        0 & 0 & 0 & 1
    \end{pmatrix}.
\end{equation}
Evolution under $U_{\textsc{swap}}(t)$ will produce i\textsc{SWAP} and $\sqrt{i\textsc{swap}}$ gates for $Jt=\pi/2$ and $Jt=\pi/4$, respectively~\cite{imamog1999quantum,schuch2003natural}. i\text{SWAP} oscillations between distant spins (Fig.~\ref{fig:iswap}) were recently demonstrated for the first time for spin qubits separated by 250 $\mu$m~\cite{dijkema2025cavity}. $\sqrt{i\textsc{swap}}$ gates could also be realized for qubits coupled to \textit{different} cavities that are themselves capacitively coupled, as analyzed by \textcite{nigg2017superconducting} for nanowire hole-spin qubits. An appealing feature of this proposal is that the holes' strong direct Rashba spin-orbit interaction provides a mechanism for electric tunability of the qubit-cavity coupling. A scalable two-dimensional architecture, as required for surface-code quantum computing, could then be realized by fabricating a square-lattice array of coupled cavities each coupled to one nanowire qubit~\cite{nigg2017superconducting}.

\subsection{Longitudinal coupling}

The effective Hamiltonians given in Eqs.~\eqref{charge-qubit-rabi} and \eqref{flopping-mode-rabi} both feature a transversal $\propto \sigma_x$ cavity-qubit coupling originating from the transition-dipole matrix element $\langle e\vert\tau_z\vert g\rangle$. Over the past decade or so, there has been significant interest in exploring the possibilities enabled by \textit{longitudinal} coupling $\sigma_z(a+a^\dagger)$, which naturally preserves the qubit eigenstates. 

Realizing a non-zero longitudinal coupling requires that the qubit eigenstates have different intrinsic dipole moments. This can be understood within the dipole approximation [Eqs.~\eqref{dipole-approximation-hamiltonian}-\eqref{dipole-approximation}], where a longitudinal coupling arises when $\langle e\vert\bm{d}\vert e\rangle\neq \langle g\vert\bm{d}\vert g\rangle$. For a qubit whose eigenstates are associated with charge configurations having a definite inversion symmetry, the qubit eigenstates are also eigenstates of the parity operator $\Pi:\bm{r}\mapsto-\bm{r}$. Since the dipole operator $\bm{d}$ [Eq.~\eqref{dipole-approximation}] is odd under inversion about $\bm{r}_0$ and $\Pi\ket{e}=\pm \ket{e}$ for an inversion-symmetric eigenstate, we have 
\begin{equation}\label{intrinsic-dipole}
    \langle e\vert\bm{d}\vert e\rangle=\langle e\vert\Pi^\dagger\bm{d}\Pi\vert e\rangle=-\langle e\vert \bm{d}\vert e\rangle,
\end{equation}
from which it follows that $\langle e\vert \bm{d}\vert e\rangle=0$, and similarly for $\langle g \vert \bm{d}\vert g\rangle$. For qubits with $\langle e(g)\vert\bm{d}\vert e(g)\rangle=0$, it immediately follows that any static coupling to the cavity is purely transverse.  This can be connected back to the charge-qubit example [cf.~Eq.~\eqref{charge-qubit-rabi}] by noting that at the charge degeneracy point $\varepsilon=0$, the qubit eigenstates are the bonding and anti-bonding states $\ket{\mathrm{L}}\pm \ket{\mathrm{R}}$, which have a definite parity under spatial inversion. Since $\tau_z=\ketbra{\mathrm{L}}-\ketbra{\mathrm{R}}$ is directly proportional to the dipole operator $\bm{d}$ for a DQD-based qubit, it immediately follows that $\langle e(g)\vert \bm{d}\vert e(g)\rangle=0$ at $\varepsilon=0$. For the flopping-mode Hamiltonian given in Eq.~\eqref{flopping-mode}, the longitudinal coupling also vanishes at the charge degeneracy point, but a finite longitudinal coupling at $\varepsilon=0$ could be achieved by instead considering a Zeeman gradient of the form $\Delta B_z\sigma_z'\tau_z$. The reason for this is that for such a coupling, the double-dot detuning depends on the spin operator $\sigma_z'$, which results in $\langle e\vert \tau_z\vert e\rangle\neq \langle g\vert\tau_z\vert g\rangle$ even at $\varepsilon=0$. By carefully selecting the placement of the nanomagnet relative to the double dot, the coupling of a flopping-mode qubit to the cavity can be made purely transverse or purely longitudinal~\cite{beaudoin2016coupling}.

A finite longitudinal coupling emerges naturally for singlet-triplet qubits, following directly from the Pauli exclusion principle. This is a consequence of the orbital wave function of the triplet being associated with a double-dot charge occupation $(1,1)$. Triplet states with $(0,2)$ and $(2,0)$ configurations are energetically disfavored by the large orbital splitting. For a non-zero tunnel coupling between the two dots, the orbital wavefunction of the singlet state will generally feature an admixture of $(1,1)$, $(2,0)$, and $(0,2)$ charge occupations since, due to Pauli exclusion, spins in a singlet state can occupy the same orbital state $\ket{\mathrm{L}}$ or $\ket{\mathrm{R}}$ while respecting the overall anti-symmetry of the total electronic wavefunction. The admixture of non-$(1,1)$ charge configurations in the singlet state leads to $\langle e \vert\bm{d}\vert e\rangle\neq \langle g \vert \bm{d}\vert g\rangle$, which in turn leads to a finite longitudinal coupling. 

Although dispersive interactions $\sigma_z a^\dagger a$ and longitudinal interactions $\sigma_z(a+a^\dagger)$ both couple the cavity to the Pauli-Z operator of the qubit, the resulting dynamics of the cavity mode are qualitatively different in the two scenarios. As discussed in Sec.~\ref{sec:dispersive-regime}, a dispersive coupling results in a shift of the cavity frequency conditioned on the state of the qubit. In a frame rotating at the bare cavity frequency $\omega_{\mathrm{c}}$, a coherent state of the cavity mode will therefore precess counterclockwise (clockwise) in phase space for a qubit in state $\ket{e}$ ($\ket{g}$). By contrast, a longitudinal coupling does not produce such a shift in the cavity frequency. Instead, the cavity mode will be \textit{displaced} in phase space in different directions depending on the qubit state. This may be seen directly from the Hamiltonian 
\begin{equation}\label{hamiltonian-longitudinal}
    H=\frac{\omega_{\mathrm{q}}}{2}\sigma_z+\omega_{\mathrm{c}}a^\dagger a + g_z\sigma_z(a+a^\dagger),
\end{equation}
where, here, the longitudinal coupling strength $g_z=g_c( \langle e \vert\tau_z\vert e\rangle-\langle g \vert\tau_z\vert g\rangle)$ depends on the difference in the intrinsic dipole moments of the qubit eigenstates. Since the position operator $\hat{x}=(a+a^\dagger)/\sqrt{2}$ generates displacements along $\hat{p}=i(a^\dagger-a)/\sqrt{2}$, a cavity-qubit coupling of the form $\sigma_z\hat{x}$ [cf.~Eq.~\eqref{hamiltonian-longitudinal}] will result in a displacement along $\pm \hat{p}$ conditioned on the qubit state. However, since the Heisenberg-picture operator $a(t)=e^{-i\omega_{\mathrm{c}}t}a$ oscillates at the cavity frequency $\omega_{\mathrm{c}}$, which typically far exceeds the strength of $g_z$, the size of this displacement is on the order of $g_z/\omega_{\mathrm{c}}\ll 1$ and would typically be neglected within a rotating wave approximation (in which terms oscillating at $\pm \omega_{\mathrm{c}}$ are neglected). 

The magnitude of the qubit-state-dependent displacement of the cavity mode can be increased by parametrically modulating the longitudinal coupling strength at the cavity frequency, $g_z(t)=\bar{g}_z+\Tilde{g}_z\cos{\omega_{\mathrm{c}}t}$. The purpose of this modulation is to introduce a term in $g_z(t)a(t)$ that does not average out within the aforementioned rotating-wave approximation. In a frame rotating at the cavity frequency, and within a rotating-wave approximation valid for $\bar{g}_z,\Tilde{g}_z\ll \omega_{\mathrm{c}}$, the Hamiltonian $H$  [Eq.~\eqref{hamiltonian-longitudinal}] with $g_z\mapsto g_z(t)$ then reads 
\begin{equation}
    H_{\mathrm{RWA}}=\frac{\omega_{\mathrm{q}}}{2}\sigma_z+\frac{\Tilde{g}_z}{2}\sigma_z(a+a^\dagger).
\end{equation}
The qubit-state-dependent displacement generated by $H_{\mathrm{RWA}}$ can be used for longitudinal parametric readout~\cite{didier2015fast,ruskov2019quantum}. Arguably, the main advantage of longitudinal parametric readout relative to dispersive readout is a better signal-to-noise ratio at short times, which, in turn, enables a faster measurement. Other advantages are that the readout is perfectly quantum non-demolition (QND) due to the absence of Purcell decay, and that the signal-to-noise ratio can be further enhanced through squeezing~\cite{didier2015fast,eddins2018stroboscopic}. The use of parametrically modulated longitudinal coupling has also been proposed to realize entangling gates between qubits coupled to a common cavity mode~\cite{royer2017fast,ruskov2021modulated}.

\begin{figure}
    \centering
    \includegraphics[width=0.6\linewidth]{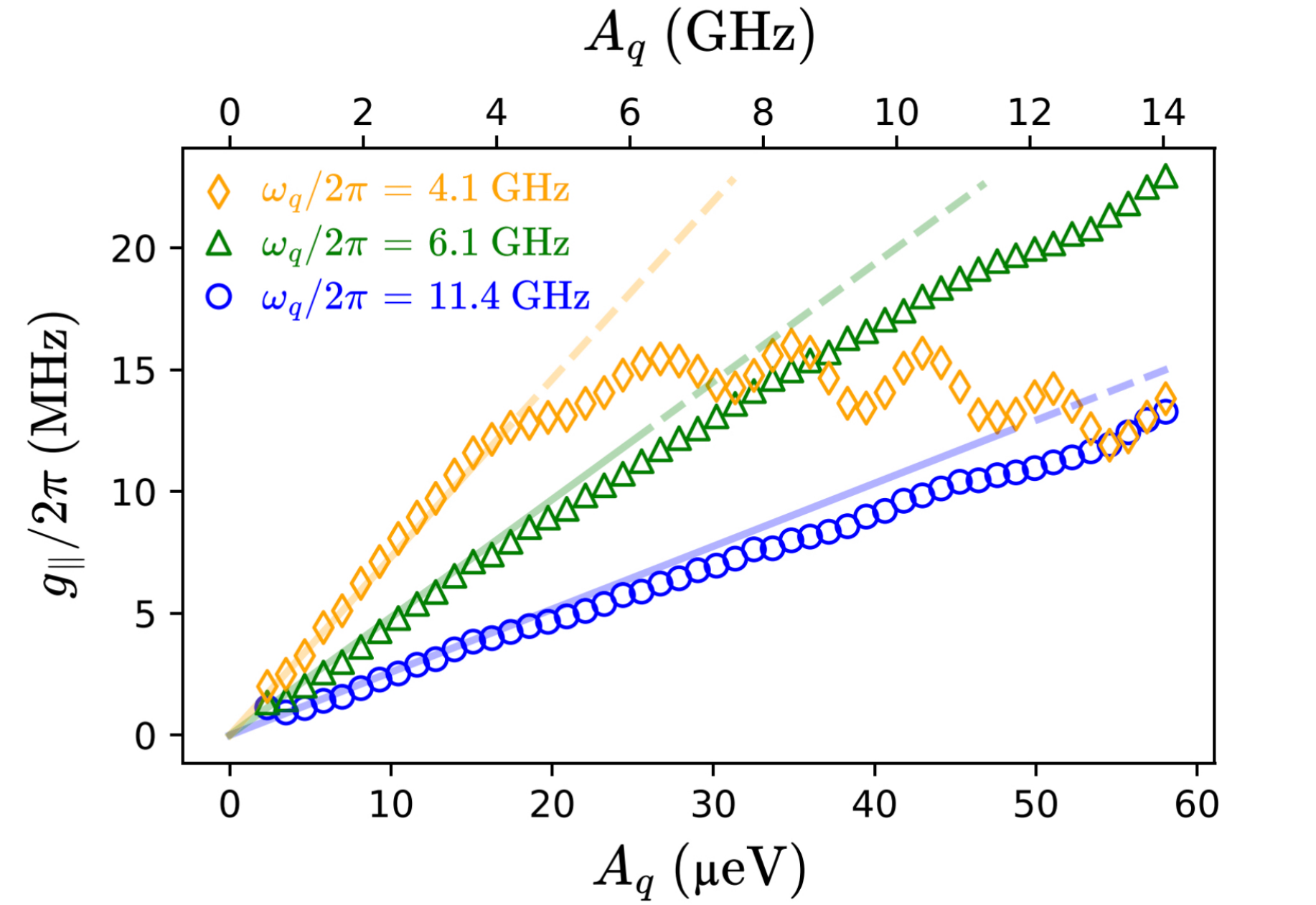}
    \caption{Longitudinal coupling strength as a function of the strength $A_q$ of the modulation at the cavity frequency. Reproduced from \textcite{champain2025parametric} with permission.}
    \label{fig:longitudinal-coupling}
\end{figure}

There are many ways to generate a parametrically modulated longitudinal coupling. The first experimental demonstrations of longitudinal parametric readout, performed with superconducting transmon qubits~\cite{ikonen2019qubit,touzard2019gated}, engineered a synthetic longitudinal coupling using a transverse cavity-qubit coupling and by driving the qubit at the frequency of the cavity. Longitudinal coupling engineered in this way has since been observed and characterized for a charge qubit in a DQD (Fig.~\ref{fig:longitudinal-coupling})~\cite{champain2025parametric,chessari2025unifying}. Another way to engineer a longitudinal coupling starting from a transverse interaction involves a bichromatic modulation $\cos{(\omega_{\mathrm{c}}t)}\cos{(\omega_{\mathrm{q}} t)}$ of the transverse coupling strength, resulting in an effective longitudinal coupling in a rotating frame~\cite{lambert2018amplified}. For hole-spin qubits, longitudinal coupling arises from the strong intrinsic spin–orbit coupling of valence-band states in materials like Si and Ge~\cite{bosco2022fully,michal2023tunable,sagaseta2025switchable}. In such systems, spin-orbit coupling allows the cavity, which couples to the hole's charge degree of freedom, to couple directly to the hole's effective Zeeman splitting without the need for external components like micromagnets. The coupling can be switched from purely longitudinal to purely transverse by changing the orientation of a magnetic field, and it can be modulated through applied gate voltages.  The longitudinal coupling naturally obtained for singlet-triplet qubits can also be modulated through applied gate voltages, as has been realized experimentally~\cite{bottcher2022parametric}; the tunability of the longitudinal couplings obtained for singlet-triplet and exchange-only flopping-mode qubits has also been explored theoretically~\cite{stastny2025singlet}. 

For a (regular) flopping-mode qubit, it is also possible to parametrically modulate the coupling of only one of the qubit eigenstates, e.g., $\ket{e}$, by modulating the tunnel coupling with the double-dot detuning chosen such that $\epsilon=-\Delta B_z$~\cite{mcintyre2024photonic}. With an asymmetric coupling Hamiltonian of the form
\begin{equation}
    H_{\mathrm{RWA}}=\frac{\omega_{\mathrm{q}}}{2}\sigma_z+\Tilde{g}_e(t)\ketbra{e}(a+a^\dagger),
\end{equation}
where $\Tilde{g}_e(t)$ is a slow envelope on top of the modulation at the cavity frequency, the cavity-qubit system can act like a quantum switch controlling the path of an incoming pulse of coherent radiation with normalized waveform $u(t)$ and amplitude $\alpha$ by tailoring $\tilde{g}_e(t)$ to satisfy $\tilde{g}_e(t)=\sqrt{\kappa/2}\alpha u(t)$~\cite{mcintyre2024photonic}. The enhancement and suppression of cavity transmission measured by \cite{corrigan2023longitudinal,harpt2025ultra} with longitudinally coupled quantum-dot hybrid qubits are consistent with this type of interference between the input and cavity fields, and could, in turn, be used for qubit readout~\cite{harpt2025ultra}. These experiments also demonstrate another important advantage of longitudinal coupling, namely the possibility of operating in the ultra-dispersive regime where the cavity and qubit are very far off resonance. In this regime, the effects of transverse coupling are strongly suppressed, but effects due to longitudinal coupling may—and do~\cite{corrigan2023longitudinal,harpt2025ultra}—remain large since longitudinal coupling strengths do not fall off with detuning like dispersive shifts. In the long run, operating in the ultra-dispersive regime could help alleviate spectral-crowding issues as devices scale up.

\section{Andreev qubits}

The strategy discussed in the previous section for coupling spin qubits to microwave cavities relied on spin-charge hybridization to achieve capacitive coupling to the cavity electric field. Another way of achieving a spin-cavity interaction is based on inductive coupling, realized through the spin-state dependence of the supercurrent flowing across a Josephson junction. This allows many of the circuit-QED strategies discussed in the previous section to be applied to these so-called Andreev qubits as well; the spin-dependent supercurrent can also be used to realize long-range inductive coupling between the qubits themselves~\cite{chtchelkatchev2003andreev}, without the need for a mediating cavity. The ability to probe a spin state via a Josephson supercurrent was first established by~\textcite{choi2000spin}. In this work, the Josephson junction consists of a DQD, each dot of which is coupled to two superconducting leads. When both dots contain one excess electron, the coupling of the dots to the superconductors induces an antiferromagnetic exchange between the spins of the two electrons. An immediate consequence of this exchange is that the supercurrent across the Josephson junction acquires a dependence on the state (singlet vs triplet) of the DQD spins~\cite{choi2000spin}.  

In this section, we review the history and recent experimental progress of qubit implementations based on Andreev bound states localized in Josephson-junction weak links: single-particle, electronic degrees of freedom with energies inside the superconducting gap~\cite{beenakker1991josephson,furusaki1991current}. Apart from quantum dots coupled to superconducting leads~\cite{choi2000spin}, another common platform for realizing (addressable) Andreev bound states involves Josephson nanowires~\cite{de2015realization,larsen2015semiconductor}. A Josephson nanowire consists of a semiconducting nanowire with a coating of superconducting material covering the entirety of the nanowire, apart from a small portion. The spatial gap in the proximitizing superconductor defines a weak link of non-superconducting material separating two superconducting regions~\cite{chang2015hard,krogstrup2015epitaxy}. Current can flow across this weak link without dissipation, even in the absence of a voltage bias, via the Josephson effect~\cite{josephson1962possible}. Although the weak links used in superconducting transmon qubits also have Andreev bound states, these links are typically formed using an insulator rather than a semiconductor. As a result, they usually exhibit a dense set of Andreev bound states clustering near the superconducting gap edge, rather than the more spectrally isolated bound states found in Josephson nanowires.

\subsection{Andreev reflection and Andreev bound states}\label{sec:abs}

The pairing potential in each superconducting region leads to a superconducting gap $\Delta$, corresponding to the energy required to create a single quasiparticle. Due to the lack of pairing potential in the normal (semiconducting) region, Andreev bound states can be understood as the electronic states of a finite square well with a height equal to $\Delta$. 

To derive their spectrum, let's consider an electron moving to the right inside the weak link with energy $E<\vert\Delta\vert$. Due to the absence of unfilled single-particle states with energy less than $\Delta$ in the superconductor, an electron incident on the normal-to-superconducting (NS) interface cannot be transmitted into the superconductor as an electron. Instead, it is reflected as a hole via Andreev reflection, while a Cooper pair with charge $-2e$ is injected into the superconductor. A similar process will occur at the left NS interface, with the hole being reflected as an electron while a Cooper pair is extracted from the superconductor. (The same logic extends to electrons moving to the left and holes moving to the right.) Andreev reflection is the mechanism responsible for mediating the flow of supercurrent across the weak link. 

For simplicity, we assume that the phase of the order parameter $\Delta$ in the left superconducting region is zero and that the order parameter $\Delta e^{i\delta}$ in the right region has a phase $\delta$. At the right NS interface, the right-moving electron with energy $E$ will acquire a phase $\delta+\phi_r$ and reflect as a left-moving hole, where here, $\phi_r=-\arccos{E/\Delta}$~\cite{sauls2018andreev}. The left-moving hole will subsequently acquire a phase $\phi_r$ while reflecting from the left NS interface. In the short-junction approximation, the dynamical phase acquired while traversing the weak link is neglected, so that the total phase acquired by the charge carrier over one cycle is given by $2\phi_r+\delta$. This approximation can be justified when the length of the weak link is short relative to the superconducting coherence length $\xi=v_\mathrm{F}/\Delta$, where $v_\mathrm{F}$ is the Fermi velocity. The spectrum of a short junction can then be found by solving the following Bohr-Sommerfeld quantization condition, requiring that the phases acquired over a full cycle add up to an integer multiple of $2\pi$:
\begin{equation}\label{constructive-interference}
    2\pi n=\pm \delta+2\phi_r,\quad n\in\mathbb{Z}.
\end{equation}
The negative sign in front of $\delta$ is relevant to the case of right-moving holes, which reflect from the superconductor with the opposite phase. Although we have not explicitly restricted $n$ above, the fact that $\arccos{x}\in[0,\pi]$ and $\delta\in[0,2\pi)$ means that Eq.~\eqref{constructive-interference} admits solutions only for $n=0$ or $n=\pm 1$, depending on the value of $\delta$. In particular, for $\delta\in(\pi,2\pi)$, we have $E=-\Delta\cos{(\delta/2)}$ coming from the positive (negative) branch of Eq.~\eqref{constructive-interference} with $n=1$ ($n=-1$), while for $\delta\in[0,\pi)$, we have $E=\Delta \cos{(\delta/2)}$ coming from Eq.~\eqref{constructive-interference} with $n=0$. These cases can be combined to give the energy of the Andreev bound state as a function of $\delta$ over the full range of $\delta\in[0,2\pi)$:
\begin{equation}\label{andreev-spectrum}
    E_{\mathrm{A}}= \Delta\bigg\lvert \cos{\left(\frac{\delta}{2}\right)}\bigg\rvert.
\end{equation}
This bound state carries a supercurrent $I_{\mathrm{A}}=(2e/\hbar )dE_{\mathrm{A}}/d\delta$ driven by the difference $\delta$ in the phases of the superconducting order parameters on either side of the weak link. This is the DC Josephson effect, whereby current flows across the junction even in the absence of an applied voltage difference~\cite{josephson1962possible,josephson1964coupled,golubov2004current}. The Andreev bound state carries a positive current across the weak link for $\delta\in (\pi,2\pi)$, corresponding to a flow of Cooper pairs from left to right, and a negative current for $\delta\in(0,\pi)$. 

In a realistic system, the zero at $\delta=\pi$ will be shifted to a finite value by elastic scattering processes in the weak link that convert right-moving electrons or holes into left-moving electrons or holes, respectively. In the presence of such scattering, the energy of the Andreev bound state is instead given by \cite{beenakker1991universal,bagwell1992suppression,golubov2004current}
\begin{equation}\label{andreev-scatter}
    E_{\mathrm{A}}=\Delta\sqrt{1-\tau \sin^2{\left(\frac{\delta}{2}\right)}},
\end{equation}
where $\tau$ is the probability of a charge carrier being transmitted from one side of the normal region to the other. 

Although most experimental work has focused on III-V semiconductors like InAs, germanium can also be used as the semiconductor component of a weak link and would enable the integration of super-semi hybrid devices on germanium or silicon~\cite{katsaros2010hybrid,maier2014majorana, scappucci2021germanium}. Proximity-induced superconductivity in Ge has been observed with Ge/Si core-shell nanowires contacted by both Al~\cite{xiang2006ge,de2018spin,sistani2019highly,ridderbos2019multiple} and NbTiN~\cite{kotekar2017quasiballistic}. Weak links have also been demonstrated in planar Ge/SiGe heterostructures contacted by Al~\cite{hendrickx2018gate,vigneau2019germanium,hendrickx2019ballistic}, with supercurrent transport having been observed through Ge channels as long as 6 $\mu$m~\cite{hendrickx2019ballistic}.

\subsection{Andreev pair qubits}

The Andreev bound states discussed in Sec.~\ref{sec:abs} are spin-degenerate and are consequently referred to as Andreev doublets. Andreev doublets host Bogoliubov quasiparticles—spin--1/2 fermions with mixed electron-hole character whose creation or annihilation changes the fermionic parity of the weak link. In the short-junction limit, a weak link hosts a single doublet having four possible states: a ground state $\ket{g}$ where there are no quasiparticles in either level of the doublet, an excited state $\ket{e}$ where both levels are occupied, and two spin states $\ket{\uparrow},\ket{\downarrow}$ corresponding to having a quasiparticle in only one of the levels. Microwave driving can only couple states having the same fermionic parity, since microwaves do not have enough energy to create or destroy fermions. Hence, in a short junction, there is a drivable transition with frequency $2E_{\mathrm{A}}$ between the ground state $\ket{g}$ and the excited state $\ket{e}$, both of which contain an even number of quasiparticles, as well as a zero-frequency transition between the odd-parity states $\ket{\uparrow}$ and $\ket{\downarrow}$. Driving the transition $\ket{\uparrow}\leftrightarrow \ket{\downarrow}$ requires that the spin degeneracy be broken, which in turn requires going beyond the short-junction regime where the Bohr-Sommerfeld quantization condition is given by Eq.~\eqref{constructive-interference}. This case is discussed in Sec.~\ref{sec:andreev-spin}.

A comment is in order concerning the way in which the states were described above. In the excitation picture of superconductivity used here, only the excited state at energy $E_{\mathrm{A}}>0$ is considered; it is implicitly understood that the branch at energy $-E_{\mathrm{A}}$ is entirely filled in the ground state. This negative-energy branch must exist due to particle-hole symmetry, and states in this band correspond to the \textit{absence} of a quasiparticle at energy $E_{\mathrm{A}}$. The state $\ket{g}$ with no quasiparticles therefore has energy $E=-E_{\mathrm{A}}$, while the state $\ket{e}$ with one quasiparticle in each spin-degenerate level has energy $E=-E_{\mathrm{A}}+2E_{\mathrm{A}}=E_{\mathrm{A}}$. The two odd-parity states have energy $E=0$ and mediate no supercurrent since their energy is independent of $\delta$. A pedagogical introduction to the different pictures of superconductivity can be found by \textcite{hays2022realizing}. 

\begin{figure}
    \centering
    \includegraphics[width=0.6\linewidth]{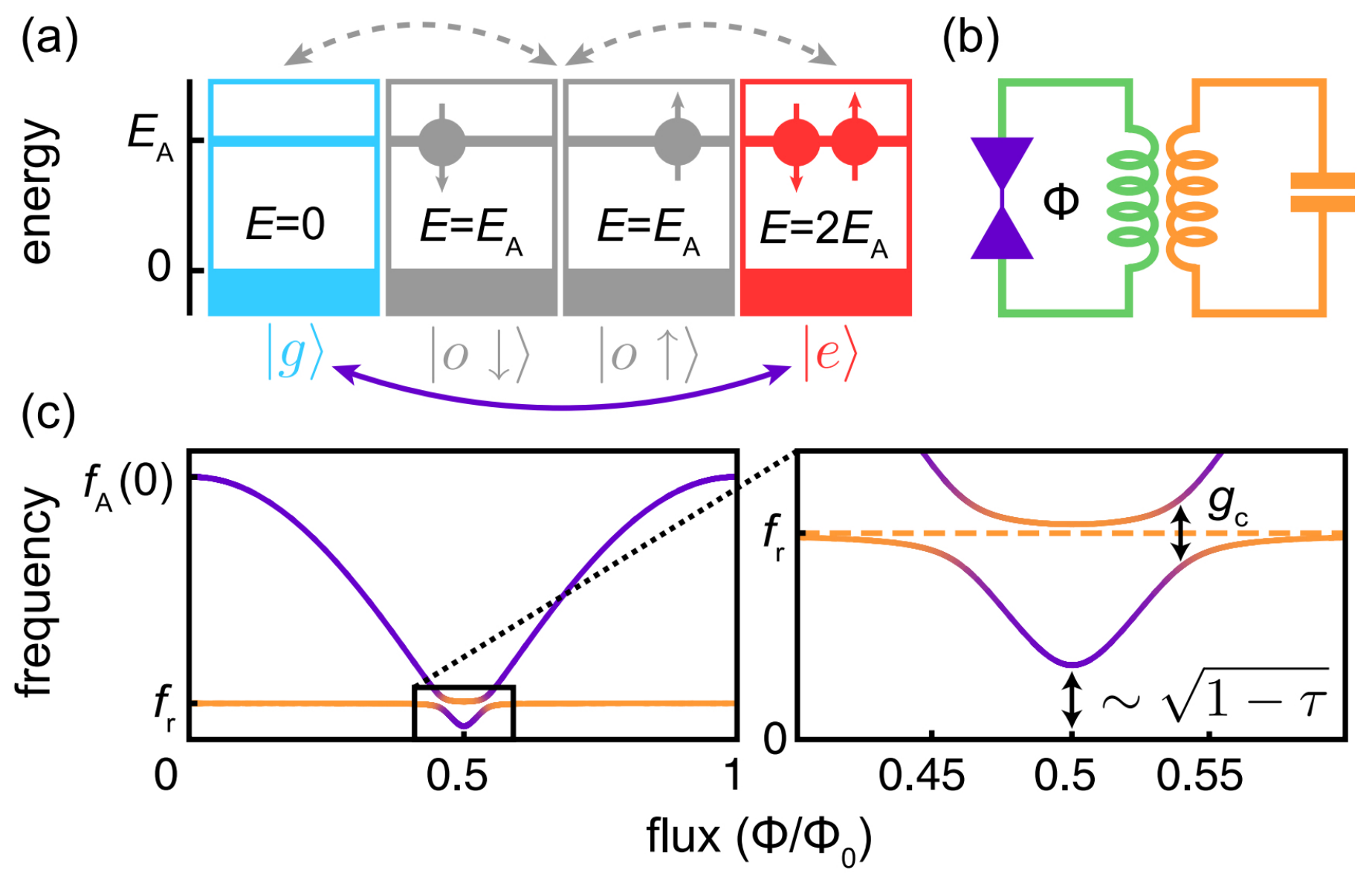}
    \caption{(a) Four states of a spin-degenerate Andreev bound state in the excitation picture of superconductivity. Andreev pair qubits are encoded in the levels $\ket{g}$ and $\ket{e}$, both of which have an even fermionic parity. (b) An Andreev pair qubit housed in a Josephson nanowire (purple) embedded in a superconducting loop (green) can be inductively coupled to and measured using a superconducting microwave resonator (orange). (c) An avoided crossing resulting from such a coupling.  Reproduced from \textcite{hays2018direct} with permission.}
    \label{fig:andreev-pair}
\end{figure}

First proposed by \textcite{zazunov2003andreev}, Andreev pair qubits (also known as Andreev level qubits) use the even-parity states $\ket{g}$ and $\ket{e}$ of a short weak link to encode the qubit, whose Hamiltonian is given by
\begin{equation}
    H_{\mathrm{A}}=E_{\mathrm{A}}\sigma_z,
\end{equation}
where $\sigma_z$ is a Pauli operator written in the $\ket{e},\ket{g}$ basis. Control over the qubit splitting $2E_{\mathrm{A}}$ can be achieved by contacting the superconducting leads of the Josephson nanowire to a superconducting loop threaded by a tunable flux $\Phi$, in terms of which $\delta=2\pi \Phi/\Phi_0$ where $\Phi_0=h/2e$ is the magnetic flux quantum. Control over the transmission parameter $\tau$ can be achieved by tuning the chemical potential in the nanowire via an applied gate voltage, providing further control over the qubit splitting~\cite{van2017microwave,janvier2015coherent}. Rotations around the $z$ axis can be achieved by varying the flux $\Phi=\bar{\Phi}+\delta\Phi$ about some DC value $\bar{\Phi}$, which, for small variations $\delta\Phi$, gives 
\begin{equation}\label{expand-about-flux}
    H=H_{\mathrm{A}}(\bar{\Phi})+ J_{\mathrm{A}}\:\delta \Phi+O(\delta\Phi^2),
\end{equation}
where $J_{\mathrm{A}}=\partial H_{\mathrm{A}}/\partial \Phi=(2\pi/\Phi_0)\partial H_{\mathrm{A}}/\partial \delta$ is the Andreev current operator. Evaluating $J_{\mathrm{A}}$ requires that the $\delta$ dependence of the Andreev levels themselves be taken into account, leading to an additional term in the derivative beyond the contribution due to $I_{\mathrm{A}}\propto d E_{\mathrm{A}}/d\delta$~\cite{zazunov2003andreev}. We quote the result, for which a detailed derivation appears in \textcite{bretheau2013localized}:
\begin{equation}\label{andreev-current-operator}
    J_{\mathrm{A}}=I_{\mathrm{A}}\left[\sigma_z+\sqrt{1-\tau}\tan{\frac{\delta}{2}}\sigma_x\right].
\end{equation}
The off-diagonal elements of the current operator can be used to drive transitions between even-parity states, as required for two-axis control.

\begin{figure}
    \centering
    \includegraphics[width=0.65\linewidth]{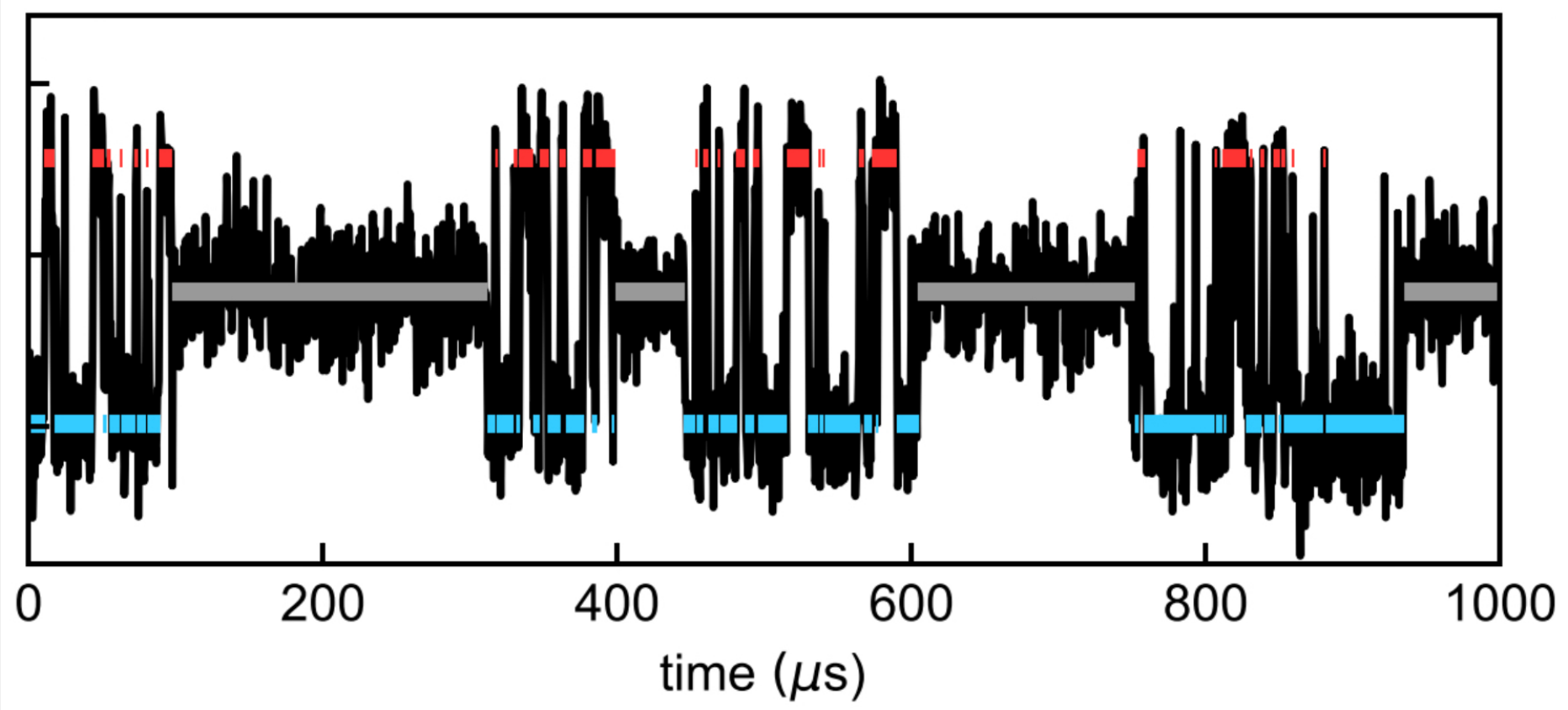}
    \caption{Parity switching measured through dispersive readout. The parity lifetime of 160 $\mu$s was attributed to quasiparticle poisoning. Reproduced from \textcite{hays2018direct} with permission.}
    \label{fig:parity-switching}
\end{figure}

The qubit can be probed through a circuit-QED setup by inductively coupling the superconducting loop to a microwave resonator (Fig.~\ref{fig:andreev-pair}). As a result of current fluctuations in the resonator, the phase $\delta$ then depends on both the applied flux $\Phi$ and the zero-point fluctuations (having amplitude $\Phi_{\textsc{zpf}}$) of the resonator magnetic field~\cite{metzger2021circuit}, which can be quantized as $\Phi_{\textsc{zpf}}(a+a^\dagger)$. Expanding in powers of $\Phi_{\textsc{zpf}}$ [similar to Eq.~\eqref{expand-about-flux}] then leads to a transversal coupling with strength $g(\delta)$ between the qubit and resonator, where
\begin{equation}
    g(\delta)=\Phi_{\textsc{zpf}}\sqrt{1-\tau}I_{\mathrm{A}}(\delta)\tan{\frac{\delta}{2}}.
\end{equation}
A coupling strength of $g/2\pi=23$ MHz has been realized and was used to probe the dispersive shift on a resonator due to an Andreev pair qubit~\cite{hays2018direct}. This experiment observed switching consistent with three possible values of the dispersive coupling, two of which are consistent with the shifts expected from the states of the Andreev pair qubit, and one of which was close to zero (Fig.~\ref{fig:parity-switching}). The latter was attributed to quasiparticle poisoning events changing the parity of the junction: For an odd-parity state, the supercurrent across the weak link is zero; the inductive coupling to the resonator is then also zero, causing the resonator to respond at its bare frequency.

Inductive coupling between two Andreev pair qubits and a common resonator mode has been realized experimentally, achieving transverse coupling strengths of $g/2\pi\approx 100{-}140$ MHz, well within the strong coupling regime~\cite{cheung2024photon}. The coupling of two qubits to the same resonator mode provides a realization of the Tavis-Cummings model~\cite{tavis1968exact}, which exhibits a dark state $(\ket{eg}-\ket{ge})\ket{0}$ where the qubits are entangled, but no photons are present in the resonator. Evidence of such a dark state has been measured using two-tone spectroscopy, demonstrating photon-mediated entanglement between Andreev pair qubits separated by a distance of 6 mm~\cite{cheung2024photon}.

\subsection{Andreev spin qubits}\label{sec:andreev-spin}

An alternate degree of freedom for encoding a qubit is provided by the spin of the quasiparticle in the odd-parity sector, as originally proposed by \textcite{chtchelkatchev2003andreev}. In the short-junction regime, these spin states are degenerate, and the transition is therefore impossible to address through microwave control. The spin degeneracy is broken in the presence of spin-orbit coupling in the semiconducting nanowire, which introduces a dependence of the Fermi velocity $v_{\mathrm{F}}$ on the spin state. Differences in the Fermi velocity become important when going beyond the short-junction approximation, as the dynamical phase acquired while propagating through the weak link can no longer be neglected. 

\begin{figure*}
    \centering
    \includegraphics[width=\linewidth]{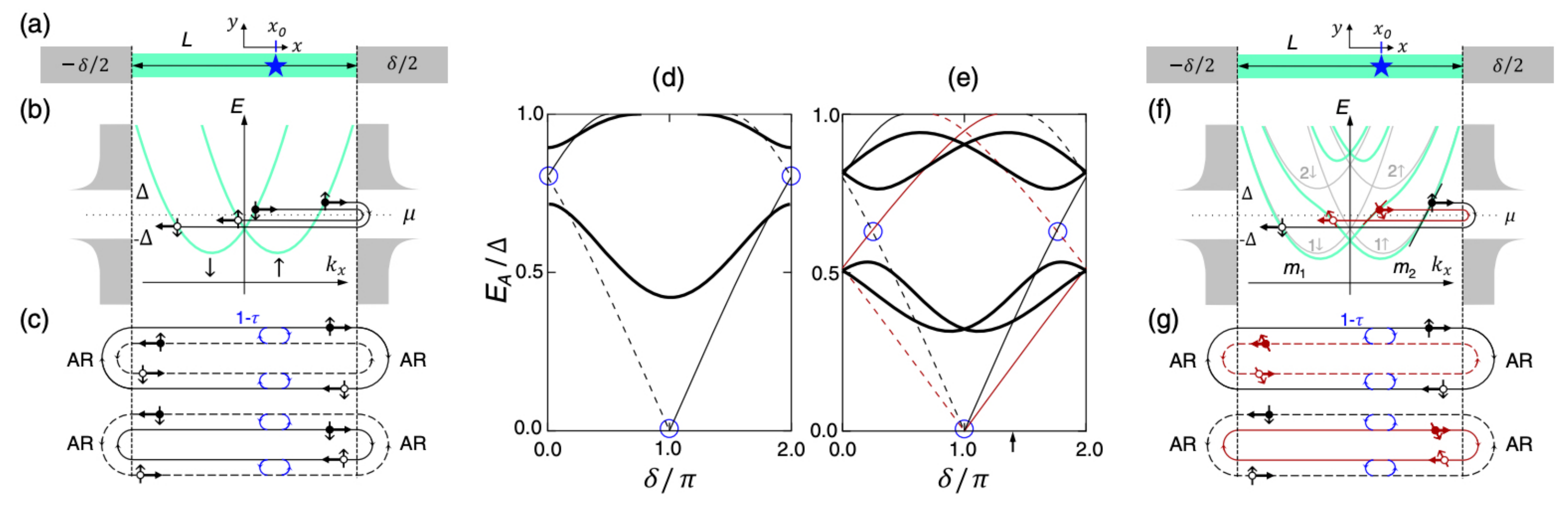}
    \caption{(a) Weak link (green) between two superconducting regions (grey) with a difference $\delta$ in the phase of their order parameters. (b) In a purely one-dimensional weak link, spin-orbit coupling shifts the momentum parabola in different directions depending on the spin state. (c) Andreev reflection at the NS interface couples electrons and holes, leading to the formation of Andreev bound states. (d) For the purely one-dimensional case, the Andreev bound states remain spin-degenerate despite the presence of spin-orbit coupling. Thick (thin) lines correspond to $\tau\neq 1$ ($\tau=1$). The higher Andreev doublet is the result of considering a longer junction (and not of spin-orbit coupling). (e) Andreev spectrum in the presence of spin-orbit coupling when the effects of the first-excited transversal subband are included. The thin and thick lines have the same meaning as in (d). (f) Spin-orbit coupling leads to avoided crossings between subbands with different spin projections, leading to a dependence of the Fermi velocity on the spin state. (g) Andreev reflection at the NS interface, where the red and black loops are associated with different velocities. Reproduced from \textcite{tosi2019spin}, licensed under CC BY 4.0.}
    \label{fig:andreev-spectrum}
\end{figure*}

The effects of spin-orbit coupling on the Andreev spectrum have been treated theoretically by \textcite{chtchelkatchev2003andreev,beri2008splitting,padurariu2010theoretical,reynoso2012spin,yokoyama2014anomalous,cayao2015sns,park2017andreev,murani2017andreev,van2017zeeman}. To understand why the Fermi velocity depends on the spin degree of freedom, we can consider a long nanowire with Rashba spin-orbit coupling, with the nanowire axis parallel to the $x$ axis [Fig.~\ref{fig:andreev-spectrum}(a)].  The Rashba spin-orbit Hamiltonian $H_{\mathrm{SO}}$ is given by
\begin{equation}
    H_{\mathrm{SO}}=\alpha_{\mathrm{R}}(\sigma_x p_y-\sigma_y p_x),
\end{equation}
where $\alpha_{\mathrm{R}}$ gives the strength of the spin-orbit coupling, $\sigma_x$ is a Pauli operator in the basis of the quasiparticle spin, and where $p_\nu=-i\partial_\nu$. Under the assumption that charge carriers are free to move along $x$ but are strongly confined along $z$ by, e.g., an electric field, we can restrict all motion to the lowest orbital state of the confinement potential along $z$. For a perfectly one-dimensional weak link (i.e., strong confinement along $y$ as well), we can neglect the effects of the term $\sim\sigma_x p_y$ in $H_{\mathrm{SO}}$ since the subbands of the confinement potential along $y$ are infinitely gapped. Treating the momentum $k_x$ along $x$ as a good quantum number, the term $\sim\sigma_y p_x$  then has the effect of displacing the parabolic energy dispersion along $k_x$ by an amount $\pm m^*\alpha_{\mathrm{R}}$ that depends on the material-dependent effective mass $m^*$ together with the state of the spin in the $\sigma_y$ basis [Fig.~\ref{fig:andreev-spectrum}(b)]. As a result, Andreev reflection couples electrons and holes of opposite spin and velocity [Fig.~\ref{fig:andreev-spectrum}(c)]. A further consequence is that for a perfectly one-dimensional weak link, spin-orbit coupling does not break the spin degeneracy [Fig.~\ref{fig:andreev-spectrum}(d)].

The situation changes drastically [Fig.~\ref{fig:andreev-spectrum}(e)] when the first-excited subband of the confinement potential along $y$ cannot be neglected (as above). We model confinement along $y$ with a harmonic-oscillator potential, such that the full Hamiltonian of the nanowire is given by 
\begin{equation}
    H=\frac{p_x^2}{2m^*}+\frac{p_y^2}{2m^*}+\frac{1}{2}m^*\omega_y^2y^2+H_{\textsc{SO}},
\end{equation}
where $\omega_y$ gives the energy difference between neighboring subbands of the transverse confinement potential.  Neglecting $H_{\mathrm{SO}}$, these subbands are spin-degenerate and disperse parabolically with $k_x$. As before, spin-orbit coupling lifts the spin degeneracy by shifting the bands in opposite directions depending on the spin state. However, in contrast to the limiting case of a perfectly one-dimensional weak link, the effects of the term $\sim \sigma_x p_y$ can no longer be neglected. This term results in avoided crossings between different subbands (since $p_y$ is linear in harmonic-oscillator creation and annihilation operators) [Fig.~\ref{fig:andreev-spectrum}(f)]. The dependence of this term on $\sigma_x$ also implies that the spin projection along $y$ is no longer a good quantum number—instead, the spin becomes entangled with the motional degree of freedom [Fig.~\ref{fig:andreev-spectrum}(g)]. Most notably, the avoided crossings lead to a change in the Fermi velocities of the Fermi points closest to $k_x=0$, while the outer Fermi points remain largely unaffected ~\cite{moroz1999effect,governale2002spin,reynoso2012spin,yokoyama2014anomalous,murani2017andreev,park2017andreev}.  

The difference in Fermi velocity will result in a difference in the dynamical phases acquired by charge carriers while propagating through the junction. To see why, we first consider a single pair of Fermi points, as is relevant to the spin-degenerate case where there is no spin-orbit coupling. Expanding about the two Fermi points at $k=\pm k_{\mathrm{F}}$, we have $E=\pm v_\mathrm{F}(k\mp k_\mathrm{F})$ with $v_{\mathrm{F}}=d E/dk\vert_{k=k_{\mathrm{F}}}$. Since the negative displacement cancels the negative sign in $k=-(k_{\mathrm{F}}+E/v_{\mathrm{f}})$ for leftward propagation, an electron traversing a normal region of length $L$ will acquire a dynamical phase $(k_{\mathrm{F}}+E/v_{\mathrm{F}})L$ independent of the direction of propagation. Holes acquire a phase $-k_{\mathrm{F}}L+E/v_{\mathrm{F}}L$ obtained by conjugating the phase and flipping $E\rightarrow -E$, resulting overall in a change in the sign of $k_{\mathrm{F}}$. Including the dynamical phase acquisition, the Bohr-Sommerfeld quantization condition [Eq.~\eqref{constructive-interference}] is now given by
\begin{equation}\label{constructive-2}
    2\pi n =\pm \delta+ 2\phi_{r}+\frac{2EL}{v_{\mathrm{F}}}.
\end{equation}

Unlike Eq.~\eqref{constructive-interference}, the Andreev spectrum can no longer be solved analytically. Insight can be gained, however, by expanding $\phi_r$ for $E\ll \Delta$, giving $2\phi_r=2E/\Delta-\pi+O((E/\Delta)^3)$. Rearranging Eq.~\eqref{constructive-2} for $E$ with this low-energy approximation gives
\begin{equation}\label{spectrum-long}
    E\simeq  \frac{\Delta}{1+L/\xi}\left[\mp \frac{\delta}{2}+ \pi\left(n+\frac{1}{2}\right)\right],
\end{equation}
where $\xi=v_{\mathrm{F}}/\Delta$ is the superconducting coherence length. As in the case of the short junction, the integer $n$ in Eqs.~\eqref{constructive-2} and \eqref{spectrum-long} cannot take on arbitrary values since bound states must have energies $E$ below the quasiparticle gap. Two important features of the Andreev spectrum for non-negligible $L/\xi$ can be inferred from the above approximation for $E$: First, the level spacing between bound states decreases as $L/\xi$ is increased, increasing the number of Andreev doublets hosted by the weak link. Second, the dispersion of $E$ with $\delta$ acquires a dependence on $\xi=v_{\mathrm{F}}/\Delta$. As discussed above, spin-orbit coupling in longer junctions gives rise to four Fermi points with different Fermi velocities $\pm v_{\mathrm{F},1}$ and $\pm v_{\mathrm{F},2}$ with $v_{\mathrm{F},1}< v_{\mathrm{F},2}$. The fact that these Fermi velocities are correlated with the spin state in turn lifts the spin degeneracy [Fig.~\ref{fig:andreev-spectrum}(e-g)].

As was the case with the short junction, disorder can scatter left-moving electrons or holes into right-moving electrons or holes. For $L\ll \xi$, the dependence of $E_{\mathrm{A}}$ on the transmission probability $\tau$ [Eq.~\eqref{andreev-scatter}] is typically derived under the assumption that right-moving electrons or holes with a given spin state scatter into left-moving electrons or holes of the same spin. In the presence of spin-orbit coupling, fast carriers moving with a velocity $v_{\mathrm{F},2}$ are, to a good approximation, in a $\sigma_y$ eigenstate. The state of slow carriers, however, cannot be associated with a definite spin projection along $y$ due to the hybridization of transversal subbands associated with different spin projections [Fig.~\ref{fig:andreev-spectrum}(f)]. Despite this, the Andreev spectra measured experimentally in long InAs nanowires~\cite{tosi2019spin} are consistent with a simplified scattering model in which excitations in the vicinity of the Fermi point with velocity $v_{\mathrm{F},1}$ ($-v_{\mathrm{F},1}$) have well-defined spin states given by the $\sigma_y$ eigenstate $\ket{\downarrow}$ ($\ket{\uparrow}$). Fast-moving excitations localized about $v_{\mathrm{F},2}$ ($-v_{\mathrm{F},2}$) have spin states $\ket{\uparrow}$ ($\ket{\downarrow}$), so within this simplified model and under the assumption that the disorder does not break time-reversal symmetry, one expects avoided crossings between, e.g., fast right-moving spin-up electrons and slow left-moving spin-up electrons [Fig.~\ref{fig:andreev-spectrum}(e),(g)]. Degeneracies at $\delta=0,\pi$ are protected by time-reversal symmetry, but as in the case of the short junction, the degeneracies are shifted in energy relative to the case with no backscattering~\cite{tosi2019spin}. 

\begin{figure}
    \centering
    \includegraphics[width=0.5\linewidth]{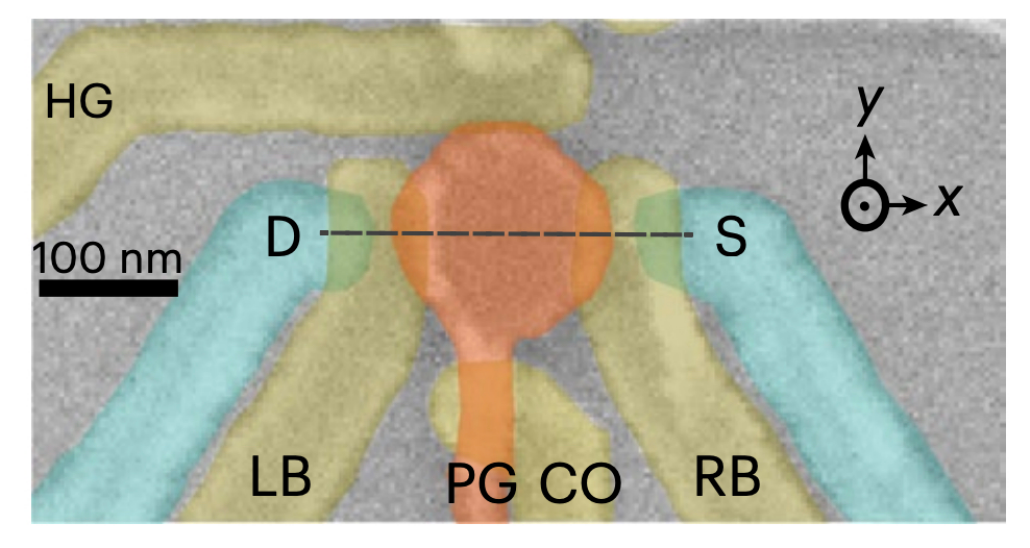}
    \includegraphics[width=0.5\linewidth]{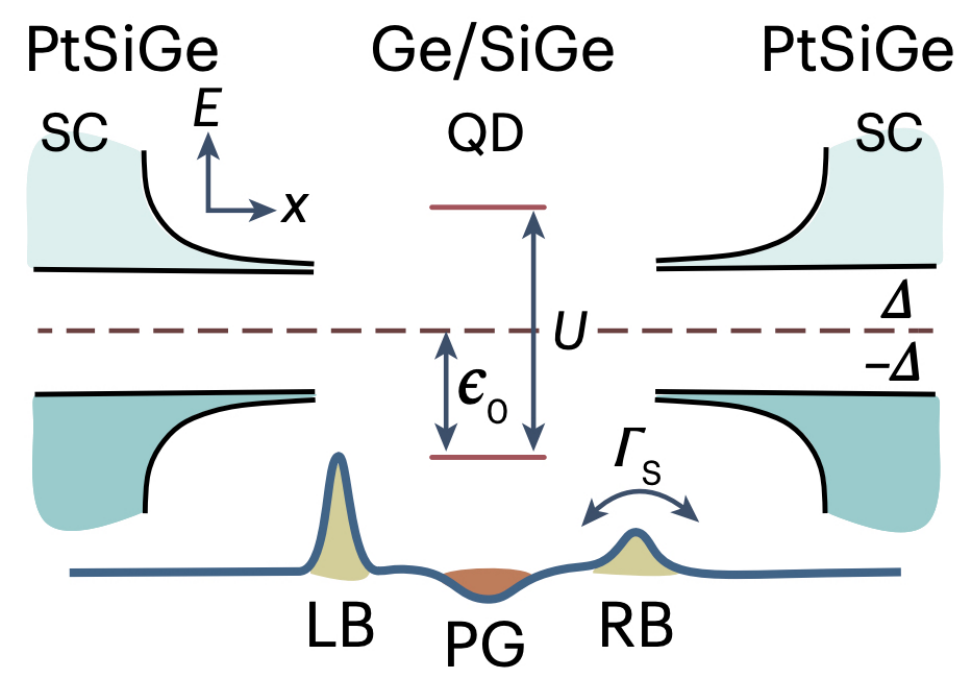}
    \caption{Quantum dot in germanium proximitized by superconducting leads S and D. The electrochemical potential $\epsilon_0$ of the dot is controlled by PG, while the coupling $\Gamma_S$ to the superconductors is controlled by LB and RB. Reproduced from \textcite{lakic2025quantum} with permission.}
    \label{fig:proximitized-dot}
\end{figure}

Even though the slow states are not associated with a definite spin projection along $y$, we nonetheless label the spin-split bands in each doublet by $\ket{\uparrow}$ and $\ket{\downarrow}$. The two lowest doublet states define the Andreev spin qubit, whose states we denote $\ket{\uparrow_{\mathrm{q}}},\ket{\downarrow_{\mathrm{q}}}$. Although the spin degeneracy has been lifted by spin-orbit coupling, it can still be challenging to couple the direct spin-flip transition to the current operator $J_{\mathrm{A}}$. For a wire with transverse mirror symmetry, driving along the nanowire axis cannot induce spin flips. While the mirror symmetry can be broken by an asymmetric coating of proximitizing superconductor or by the presence of gates~\cite{metzger2021circuit,hays2021coherent}, the qubit can also be manipulated and measured by taking advantage of higher (ancilla) doublets in the weak link, whose spin states we denote $\ket{\uparrow_{\mathrm{a}}}$ and $\ket{\downarrow_{\mathrm{a}}}$~\cite{hays2020continuous,hays2021coherent}. As demonstrated by \textcite{hays2020continuous}, the quasiparticle spin can be measured in a circuit QED setup by realizing a large dispersive coupling between a microwave resonator and the spin-preserving transitions $\ket{\uparrow_{\mathrm{q}}(\downarrow_{\mathrm{q}})}\leftrightarrow \ket{\uparrow_{\mathrm{a}}(\downarrow_{\mathrm{a}})}$. Manipulation of the qubit state has also been achieved by taking advantage of spin-flip transitions $\ket{\uparrow_{\mathrm{q}}(\downarrow_{\mathrm{q}})}\leftrightarrow \ket{\downarrow_{\mathrm{a}}(\uparrow_{\mathrm{a}})}$ between the qubit and ancilla manifolds~\cite{hays2021coherent}. With control over both spin-preserving and spin-flip transitions, a $\Lambda$ system can be realized by simultaneously driving one spin-preserving and one spin-flip transition involving the same ancilla state. By detuning the drives by the same amount from their respective transitions, population can be transferred between the two qubit states through a Raman transition without significantly populating the ancilla state. This strategy was used to perform Ramsey and Hahn-echo experiments that revealed coherent oscillations between $\ket{\uparrow_{\mathrm{q}}}$ and $\ket{\downarrow_{\mathrm{q}}}$ decaying on timescales of $T_{2\mathrm{R}}=18$ ns and $T_{2\mathrm{E}}=52$ ns, respectively \cite{hays2021coherent}.

Parity lifetimes on the order of tens of microseconds were measured by \textcite{hays2020continuous, hays2021coherent}, limited by quantum jumps between the even and odd-parity subspaces. Such parity switches are detrimental to the operation of Andreev qubits and can be mitigated by hosting the Andreev spin qubit in a quantum dot embedded in the Josephson junction~\cite{choi2000spin,meng2009self,lee2014spin,lee2017scaling,fatemi2022microwave,kurilovich2021microwave,bargerbos2022singlet,matute2022signatures,sahu2024ground,kurilovich2024demand}. With such a setup, ground state properties are determined by a competition between the coupling to the superconductors and the charging energy $U$ of the quantum dot: A large charging energy favors single occupancy, leading to an odd-parity, spin-1/2 ground state, while a strong coupling to the superconductors favors a singlet ground state. This interplay has been measured in an Al-proximitized InAs nanowire~\cite{bargerbos2022singlet}. A change in the ground-state parity of a proximitized quantum dot in germanium has also been measured experimentally using control over the dot's coupling to the superconducting leads (Fig.~\ref{fig:proximitized-dot}) \cite{lakic2025quantum}. The effects of Coulomb repulsion can therefore be used to prepare the system in the computational subspace of the Andreev spin qubit. Tuning ground-state properties in this manner also protects against leakage to non-computational states, resulting in significantly longer parity-switching times on the order of ms~\cite{pita2023direct}.  The quantum-dot confinement allows driving through electric-dipole spin resonance, enabling direct spin-flip transitions between qubit states without the need for ancilla doublets as in \textcite{hays2021coherent}. By embedding the Josephson junction containing the Andreev spin qubit in a transmon circuit~\cite{pavevsic2024generalized,gungordu2025quantum}, the transmon frequency can be made sensitive to the spin-dependent supercurrent associated with the qubit states. Using this approach, spin-transmon coupling with a strength of $2g/2\pi=104$ MHz has been realized~\cite{pita2023direct}. Coupling of (conventional) spin qubits to Andreev spin qubits, themselves coupled to resonators, has also been proposed as a way of measuring spin qubits using the resonator response to a probe tone~\cite{jakob2025fast}.

\begin{figure}
    \centering
    \includegraphics[width=0.55\linewidth]{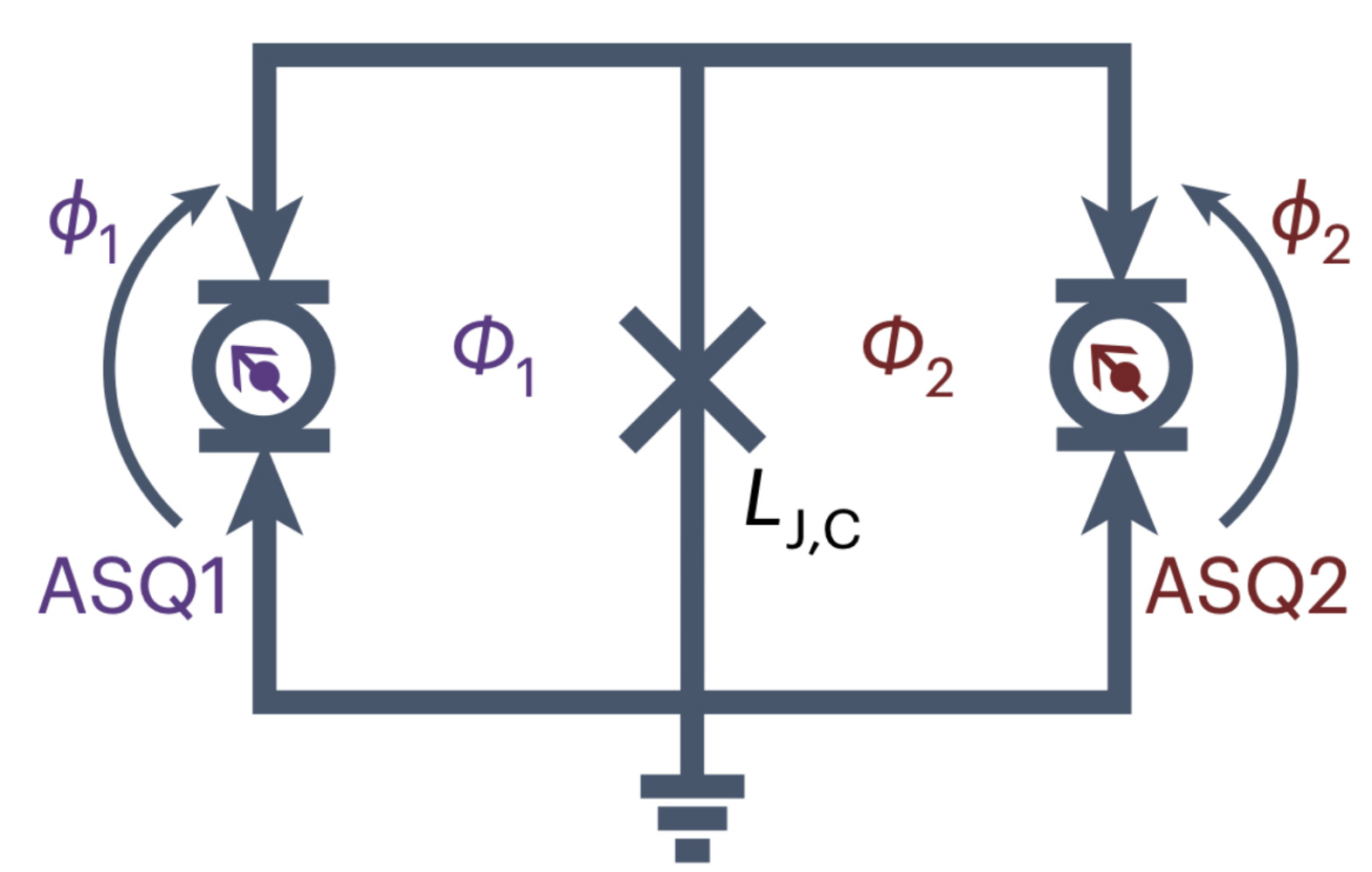}
    \caption{Two Andreev spin qubits ASQ1 and ASQ2 embedded in superconducting loops connected in parallel with a Josephson junction having a tunable inductance $L_{\mathrm{J,C}}$. Reproduced from \textcite{pita2024strong} with permission.}
    \label{fig:andreev-2Q}
\end{figure}

The spin-dependent supercurrent can also be used to realize tunable long-range coupling between distant qubits. In contrast with spin-spin interactions mediated by virtual photons~\cite{dijkema2025cavity}, which arise from a second-order process, the spin-dependence of the supercurrent enables direct two-qubit interactions.  A ZZ coupling $J\sigma_{z_1}\sigma_{z_2}$ between Andreev spin qubits separated by 25 $\mu$m has been realized by connecting two superconducting loops, each hosting a spin qubit, in parallel with a shared Josephson junction whose inductance can be tuned by a gate voltage (Fig.~\ref{fig:andreev-2Q}) ~ \textcite{pita2024strong}. In such a setup, the spin-dependent supercurrent associated with one qubit flows across one arm of the loop containing the other qubit, thereby coupling to its transition frequency. Coupling strengths $J/2\pi \approx 170$ MHz have been realized in this way~\cite{pita2024strong}, exceeding by roughly an order of magnitude the values achieved so far based on capacitive coupling~\cite{dijkema2025cavity}. An architecture for realizing tunable all-to-all connectivity of multiple Andreev spin qubits based on supercurrent-mediated coupling was proposed by \cite{pita2025blueprint}.

Inductive coupling is not the only mechanism by which Andreev spin qubits can interact. As mentioned at the beginning of this section, spins housed in normal-region quantum dots coupled to the same superconducting leads can also interact via the superconductor. In this case, the coupling arises from direct wavefunction overlap~\cite{choi2000spin,hassler2015exchange,kornich2019fine,gonzalez2021long}, which will exist provided the dots are separated by a distance smaller than the superconductor's coherence length—roughly 0.7 $\mu$m for Al-proximitized InAs~\cite{mayer2019superconducting}. This distance exceeds the tens of nanometers over which non-superconducting spin qubits typically interact through the exchange interaction~\cite{xue2022quantum}.  For Andreev spin qubits interacting through wavefunction overlap, the impact of spin-orbit coupling on the spin-spin interaction has been analyzed theoretically, and in particular, it was found that by tuning the superconducting phase difference $\delta$, the interaction can be made Ising-like, thereby enabling CZ gates~\cite{spethmann2022coupled}.

\section{Shuttling spins}

While circuit QED provides a photonic route to mediating long-range interactions, a fully spin-based approach appearing in the original proposal by \textcite{loss1998quantum} involves transferring a quantum state from one dot to the next via sequential SWAP gates, realized by pulsing the exchange interaction between spins in neighboring dots. Since sequential SWAP gates lead to errors that scale linearly with distance, an alternate strategy more amenable to transport over longer distances involves moving (``shuttling'') the spins themselves~\cite{taylor2005fault}. This approach has gained significant traction in recent years: With regards to scalability, it is widely accepted that sparse qubit arrays equipped with shuttling~\cite{boter2022spiderweb,kunne2024spinbus} could help alleviate the inter-qubit crosstalk~\cite{undseth2023nonlinear} and electrode-crowding issues liable to challenge the scalability of denser architectures based on direct exchange-based coupling. Notably, sparse arrays remain compatible with exchange-based two-qubit interactions, realized by temporarily shuttling electrons into close proximity, but are not constrained by a fixed qubit layout as in static architectures. As an additional hardware benefit, open spaces produced by the sparse layout could be used to integrate the classical electronics needed for control and readout~\cite{vandersypen2017interfacing}.  

The two primary modes of shuttling being pursued at the time of writing are known as bucket-brigade and conveyor-mode shuttling. Bucket-brigade shuttling proceeds via sequential adiabatic charge transfers between neighboring quantum dots, achieved by varying the inter-dot detuning. This approach can be used as the basis for single-qubit logic, but requires an increasing number of control pulses as the distance is increased. Conveyor-mode shuttling achieves spin transport by trapping the electron in a minimum of a moving potential waveform and can be realized with a number of control pulses that is independent of distance. While this approach reduces the control complexity, it is at the cost of the fine-tuned, localized control afforded by bucket-brigade shuttling. The strategies are therefore widely regarded as complementary, with bucket-brigade and conveyor-mode shuttling better suited to shorter and longer distances, respectively.

The flexible, reconfigurable connectivity enabled by shuttling is naturally compatible with non-planar quantum error correction codes, such as the class of bivariate bicycle (BB) quantum low-density parity check (LDPC) codes~\cite{mackay2004sparse,bravyi2024high}.  In this respect, the development of shuttling-based architectures for spin qubits parallels recent advances with reconfigurable arrays involving trapped ions~\cite{pino2021demonstration,sterk2022closed} and neutral atoms~\cite{bluvstein2022quantum,bluvstein2024logical}.

\subsection{Surface acoustic waves} 

Early shuttling demonstrations were achieved using surface acoustic waves (SAWs) — acoustic vibrations confined near the surface of a piezoelectric material. In such devices, metallic gates are patterned above a two-dimensional electron gas in a semiconductor heterostructure (typically GaAs/AlGaAs) to define an effective one-dimensional channel. An alternating RF voltage applied to the electrodes of an interdigitated transducer generates an oscillating electric field which, due to the piezoelectricity of the material, excites a propagating mechanical surface wave accompanied by a phase-locked electrostatic potential. Combined with the static one-dimensional confinement, this traveling potential modulation forms a train of moving quantum dots in which electrons can be captured and transported through the channel.

Evidence of electron trapping in SAWs was first observed in the context of current measurements in the 1990s~\cite{shilton1996high,talyanskii1997single}. Building on these early demonstrations of SAW-driven transport, later experiments achieved the transfer of single electrons between quantum dots separated by a few micrometers~\cite{mcneil2011demand, hermelin2011electrons}, and the electron analog of a beamsplitter has been demonstrated for an electron trapped in a SAW minimum~\cite{takada2019sound}. SAW-based, spin-preserving transfer of electrons has also been demonstrated over distances of a few micrometers~\cite{bertrand2016fast,jadot2021distant}. Coherent spin rotations have been realized via transport through the channel, enabled by the effective magnetic field induced by the large spin-orbit coupling in GaAs~\cite{jadot2021distant}. 

The SAW velocity is largely fixed by the material and has limited tunability. In addition, SAW-based shuttling in non-piezoelectric materials like silicon would likely require the incorporation of piezoelectric films, as realized in Refs.~\textcite{barros2012ambipolar,buyukkose2013ultrahigh}. Other shuttling techniques, namely bucket-brigade and conveyor-mode shuttling, have emerged in recent years and achieve effects similar to those of SAW-based transport, but with greater flexibility in material selection and control of the shuttling speed.

\subsection{Bucket-brigade shuttling}

\begin{figure}
    \centering
    \includegraphics[width=0.8\linewidth]{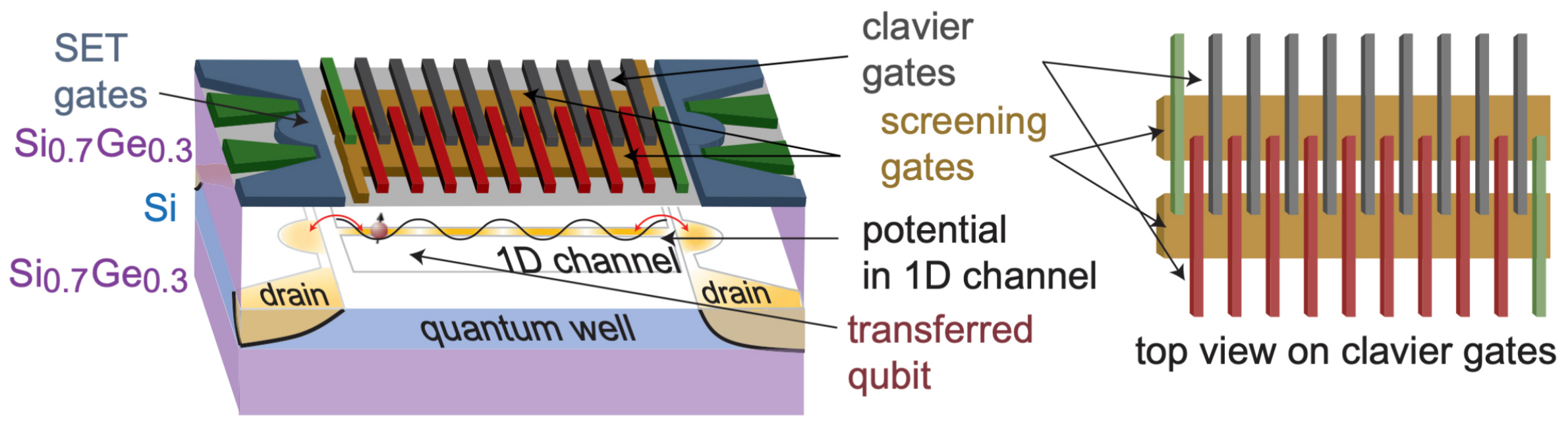}
    \caption{Gate-defined shuttling channel in a semiconductor heterostructure: A one-dimensional channel is formed in a two-dimensional electron gas using a pair of screening gates. Control over the shuttling process is provided by a number of interdigitated clavier gates on either side of the channel. Reproduced from \textcite{langrock2023blueprint}, licensed under CC BY 4.0.}
    \label{fig:clavier-gates}
\end{figure}

Greater control over the shuttling speed can be achieved using bucket-brigade transfer, albeit at the expense of increased control complexity. In this approach, a pair of screening gates fabricated on the surface of a semiconductor heterostructure defines a one-dimensional transport channel in the underlying two-dimensional electron gas (Fig.~\ref{fig:clavier-gates}). Additional gates are then patterned across the channel to form a series of quantum dots. By modulating the gate voltages in a phased sequence, the potential minima of adjacent dots can be shifted relative to one another, enabling a controlled sequence of adiabatic charge transfers. This stepwise process allows electrons to be transferred along the channel, with the fidelity of each transfer determined by the Landau–Zener dynamics between neighboring sites.

The transfer of charge between adjacent quantum dots indexed $(n,n+1)$ can be described by the Hamiltonian
\begin{equation}
    H_n=\frac{\epsilon_n}{2}\sigma_z+t_n\sigma_x,
\end{equation}
where $\epsilon_n$ is the detuning between the ground states of dots $n$ and $n+1$, $t_n$ is the tunnel coupling between them, and where $\sigma_z$ and $\sigma_x$ are Pauli matrices acting in the space of the two ground states. Varying $\epsilon_n$ from negative to positive will take the system through an anti-crossing with a minimum gap of $2t_n$ at $\epsilon_n=0$. The probability $P_{\mathrm{LZ}}$ of remaining in the ground state of $H_n$ throughout the detuning sweep is then given by the well-known Landau-Zener formula $P_{\mathrm{LZ}}=e^{-2\pi t_n^2/v}$~\cite{shevchenko2010landau}, where $v$ is the sweep rate. Theoretical studies of adiabatic transfer between a pair of quantum dots have analyzed errors due to spin-orbit coupling and valley states~\cite{li2017intrinsic,ginzel2020spin}, as well as errors due to 1/$f$ noise~\cite{krzywda2020adiabatic}.  

The first demonstrations of bucket-brigade shuttling were achieved in GaAs~\cite{baart2016single,fujita2017coherent,flentje2017coherent}. Following the demonstration of spin-preserving shuttling over effective distances of 80 $\mu$m in a three-dot array~\cite{baart2016single}, the effect of local Zeeman splittings  on the preservation of an initial spin singlet was investigated in a four-dot array by \textcite{fujita2017coherent}. In this scenario~\cite{fujita2017coherent}, one spin constituting half of a singlet state was shuttled to the end of the array and back, with the singlet return probability then measured via Pauli spin blockade. Differences in the Zeeman splittings of different dots led to the observation of singlet-triplet oscillations—evidence of coherence-preserving spin transfer. Spin coherence under shuttling has also been explored in a three-dot system arranged in a triangular geometry~\cite{flentje2017coherent}. With shuttling, coherence times eight times longer than the static case were observed at 200 mT~\cite{flentje2017coherent}, pointing to a significant reduction of hyperfine-induced dephasing due to motional narrowing~\cite{huang2013spin}. 

Bucket-brigade shuttling has also been achieved in linear dot arrays in silicon, initially with a charge degree of freedom~\cite{mills2019shuttling}. \textcite{noiri2022shuttling} provided a proof-of-principle demonstration of how two-qubit gates could be realized for qubits in distant processors: By shuttling distant spins into neighboring quantum dots, two-qubit gates could be performed using exchange interactions, and the interactions controlled with a large on-off ratio by simply separating the spins afterwards~\cite{noiri2022shuttling}. Shuttling over cumulative distances of 80 $\mu$m, equivalent in this case to 1000 charge transfers, has been demonstrated in a four-dot array with a spin-flip probability of 0.01\% per hop~\cite{zwerver2023shuttling}. Shuttling-induced errors were also investigated by~\textcite{foster2025dephasing} for a singlet-triplet qubit in a triple quantum dot, where it was observed that shuttle-induced spin-flip mechanisms became more pronounced in the regime of $>10^3$ shuttles. A study of the interplay between Zeeman splitting and tunnel coupling during bucket-brigade shuttling observed a degradation of shuttling fidelity when the average Zeeman splitting between the two dots was comparable to $2t_n$, emphasizing the importance of working with dots that are strongly tunnel-coupled~\cite{lin2025interplay}, as well as the importance of accounting for the Zeeman splitting of the spin state in addition to the Landau-Zener physics of the charge transfer itself~\cite{feng2023control}. This regime can be avoided by lowering the magnetic field, and in this way, shuttling fidelities of $99\%$ were achieved by~\cite{lin2025interplay}.

\begin{figure}
    \centering
    \includegraphics[width=0.7\linewidth]{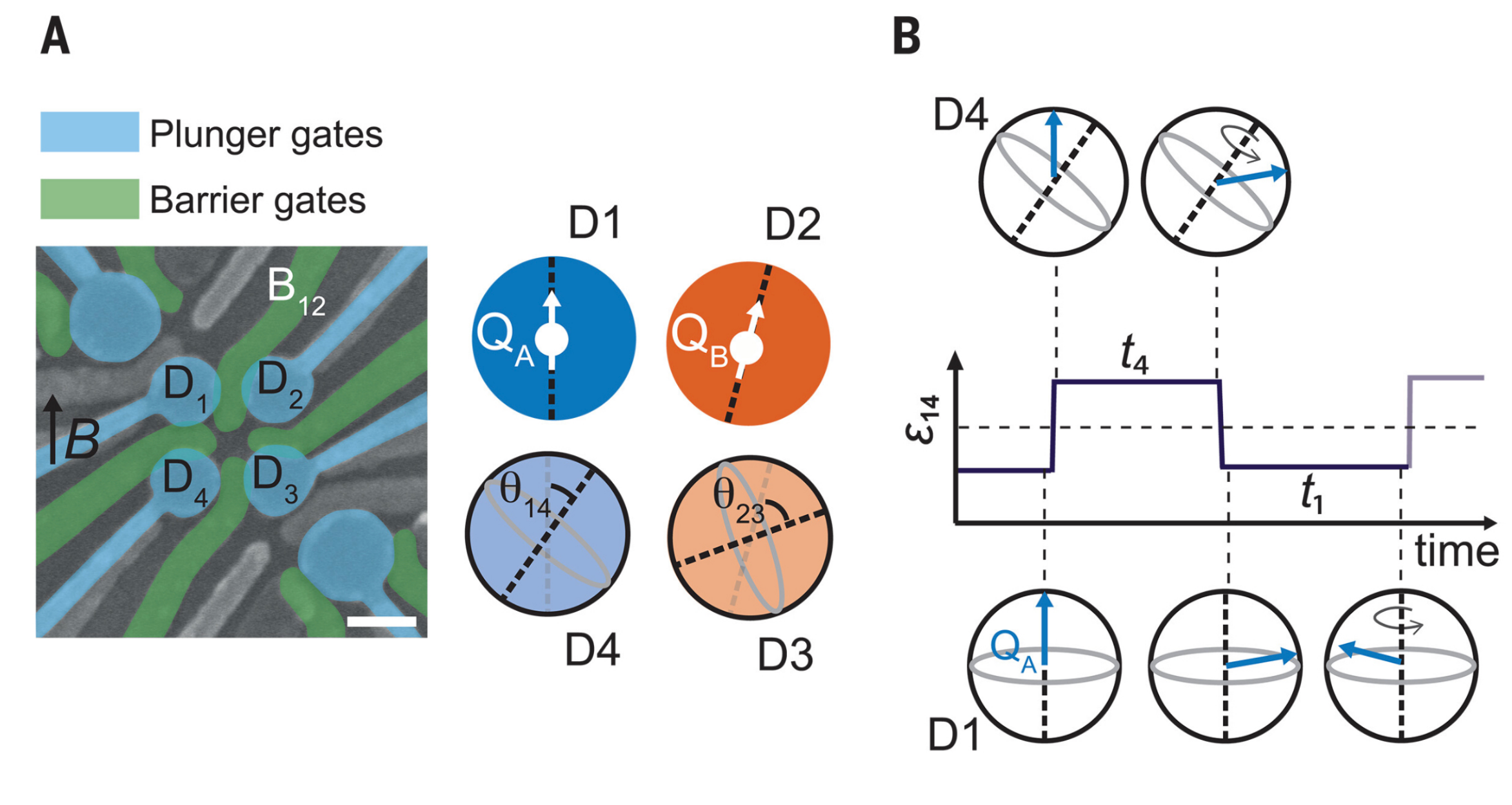}
    \caption{Single-qubit gates can be realized by physically transporting spins between quantum dots and taking advantage of site-dependent differences in the qubit quantization axis. Reproduced from \textcite{wang2024operating} with permission.}
    \label{fig:hopping}
\end{figure}

Adiabatic charge transfers (``hops'') between neighboring quantum dots can also be used to implement single-qubit gates. This idea can be traced back to the original Loss-DiVincenzo proposal~\cite{loss1998quantum}, which proposed that two-axis control be realized by taking advantage of site-dependent differences in qubit quantization axes. Hopping-based gates with fidelities exceeding $99\%$ have been achieved in $2\times 2$ dot arrays (Fig.~\ref{fig:hopping}) using both hole spins in germanium~\cite{wang2024operating,van2024coherent} and electron spins in silicon~\cite{unseld2025baseband}, where site-dependent variations in the qubit quantization axis resulted from intrinsic spin-orbit coupling and the field gradient due to a micromagnet, respectively. Hopping-based gates offer the advantage of operating with baseband voltage control rather than microwave-frequency driving, as is required for electric-dipole spin resonance (EDSR); this could help mitigate the impacts of heating and unwanted crosstalk between qubits. Hopping spins have also been used to characterize the site-dependent Larmor frequencies and dephasing times across a 10-dot array, providing an avenue towards the characterization of large-scale systems~\cite{wang2024operating}. Potential drawbacks of hopping-based gates are the high levels of timing precision required, on the order of tens of picoseconds~\cite{unseld2025baseband}, and their incompatibility with virtual phase gates~\cite{mckay2017efficient}.

For long-range transfer, bucket-brigade shuttling imposes stringent requirements on material uniformity in architectures where charge transfer must be orchestrated through a limited set of control lines addressing only a few gate sets. It has been estimated, for instance, that achieving adiabatic transfer through 100 tunnel-coupled quantum dots in a total time of 1 $\mu$s and with errors due to non-adiabaticity on the order of $10^{-3}$ would require tunnel couplings of $t_n>35$ $\mu$eV~\cite{langrock2023blueprint}. Achieving uniformly large tunnel couplings with only common voltage control may be challenging in, e.g., Si/SiGe with state-of-the-art disorder~\cite{langrock2023blueprint}, leading to increased probabilities that the conditions for adiabatic transfer are not satisfied for some pairs of dots along the channel.

\subsection{Conveyor-mode shuttling}

\begin{figure}
    \centering
    \includegraphics[width=0.8\linewidth]{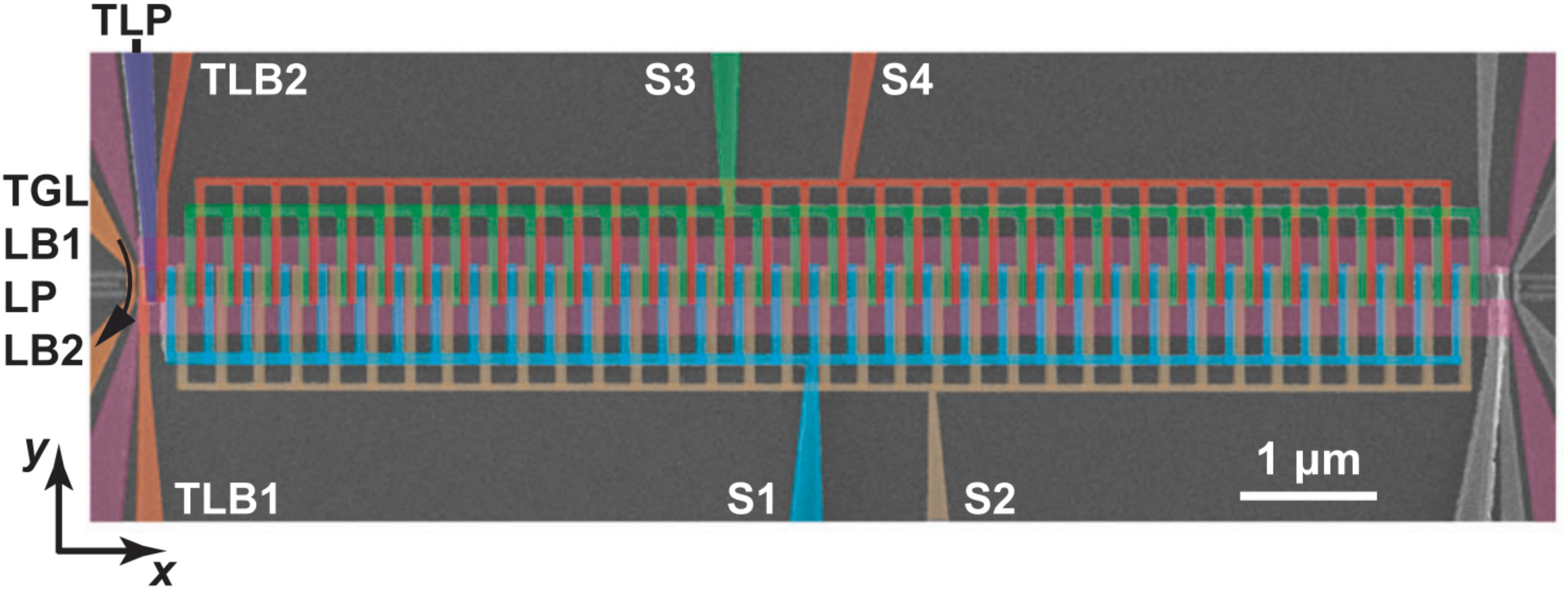}
    \caption{Channel used for conveyor-mode shuttling over distances of $10$ $\mu$m. The travelling potential was created by applying four phase-offset voltage modulations to the four connected sets of clavier gates coupled to S1-S4. Reproduced from \textcite{xue2024si}, licensed under CC BY 4.0.}
    \label{fig:long-channel}
\end{figure}

An alternate, more scalable approach is provided by conveyor-mode shuttling, for which the control requirements are independent of the shuttling distance.  For conveyor-mode shuttling, localized control at individual charge anticrossings between tunnel-coupled quantum dots is not required. Instead, a small number of phase-shifted sinusoidal voltage signals are applied to connected clavier-gate sets to create a moving potential in which an electron can be trapped. Conveyor-mode shuttling is thus conceptually similar to shuttling based on surface acoustic waves but comes with increased tunability of the shuttling speed. With four connected gate sets, the moving potential can be created by applying four phase-offset voltage modulations given (up to DC biases) by~\cite{seidler2022conveyor,kunne2024spinbus}
\begin{equation}
    V_i=A_S\cos{[2\pi f t-\pi/2(i-1)]},\quad i =1,\dots,4.
\end{equation}
Since the main requirement is that the applied voltages produce a moving quantum dot capable of trapping the charge, a quantum dot with a depth exceeding the typical variation of the electrostatic disorder potential on the  scale of the dot is expected to provide some robustness to material non-uniformity~\cite{langrock2023blueprint}. 

If spin-orbit coupling or Overhauser fields lead to spatial variation in the spin splitting $\omega_{\mathrm{q}}(x)$, then a spin will precess at different frequencies depending on its position in the channel. Fluctuations in $\omega_{\mathrm{q}}(x)$ due to, e.g., nuclear-spin dynamics or 1/$f$ noise will generally dephase the spin; however, this dephasing can be mitigated by shuttling at an increased velocity $v$ in order to take advantage of motional-narrowing effects. A theoretical analysis in terms of dynamical-decoupling filter functions was performed by \textcite{bosco2024high}, where it is shown that for realistic device parameters, spectral weight in the filter function can be shifted to tens of MHz by the motion of the dot—away from the low-frequency noise that typically limits spin qubits.  The velocity $v$ cannot be increased arbitrarily, however, without triggering nonadiabatic transitions involving orbital and valley degrees of freedom, which generally lead to level-dependent spin precession and additional dephasing~\cite{langrock2023blueprint}. An additional concern related to the presence of valleys is the possibility of encountering a spin-valley hotspot $E_\mathrm{Z}\approx E_{\mathrm{VS}}$ during shuttling, where the Zeeman splitting $E_\mathrm{Z}$ is approximately equal to the valley splitting $E_{\mathrm{VS}}$~\cite{volmer2025reduction}. At such a hotspot, spin-valley hybridization leads to enhanced spin relaxation. This error mechanism can be suppressed by operating at magnetic fields chosen so that the Zeeman splitting is either well below or well above the typical range of valley splittings along the shuttling channel. For instance, it has been estimated~\cite{langrock2023blueprint} that for a magnetic field of 20 mT, the Zeeman splitting of $E_\mathrm{Z}\approx 2$ $\mu$eV is well below the minimum valley splitting of $E_{\mathrm{VS}}\approx 10$ $\mu$eV reported by \textcite{kawakami2014electrical} for Si/SiGe.

A proof-of-principle demonstration of conveyor-mode shuttling with fidelity $>99\%$ was achieved in 2022 in a 420-nm channel in Si/SiGe~\cite{seidler2022conveyor}. The fidelity reported in this experiment is a charge-shuttling fidelity quantifying the number of successful shuttles, as measured by a nearby single-electron transistor acting as a charge sensor. A charge-shuttling fidelity of 99.7\% was later demonstrated in a 10-$\mu$m long channel (Fig.~\ref{fig:long-channel})~\textcite{xue2024si}, which, notably, is a distance that comes with improved prospects for scalability~\cite{kunne2024spinbus}. This work also showed that 34 electrons could be loaded into different minima of the traveling potential, allowing them to be shuttled through the channel simultaneously~\cite{xue2024si}.  Spin-preserving conveyor-mode shuttling with a fidelity of 99.3\% has also been demonstrated~\cite{struck2024spin}. This was achieved by initializing a singlet at one end of the channel, shuttling one of the spins, then measuring the singlet-return probability through Pauli spin blockade~\cite{struck2024spin}. Observed increases in $T_2^*$ times achieved for longer shuttling distances are consistent with predicted enhancements due to motional narrowing~\cite{struck2024spin}.

A comparison of spin-shuttling fidelities for bucket-brigade and conveyor-mode shuttling was performed using a linear six-dot array in Si/SiGe~\cite{de2025high}.  This work reported phase-flip probabilities of 0.2\% per hop for bucket-brigade operation, while the phase-flip probabilities obtained for conveyor-mode shuttling were lower by an order of magnitude. While previous works used phase-shifted voltage signals with a single frequency, \textcite{de2025high} reduced charge leakage using a two-frequency scheme designed to engineer destructive interference at alternate minima in the conveyor-mode potential. This modification enabled spin-preserving shuttling over distances of 10 $\mu$m in less than 200 ns while maintaining high fidelities of 99.5\%, corresponding to an impressive speed of approximately 50 m/s. 

% \begin{figure}
 %   \centering
  %  \includegraphics[width=0.7\linewidth]{shuttle-gate.pdf}
   % \caption{Spins in separate minima of a conveyor-mode potential can be brought close together to realize an entangling gate. The exchange coupling $J$ between the two spins can be turned on and off using a barrier-gate voltage. Figure from \textcite{matsumoto2025two}.}
  %  \label{fig:shuttle-gate}
%\end{figure}

The same two-frequency conveyor mode was used to demonstrate CZ gates between shuttled spins~\cite{matsumoto2025two}. The gate was realized by moving two spins in different conveyor-mode minima towards each other and precisely controlling their interaction through a barrier-gate voltage. This shuttling-based CZ gate was then used to realize quantum teleportation across a six-dot array~\cite{matsumoto2025two}. Due to the inability of the measurement to distinguish all four Bell states, the teleportation succeeded only probabilistically, motivating future improvements in readout techniques. Going beyond CZ gates, it has also been shown theoretically that a wide class of two-qubit gates, including fermionic simulation gates, could be implemented by taking advantage of an intrinsic or engineered spin-orbit interaction during shuttling~\cite{fernandez2025spin}. In the case of hole spins in Ge, \textcite{ademi2025distributing} have demonstrated a protocol to compensate for such spin-orbit-induced rotations during shuttling and have realized shared entanglement between micrometer-separated spins using shuttling in combination with local control.

\subsection{Prospects for quantum error correction}

In static, two-dimensional architectures, the spatial layout of the qubits typically imposes constraints on the quantum error-correcting codes that can be implemented. The leading candidate in this setting is the surface code~\cite{fowler2012surface}, which only requires stabilizer measurements of small groups of neighboring qubits. The surface code is a member of the Calderbank-Shor-Steane (CSS) class of codes and, consequently, admits a transversal CNOT gate~\cite{gottesman1998theory}. Transversal two-qubit gates are fault-tolerant in the sense that an error on one code block (i.e., a group of physical qubits encoding a logical qubit) will propagate to a bounded number of physical qubits in the second code block. Implementing a transversal two-qubit gate between spatially separated surface-code qubits would be very challenging in a static architecture due to the need to have all physical qubits from one code block interact with all physical qubits from the other. For this reason, alternative techniques such as defect-based methods~\cite{raussendorf2007topological,raussendorf2007fault,fowler2012surface} or lattice surgery~\cite{horsman2012surface} are typically considered instead, both of which must be implemented over a large number $O(d)$ of code cycles, where here $d$ is the code distance. This can be contrasted with the single cycle required for a transversal two-qubit gate.

\begin{figure}
    \centering
    \includegraphics[width=0.7\linewidth]{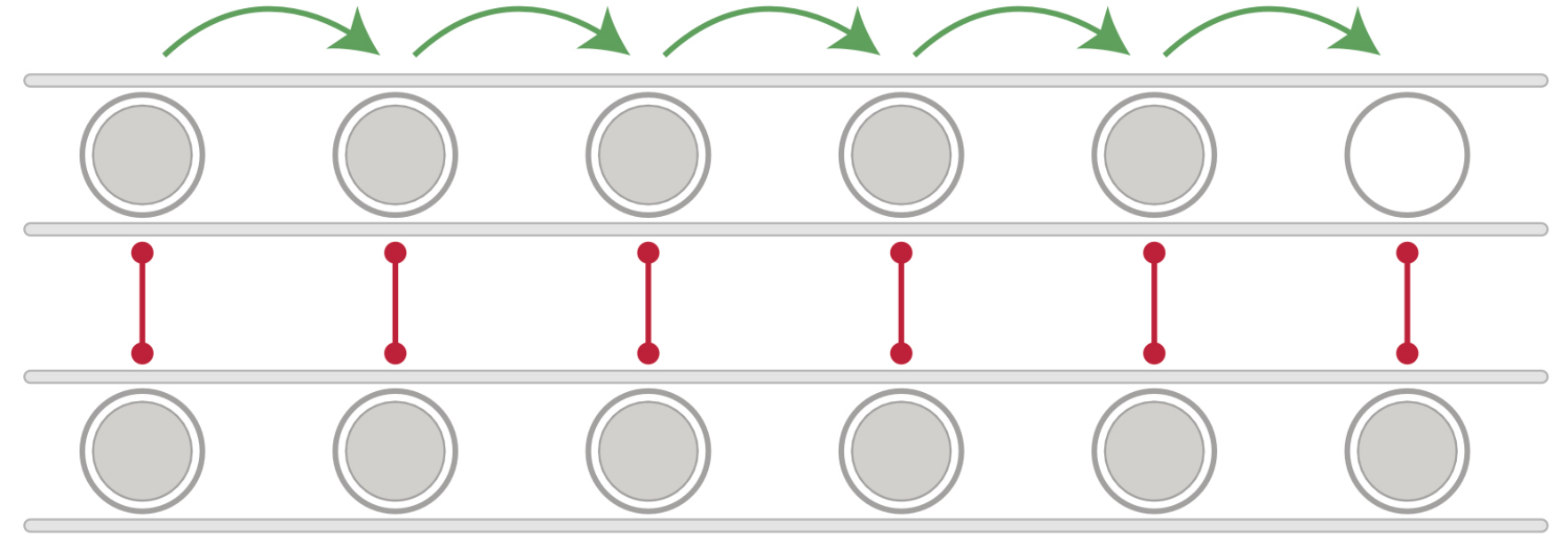}
    \caption{A $2\times N$ array of qubits equipped with shuttling can be used to realize different bipartite connectivity graphs, as required for running quantum error-correcting codes. The ancilla qubits are placed in the bottom row, while the physical code qubits are shuttled back and forth in the top row. Reproduced from \textcite{siegel2024towards}, licensed under CC BY 4.0.}
    \label{fig:2xN}
\end{figure}

The ability to physically transport qubits would allow them to be dynamically repositioned for syndrome extraction and two-qubit gates between logical qubits, without the need for a dense network of statically patterned mediating couplers. To help mitigate the large spatial footprint of the readout components relative to the qubits themselves, \textcite{chadwick2025manufacturable} proposed surface-code computing with serialized ancilla qubit readout, facilitated by the ability to shuttle the ancilla qubits to a smaller number of readout zones than would otherwise be required for fully parallelized syndrome readout. Spin shuttling is also particularly well suited for use with non-planar codes, such as BB qLDPC codes~\cite{bravyi2024high}, where long-range parity checks are required. Early demonstrations of qLDPC codes with two-dimensional arrays of superconducting qubits have required the use of air bridges to allow for the crossing of long-range couplers~\cite{wang2025demonstration}, which could likely be avoided in an architecture compatible with qubit shuttling. Small qLDPC codes compatible with near-term implementation were constructed by \textcite{andersen2025small}, where a detailed shuttling-based implementation of the required parity-check circuits is also given. The problem of embedding quantum error correction codes, including surface and qLDPC codes, onto geometrically constrained near-term $2\times N$ arrays (Fig.~\ref{fig:2xN}) equipped with shuttling was considered by \textcite{siegel2024towards}. In that work, it was proposed that one row of the array contains the ancilla qubits used for stabilizer measurements, while the code qubits are contained in the other. Arbitrary bipartite connectivity graphs  could then be realized by shuttling one row of qubits past the other. In practice, however,  the size of the code would likely be limited by a practical upper bound on the number of shuttling steps. 

To achieve surface-code quantum computing in a shuttling-based architecture, \textcite{siegel2025snakes} proposed a setting where logical qubits are shuttled past one another as one-dimensional chains. This would enable all-to-all connectivity at the logical level, as well as transversal two-qubit gates. However, a notable challenge confronting this approach is the possibility of introducing massively correlated errors on logical qubits caused by shuttling all of their constituent physical qubits along the same channel. The proposed solution combines lattice surgery and error detection enabled by ``monitor qubits'', which are incorporated into the shuttled logical qubit and later measured to infer the presence of such a correlated error~\cite{siegel2025snakes}. A shuttling-based architecture that maintains the spatial separation of physical qubits belonging to the same logical qubit is provided by the looped pipeline architecture of \textcite{cai2023looped}.

\section{Spin qubits and topological spin textures}

There have also been proposals for connecting distant spin qubits using nanoscale topological spin textures acting as solid-state flying qubits~\cite{zou2025topological}. For long-distance entanglement via magnons, see also \textcite{trifunovic2013longdistance,flebus2019entangling}. These flying qubits are encoded in the chirality of a magnetic domain wall in a quasi one-dimensional ferrimagnetic nanowire acting as a magnetic racetrack~\cite{parkin2008magnetic}. Domain walls have widespread applications in the field of spintronics; see \textcite{yang2021chiral} for a review. Although many experiments operate in the classical regime, there has also been significant progress in stabilizing and manipulating nanoscale domain walls~\cite{ryu2013chiral,yang2015domain}, where quantum effects become important. 

\begin{figure}
    \centering
    \includegraphics[width=0.43\linewidth]{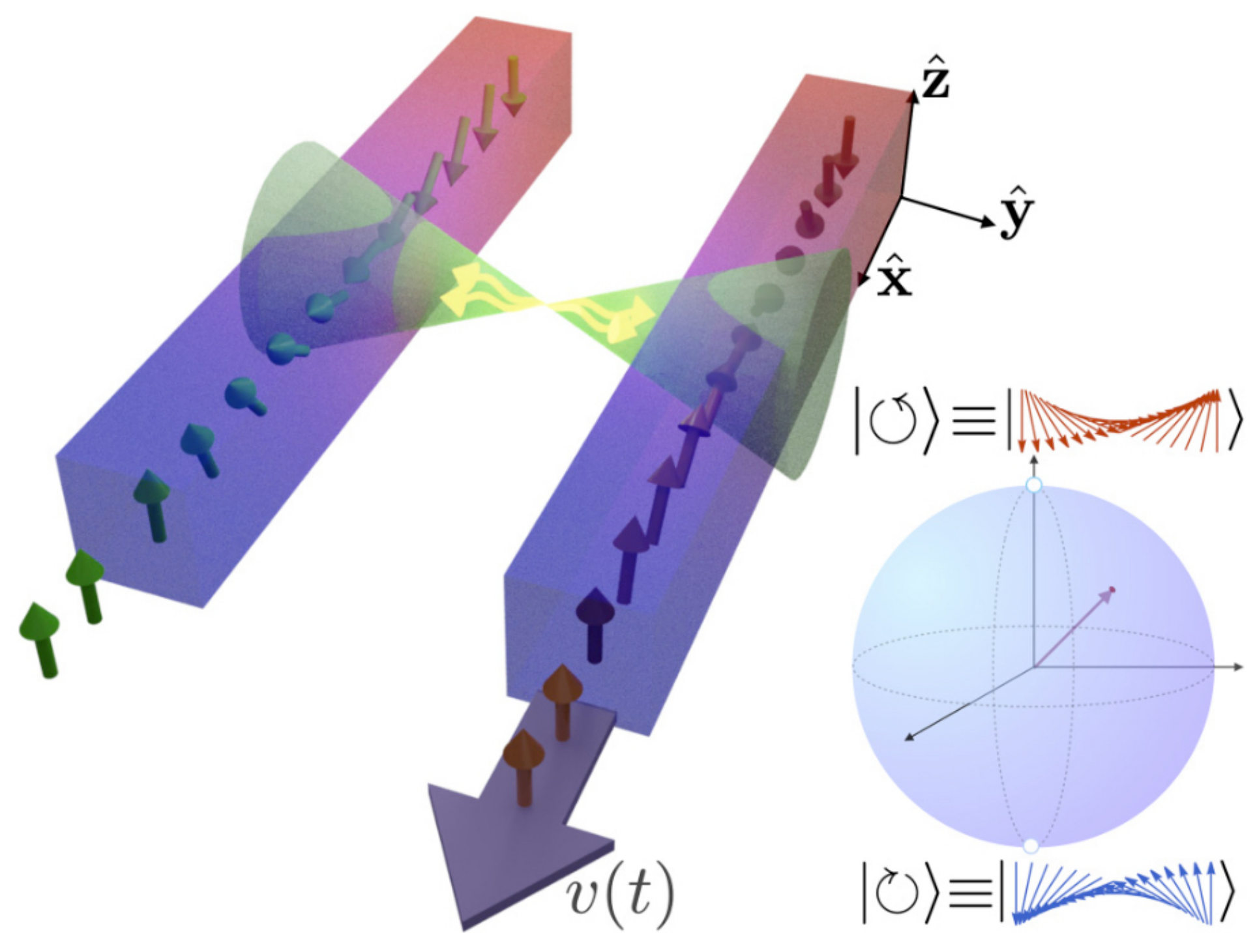}\hspace{1cm}
    \includegraphics[width=0.5\linewidth]{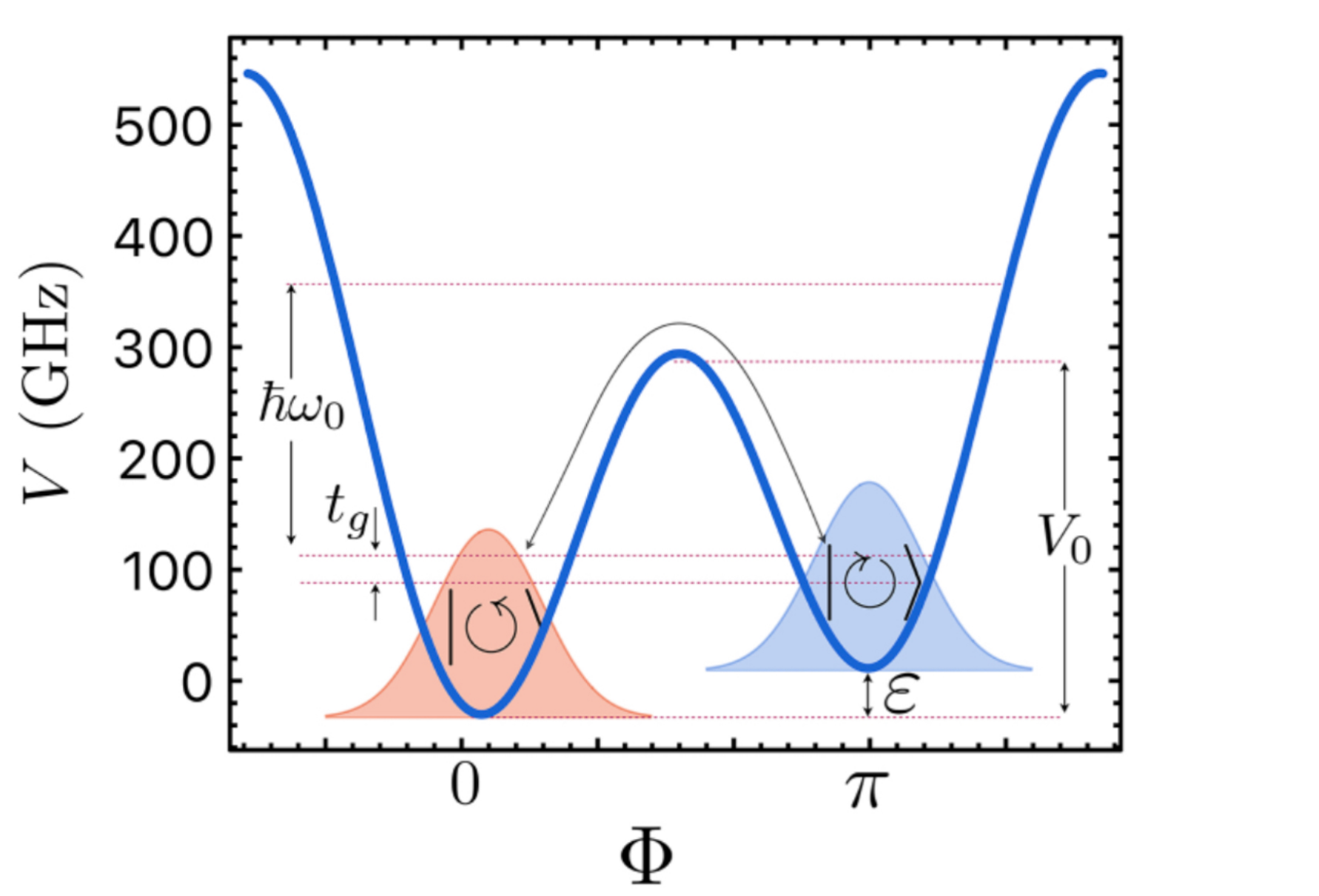}
    \caption{Left: Sketch of two domain-wall qubits on parallel ferrimagnetic racetracks. The domain wall on the right (left) is in the chirality state $\ket{\circlearrowleft}=\ket{\Phi=0}$ ($\ket{\circlearrowright}=\ket{\Phi=\pi}$). The individual spins depicted here represent the excess spin in the ferrimagnetic unit cell. Right: In the presence of an easy-plane anisotropy, the potential energy $V(\Phi)$ has a double-well shape with minima corresponding to $\ket{\circlearrowleft}$ and $\ket{\circlearrowright}$. Reproduced from \textcite{zou2023quantum}, licensed under CC BY 4.0.}
    \label{fig:domain-wall}
\end{figure}

A quasi one-dimensional ferrimagnetic nanowire can host domain-wall spin textures described by the continuous N\'eel vector field $\bm{n}(x)$, where $x$ is a spatial coordinate along an axis parallel to the nanowire. For a system with only an easy $z$-axis anisotropy (making spin alignment along $z$ energetically favorable), the components of the N\'eel vector are given by~\textcite{kim2023mechanics}
\begin{align}
    n_x(x)+in_y(x)&=e^{i\Phi}\mathrm{sech}(x-X),\,\,\,n_z(x)=\tanh{(x-X)}.
\end{align}
The quantities $X$ and $\Phi$ denote the real-space position of the domain wall and its azimuthal angle in spin space, respectively, with the latter  corresponding to the direction along which the spins point at the center of the wall. Since $X$ and $\Phi$ are both zero modes, variations of $X$ and $\Phi$ leave the energy of the system invariant. Introducing an easy $xz$-plane anisotropy breaks the $U(1)$ symmetry associated with $\Phi$-invariance, reducing it to a $\mathbb{Z}_2$ symmetry where the chirality states $\ket{\circlearrowleft}=\ket{\Phi=0}$ and  $\ket{\circlearrowright}=\ket{\Phi=\pi}$ are energetically favored. The two states spanning the domain-wall qubit subspace are therefore N\'eel-type walls in which the spins rotate in-plane (Fig.~\ref{fig:domain-wall}). 

In the absence of an external magnetic field, the two chirality states $\ket{\circlearrowleft}$ and $\ket{\circlearrowright}$ are degenerate; however, this $\mathbb{Z}_2$ symmetry can be further broken by applying an external field $\bm{h}$ in the $xy$ plane. For nanoscale domain walls, in which the magnetization flips over a length scale of $N\approx 10^2$ spins, quantum effects are non-negligible~\cite{chiolero1997macroscopic,qu2025density}, and tunneling leads to the hybridization of the two chirality states with a tunnel splitting $t_g\approx 4\omega_0\sqrt{S_{\mathrm{inst}}/2\pi}e^{-S_\mathrm{inst}}$, where $S_\mathrm{inst}\approx 4V_0/\omega_0$ is the instanton action written in terms of the tunneling barrier $V_0$ and level spacing $\omega_0$ (set by the curvature of the potential minimum). Both $V_0$ and $\omega_0$ can be tuned by the external magnetic field $h_y$ along $y$. For an external field $h_y$ that is weak relative to the strength of the easy $xz$-plane anisotropy, the ratio $\omega_0/t_g\propto e^{4V_0/\omega_0}$ is $\gg 1$, and the subspace spanned by $\ket{\circlearrowleft}$ and $\ket{\circlearrowright}$ is well isolated from higher energy levels. Projecting onto this low-energy subspace, one obtains an effective Hamiltonian $H_{\textsc{dw}}$ describing the domain-wall qubit, given by~\textcite{zou2023quantum} 
\begin{equation}
    H_{\textsc{dw}}=\frac{\epsilon}{2}\tau_z-\frac{t_g}{2}\tau_x.
\end{equation}
Here, $\tau_i$ are Pauli matrices in the basis of chirality states ($\tau_z=\ketbra{\circlearrowright}-\ketbra{\circlearrowleft}$), and $\epsilon\propto  h_x$ is a detuning that depends on the strength of the external field along $x$. 

The external magnetic field also creates a net magnetization in the ferrimagnetic nanowire. The modes $X$ and $\Phi$ are consequently coupled under the motion of the domain wall by the spin Berry phase arising from the excess spin in the ferrimagnetic unit cell. This leads to an effective spin-orbit interaction between $X$ and $\Phi$ that can be used to perform operations on the domain-wall qubit by controlling its motion along the racetrack. This spin-orbit interaction causes the domain wall to flip chirality over a distance $\ell_{\mathrm{so}}$ controlled by both the magnetic field $h_y$ along $y$ and the excess spin per unit cell. An effective Hamiltonian accounting for the spatial motion of the domain wall can be derived by starting from the Hamiltonian for $X$ and $\Phi$, projecting onto the qubit subspace spanned by $\ket{\circlearrowleft},\ket{\circlearrowright}$, transforming to a frame co-moving with the domain wall, and eventually projecting onto the ground state of the orbital motion. Assuming $\ell_{\mathrm{so}}$ on the order of a nanometer, this procedure gives~\cite{zou2023quantum}
\begin{equation}
    H_{\textsc{dw}}(t)=\frac{\tilde{\epsilon}(t)}{2}\tau_z-\frac{\tilde{t}_g}{2}\tau_x,
\end{equation}
where $\tilde{\epsilon}(t)=2v(t)/\ell_{\mathrm{so}}+
\epsilon$ and $\tilde{t}_g=t_g e^{-\ell_p^2/\ell_{\mathrm{so}}^2}$ with $\ell_p$ denoting the characteristic length scale of the harmonic-oscillator potential associated with the quantized orbital motion. The motion of the domain wall consequently leads to a suppression of the tunneling rate $t_g$ between chirality states, as well as to an additional term in the detuning that depends on the domain wall velocity $v(t)=\partial_t\mathcal{X}(t)$, where $\mathcal{X}(t)=\langle X\rangle$ is the expectation value of the position operator of the domain wall with respect to the orbital ground state.

Distant spin qubits could then be entangled by taking advantage of the domain wall motion: By allowing the ferrimagnetic racetrack to interact with a quantum-dot spin $\bm{\sigma}$ via the exchange interaction, a time-dependent coupling of the form~\cite{zou2025topological} 
\begin{equation}
    V(t)=J[\mathcal{X}(t)]\tau_z\sigma_x
\end{equation}
can be engineered between the spin qubit and the domain-wall qubit. Notably, the strength of the coupling depends on the spatial position $\mathcal{X}(t)$ of the domain wall. With control over the domain-wall motion, such a coupling could therefore be used to realize a $\sqrt{i\textsc{swap}}$ gate between the domain-wall qubit and a first spin qubit. The domain wall can then be routed towards a second spin qubit, with which it is made to undergo an $i\textsc{swap}$ gate. This sequence of steps generates entanglement between the two spin qubits without requiring a measurement of the domain-wall qubit itself.   

\section{Conclusion}

Over the past quarter century or so, spin qubits have emerged as a leading platform for quantum computing. In addition to having demonstrated significant improvements in coherence times over the years~\cite{stano2022review}, the compatibility of spin qubits with the existing semiconductor industry provides a realistic pathway towards the realization of large-scale qubit arrays~\cite{zwerver2022qubits,neyens2024probing}. As devices scale to larger numbers of qubits, advances in automated tuning and calibration~\cite{teske2019machine,zwolak2020autotuning,ziegler2022toward,schuff2026fully}, as well as machine-learning-driven device characterization~\cite{lennon2019efficiently,chatterjee2021semiconductor,schuff2023identifying,ziegler2023automated,craig2024bridging},  will become increasingly important in ensuring that all qubits are simultaneously tuned to favorable parameter regimes. 

Current research continues to explore a broad range of materials, carrier-types (electron, hole, nucleus), and qubit encodings, all of which come with their own favorable characteristics and challenges. Although the original Loss-Divincenzo qubit~\cite{loss1998quantum} remains the workhorse of the field, alternative qubit encodings and the integration of nuclear spins as long-lived memories may provide complementary advantages, helping to balance coherence, control requirements, and scalability. The need to integrate large amounts of classical electronics into a small region of space also motivates the development of long-range interconnects for linking distant qubits or quantum-processing modules, which could help alleviate the wiring bottleneck (``fan out'') problem as devices scale up~\cite{vandersypen2017interfacing}. Spin shuttling, spin-circuit QED, and super-semi hybrid devices have all demonstrated significant experimental progress over the past 5-10 years and all provide viable avenues towards reaching this goal.

Single- and two-qubit gates with fidelities exceeding the 99\% surface-code threshold have been demonstrated for phosphorus-31 nuclear spins~\cite{madzik2022precision} and electron spins in silicon~\cite{noiri2022fast, xue2022quantum, mills2022two}. More recently, simultaneous single-qubit gates with 99.999\% fidelity have been achieved in a five-qubit device~\cite{wu2025simultaneous}. That being said, the fault-tolerance thresholds typically quoted for, e.g., the surface code~\cite{fowler2012surface} are found under simplified noise models that only provide a first-order approximation of the far more complicated noise environments found in solid-state systems. For instance, it is known that spatially correlated noise can emerge due to two-level fluctuators~\cite{rojas2023spatial}, and although early investigations suggest that such correlated noise does not destroy the suppression of logical errors with increasing code distance~\cite{rojas2026scaling}, convincingly demonstrating that spin qubits have passed the error-correction threshold will eventually require that these gates be used to extend the lifetime of a logical qubit in the presence of this more complicated noise. An additional challenge facing spin qubits in this respect is that state preparation and measurement (SPAM) errors tend to exceed gate errors; the longer duration of measurements relative to gates also sets a lower bound on the speed at which error correction can be applied. 

Overall, the rapid ongoing progress in the field of spin qubits highlights the strong potential of this platform for semiconductor-based quantum computing. Continued advances in materials engineering, device fabrication, and the theoretical understanding of these systems, together with advances in adjacent areas like quantum error correction, are expected to lead to further improvements in qubit coherence, control, and noise resilience. Building on all of the progress in the field, some of which was covered in this review, it is likely that exciting developments are still to come.

\vspace{2ex}
\noindent\textit{Acknowledgements}---This work was supported as part of NCCR SPIN, a National Center of Competence in Research, funded by the Swiss National Science Foundation (grant number 225153). D.L.~acknowledges the Deanship of Research and the Quantum Center at KFUPM for the support received under Grant no.~CUP25102 and no.~INQC2600, respectively.

%apsrmp4-2.bst 2018-12-27 (MD) hand-edited version of apsrmp4-1.bst
%Control: author (75) reversed first initials jnrlst

\end{document}